\def\ftmagnification{1200}
\def\spacingNumerator{5}
\def\spacingDenominator{4}

\def\ifundefined#1{\expandafter\ifx\csname#1\endcsname\relax}
\ifundefined{ftmagnification}  \def\ftmagnification{1200} \fi
\ifundefined{spacingNumerator}  \def\spacingNumerator{5} \fi
\ifundefined{spacingDenominator}  \def\spacingDenominator{4} \fi


\magnification\ftmagnification
\tolerance=10000
\hsize=17truecm\vsize=23truecm

\parindent=40pt
\mathsurround=0pt
     \multiply\baselineskip by \spacingNumerator
     \divide \baselineskip by \spacingDenominator 

%
%
\def\today{\ifcase\month\or January\or February\or March\or April\or
     May\or June\or July\or August\or September\or October\or November\or
     December\fi\space\number\day, \number\year}
%
%
\def\dst{\displaystyle}
\def\sst{\scriptstyle}
\def\tst{\textstyle}
%
%
\def\frac#1#2{\dst {#1\over#2}}     
\def\sfrac#1#2{{\tst{#1\over#2}}}   

\def\deqalign#1{\vcenter{\openup1\jot \mathsurround=0pt \ialign{
                \strut\hfil$\displaystyle{##}$&&$\displaystyle{{}##}$\hfil
                \crcr
                #1\crcr}}}         


%
%

\def\ga{\gamma}
\def\de{\delta}
\def\ep{\epsilon}
\def\ze{\zeta}

\def\la{\lambda}

\def\si{\sigma}

\def\om{\omega}
\def\Ga{\Gamma}

\def\Si{\Sigma}

\def\Om{\Omega}   
%
%
\def\pmb#1{\setbox0=\hbox{#1}       
     \kern-.025em\copy0\kern-\wd0
     \kern.05em\copy0\kern-\wd0
     \kern-.025em\box0}             
\def\0{{\bf 0}}

\def\k{{\bf k}}

\def\t{{\bf t}}

\def\q{{\bf q}}
\def\x{{\bf x}}
\def\y{{\bf y}}

\def\p{{\bf p}}

\def\cB{{\cal B}}
\def\cE{{\cal E}}

\def\cG{{\cal G}}

\def\cO{{\cal O}}

%
%
\font\tenfrak                 = eufm10
\font\sevenfrak               = eufm7
\font\fivefrak                = eufb5
\newfam\frakfam
     \textfont\frakfam=\tenfrak
     \scriptfont\frakfam=\sevenfrak   
     \scriptscriptfont\frakfam=\fivefrak
\def\frak{\fam\frakfam\tenfrak}
\font \tensans                = cmss10
\font \fivesans               = cmss10 at 5pt
\font \sevensans              = cmss10 at 7pt
\newfam\sansfam
     \textfont\sansfam=\tensans
     \scriptfont\sansfam=\sevensans
     \scriptscriptfont\sansfam=\fivesans
\def\sans{\fam\sansfam\tensans}
%
%
\def\bbbr{{\rm I\!R}}  
\def\bbbn{{\rm I\!N}}

\def\bbbc{{\mathchoice {\setbox0=\hbox{$\displaystyle\rm C$}\hbox{\hbox 
to0pt{\kern0.4\wd0\vrule height0.9\ht0\hss}\box0}}
{\setbox0=\hbox{$\textstyle\rm C$}\hbox{\hbox
to0pt{\kern0.4\wd0\vrule height0.9\ht0\hss}\box0}}
{\setbox0=\hbox{$\scriptstyle\rm C$}\hbox{\hbox
to0pt{\kern0.4\wd0\vrule height0.9\ht0\hss}\box0}}
{\setbox0=\hbox{$\scriptscriptstyle\rm C$}\hbox{\hbox
to0pt{\kern0.4\wd0\vrule height0.9\ht0\hss}\box0}}}}
\def\bbbq{{\mathchoice {\setbox0=\hbox{$\displaystyle\rm               
Q$}\hbox{\raise
0.15\ht0\hbox to0pt{\kern0.4\wd0\vrule height0.8\ht0\hss}\box0}}
{\setbox0=\hbox{$\textstyle\rm Q$}\hbox{\raise
0.15\ht0\hbox to0pt{\kern0.4\wd0\vrule height0.8\ht0\hss}\box0}}
{\setbox0=\hbox{$\scriptstyle\rm Q$}\hbox{\raise
0.15\ht0\hbox to0pt{\kern0.4\wd0\vrule height0.7\ht0\hss}\box0}}
{\setbox0=\hbox{$\scriptscriptstyle\rm Q$}\hbox{\raise
0.15\ht0\hbox to0pt{\kern0.4\wd0\vrule height0.7\ht0\hss}\box0}}}}
\def\bbbz{{\mathchoice {\hbox{$\sans\textstyle Z\kern-0.4em Z$}}       
{\hbox{$\sans\textstyle Z\kern-0.4em Z$}}
{\hbox{$\sans\scriptstyle Z\kern-0.3em Z$}}
{\hbox{$\sans\scriptscriptstyle Z\kern-0.2em Z$}}}}
%
%
\def\const{{\rm const}\,}

\def\half{\sfrac{1}{2}}

\def\optbar#1{\vbox{\ialign{##\crcr\hfil${\scriptscriptstyle(}\mkern -1mu
         \vrule height 1.2pt width 3pt depth -.8pt
         {\scriptscriptstyle)}$\hfil\crcr
          \noalign{\kern-1pt\nointerlineskip}$\hfil\displaystyle{#1}\hfil$\crcr}}}
\def\<{\left<}
\def\>{\right>}

\def\smprod{\mathop{\textstyle\prod}}
\def\smsum{\mathop{\textstyle\sum}}
\def\set#1#2{\big\{ \ #1\ \big|\ #2\ \big\}}
\def\eval#1{\big|\lower4pt\hbox{$\displaystyle\sst #1$}}
%
%
\font \tafontt                = cmbx10 scaled\magstep2
\font \tbfontt                = cmbx10 scaled\magstep1
\def\titlea#1{\centerline{\tafontt #1 }\vskip.5truein}
\def\titleb#1{\removelastskip\vskip.3truein%
\noindent{\tbfontt #1 }\vskip.25truein}

%
%
\def\newenvironment#1#2#3#4{\long\def#1##1##2{%
\removelastskip\penalty-100\vskip\baselineskip%
\noindent{#3#2\if!##1!.\else\unskip\ \ignorespaces
##1\unskip\fi\ }{#4\ignorespaces##2\vskip\baselineskip}}}
\newenvironment\lemma{Lemma}{\bf}{\it}
\newenvironment\proposition{Proposition}{\bf}{\it}
\newenvironment\theorem{Theorem}{\bf}{\it}
\newenvironment\corollary{Corollary}{\bf}{\it}
\newenvironment\example{Example}{\bf}{\rm}
\newenvironment\problem{Problem}{\bf}{\rm}
\newenvironment\definition{Definition}{\bf}{\rm}
\newenvironment\remark{Remark}{\bf}{\rm}
\newenvironment\hypothesis{Hypothesis}{\bf}{\it}
\newenvironment\convention{Convention}{\bf}{\it}

\def\Item{\vskip.1in\noindent}

%
%
\long\def\proof#1{\removelastskip\penalty-100\vskip\baselineskip\noindent{\bf
            Proof\if!#1!\else\ \ignorespaces#1\fi:\ }\ \ \ignorespaces}
\long\def\prf{\removelastskip\penalty-100\vskip\baselineskip\noindent{\bf
            Proof:\ }\ \ \ignorespaces}
\def\endproof{\hfill\vrule height .6em width .6em depth 0pt\goodbreak\vskip.25in }

\ifundefined{warnForwardRef}  \def\warnForwardRef{n} \fi
\newcount\chapno
\newcount\sectno
\newcount\equano
\newcount\theono
\newcount\probno

\def\IgNoRe#1{}

\chapno=0
\sectno=0
\equano=0
\theono=0
\probno=0
\def\eqhead{}
\def\frefwarning{\if\warnForwardRef y\immediate\write16{   Forward reference on line \the\inputlineno}\fi}
\def\qqqrefwarning{\immediate\write16{   ??? reference on line \the\inputlineno}}

\def\chap#1{\equano=0\sectno=0\theono=0\probno=0\global\advance\chapno by 1%
\def\eqhead{\ifcase\chapno\or I\or II\or III\or IV\or V\or VI\or VII\or
VIII\or IX\or X\or XI\or XII\or XIII\or XIV\or XV\or XVI\or XVII\or XVIII\or
XIX\or XX\or XXI\or XXII\or XXIII\or XXIV\or XXV\or XXVI\or XXVII\or XXVIII\or XXIX\or XXX\or XXXI\or XXXII\or XXXIII\or XXXIV\or XXXV\or XXXVI\or XXXVII\or XXXVIII\or XXXIX\fi.}%
\titlea{\eqhead \hglue 5pt #1}%
}

\def\sect#1{\global\advance\sectno by 1%
\titleb{\eqhead\number\sectno  \hglue 5pt #1}%
}%

\def\appendix#1#2{\equano=0\sectno=0\theono=0\probno=0\def\eqhead{#1.}
\titlea{Appendix #1: #2}%
}

\def\:#1{\def\temp{\expandafter\IgNoRe\string#1}%
\expandafter\ifx\csname\temp\endcsname\relax%
\expandafter\gdef#1{\qqqrefwarning ???}\fi#1}

\def\Eqn{{\hbox{\global\advance\equano by 1}}%
\eqno ({\rm \eqhead\number\equano})}%

\def\Eqno{{\hbox{\global\advance\equano by 1}}%
 ({\rm \eqhead\number\equano})}%

\def\EQN#1{\Eqn\edef\Zwi{\eqhead\number\equano}%
\global\let #1=\Zwi
}

\def\EQNO#1{\Eqno\edef\Zwi{\eqhead\number\equano}%
\global\let #1=\Zwi
}

\def\STM#1{{\global\advance \theono by 1}%
\eqhead\number\theono
\edef\Zwi{\eqhead\number\theono }
\global\let#1=\Zwi
}

\def\PRB#1{{\global\advance \probno by 1}%
\eqhead\number\probno
\edef\Zwi{\eqhead\number\probno }
\global\let#1=\Zwi
}

\def\PG#1{\def\Zwi{\number\pageno }
\global\let#1=\Zwi
}

\def\Stm{{\global\advance \theono by 1}%
\eqhead\number\theono
}

\def\Prb{{\global\advance \probno by 1}%
\eqhead\number\probno
}

\def\EDEF#1#2{
\def\tEmP{#1}\expandafter\gdef\tEmP{#2}
}



\def\suffix{ps}
\newcount\system
\global\system=3   

\def\ifundefined#1{\expandafter\ifx\csname#1\endcsname\relax}
\ifundefined{figdir}\def\figdir{}\fi
%
%
\newcount\firstline
\newdimen\pswidth  \newdimen\xleft
\newdimen\psheight \newdimen\ytop \newdimen\ybot
\newcount\justx \newcount\justy
\global\justx=0 \global\justy=0
\newdimen\vpos \newtoks\labeL 
\newread\labeLfile \newdimen\xcoord \newdimen\ycoord
\newif\ifdoit 
\newbox\labox
\newdimen\xdvikwid 
\newdimen\xdvikht
\newdimen\pspoints
\newdimen\rwi
\pspoints=1bp
\newcount\temp
\def\readdim#1{\global\read\labeLfile to \temp
\global #1=\temp pt}
%
%
%
%
\def\figcrop#1{\par
\openin\labeLfile=\figdir#1.lbl                                              
\global\read\labeLfile to\firstline\message{#1}               
\global\read\labeLfile to\temp
\readdim{\ybot}
\readdim{\xleft}
\readdim{\ytop}
\global\read\labeLfile to\justx
\global\read\labeLfile to\justy
\global\read\labeLfile to\labeL
\readdim{\pswidth}
\global\advance\pswidth by -\xleft
\readdim{\psheight}
\global\advance\ybot by -\psheight
\global\advance\psheight by -\ytop
\global\read\labeLfile to\justx
\global\read\labeLfile to\justy
\global\read\labeLfile to\labeL
\vbox to\psheight{\vfill
\ifnum\system=1
\fi 
\ifnum\system=2
\fi 
\ifnum\system=3
                                                 \fi         
\ifnum\system=4
\fi         
\ifnum\system=1
\hbox to \pswidth{\kern-\xleft\special{postscriptfile \figdir#1.\suffix }\hfil}\fi
\ifnum\system=2
\hbox to \pswidth{\kern-\xleft\special{ps: plotfile \figdir#1.\suffix }\hfil}\fi
\ifnum\system=3
\hbox to \pswidth{\kern-\xleft\includegraphics{\figdir#1.\suffix}\hfil}\fi
\ifnum\system=4
\hbox to \pswidth{\kern-\xleft\includegraphics{\figdir#1.\suffix}\hfil}\fi
\ifnum\system=5
\hbox to \pswidth{\kern-\xleft\includegraphics{\figdir#1.\suffix}\hfil}\fi 
\ifnum\system=6
   \xdvikwid=\pswidth
   \xdvikht=\psheight
   {\global\divide\xdvikwid by \pspoints}
   {\global\divide\xdvikht by \pspoints}
   \rwi=\xdvikwid
    {\global\multiply\rwi by 10}
\hbox to \pswidth{\kern-\xleft\includegraphics{\figdir#1.\suffix\space}\hfil}\fi                   
\vskip -\baselineskip
\vskip -\ybot 
\vskip-\psheight %
\hbox to\pswidth  {\hss}%
\parindent=0pt\offinterlineskip                                       
\vpos=0 pt%
\loop\readdim{\xcoord}                                 
\ifdim \xcoord < -999pt \doitfalse\else\doittrue\fi                        
\ifdoit \advance \xcoord by -\xleft
\readdim{\ycoord}
\advance \ycoord by -\ytop                              
\global\read\labeLfile to\justx                                       
\global\read\labeLfile to\justy                                       
\global\read\labeLfile to\labeL
\global\setbox\labox=\hbox{\labeL\hskip-0.3em}%
\advance\vpos by-\ycoord                                              
\vskip-\vpos \vpos=\ycoord                                         
\hbox to\pswidth{\hskip\xcoord %
\hbox to 0pt{\ifnum\justx>0\hss\fi%
\vbox to0pt{%
\ifnum\justy<2\vss\fi%
\copy\labox\kern0pt%
\ifnum\justy>0\vss\fi}%
\ifnum\justx<2\hss\fi}%
\hss}%
\repeat%
\advance\vpos by-\psheight%
\vskip-\vpos %
}\closein\labeLfile}
%
%
%
\def\figplace#1#2#3{
\openin\labeLfile=\figdir#1.lbl
\ifeof \labeLfile
       \immediate\write16{***Can't find \figdir#1.lbl; Skipping it.***}
\else  \closein\labeLfile
       \null\hskip#2\raise #3 \hbox{\figcrop{#1}}
\fi
}
%
%
%
%
\def\figput#1{
\openin\labeLfile=\figdir#1.lbl
\ifeof \labeLfile
       \immediate\write16{***Can't find \figdir#1.lbl; Skipping it.***}
\else  \closein\labeLfile
       \hbox{\figcrop{#1}}
\fi
}


\font\tenscript                 = pzcmi at 10pt
\font\sevenscript               = pzcmi at 7pt
\font\fivescript                = pzcmi at 5pt
\newfam\scriptfam
     \textfont\scriptfam=\tenscript
     \scriptfont\scriptfam=\sevenscript   
     \scriptscriptfont\scriptfam=\fivescript
\def\script{\fam\scriptfam\tenscript}

    \newenvironment\notation{Notation}{\bf}{\rm}

          \def\stoday{\number\day\space\ifcase\month\or Jan\or Feb\or 
                      Mar\or Apr\or May\or Jun\or Jul\or Aug\or Sep\or 
                      Oct\or Nov\or Dec\fi, \number\year}

         \def\squiggle{\raise2pt\hbox{${\scriptstyle\sim}$}}

    \def\form{{\ssst\script form}}
    \def\il{\jbar}
    \def\IR{{\rm\ssst IR}}
    \def\ren{{\rm ren}}
    \def\dunion{\cup\kern-0.7em\cdot\kern0.45em}
    
    \def\ib{{\rm b}}
    \def\tb{{b}}
    
    \def\vi{\fl^{\raise2pt\hbox{$\scriptscriptstyle{1/n_0}$}}}
    \def\tanV{\vec {\rm t}}
    \def\normV{\vec {\rm n}}
    
    \def\inp{{\rm in}}
    \def\out{{\rm out}}


    \def\abcst{{\sst const}}

     \def\ssst{\scriptscriptstyle}
     \def\bde{{\mathchoice{\pmb{$\de$}}{\pmb{$\de$}}
                              {\pmb{$\sst\de$}}{\pmb{$\ssst\de$}}}}
    \def\jbar{{\mathchoice
                   {{\smash{\lower1ex\hbox{$\mathchar'26$}}\mkern-9mu j}}
                   {{\smash{\lower1ex\hbox{$\mathchar'26$}}\mkern-9mu j}}
                   {{\smash{\lower1.2ex\hbox{$\mathchar'26$}}\mkern-10.2mu j}}
                   {{\smash{\lower1.2ex\hbox{$\mathchar'26$}}\mkern-10.2mu j}}}}

    \def\cD{{\cal D}}
    \def\cK{{\cal K}}

    \def\cU{{\cal U}}
    \def\cV{{\cal V}}
    \def\cW{{\cal W}}

    \def\fl{{\frak l}}

    \def\fK{{\frak K}}

     \def\tv{\kern8pt\tilde{\kern-8pt\pmb{$\vert$}}}
     \def\tV{\kern8pt\tilde{\kern-8pt\pmb{$\big\vert$}}}
     \def\tVV{\kern8pt\tilde{\kern-8pt\pmb{$\Big\vert$}}}

    \def\tn{|\kern-1pt|\kern-1pt|}
    \def\TN{\big|\kern-1.5pt\big|\kern-1.5pt\big|}
    \def\TTN{\Big|\kern-2pt\Big|\kern-2pt\Big|}

    \def\trn{|\kern-1pt|\kern-1pt|^{\,\tilde{\,}}}
    \def\TRN{\big|\kern-1.5pt\big|\kern-1.5pt\big|^{\,\tilde{\,}}}
    \def\TTRN{\Big|\kern-2pt\Big|\kern-2pt\Big|^{\,\tilde{\,}}}

     \def\tnorm{\kern8pt\tilde{\kern-8pt\|}}
     \def\Tnorm{\kern8pt\tilde{\kern-8pt\big\|}}
     \def\TNorm{\kern8pt\tilde{\kern-8pt\Big\|}}
     \def\TNOrm{\kern8pt\tilde{\kern-8pt\bigg\|}}

    \def\rw{\mathclose{:}}
    \def\lw{\mathopen{:}}
    \def\lW{\mathopen{{\tst{\hbox{.}\atop\raise 2.5pt\hbox{.}}}}}
    \def\rW{\mathclose{{\tst{{.}\atop\raise 2.5pt\hbox{.}}}}}
    \def\lww{\mathopen{{\tst{\raise 1pt\hbox{.}\atop\raise 1pt\hbox{.}}}}}
    \def\rww{\mathclose{{\tst{\raise 1pt\hbox{.}\atop\raise 1pt\hbox{.}}}}}

   \font\sixrm=cmr6   \font\eightrm=cmr8  
   \font\sixi=cmmi6   \font\eighti=cmmi8  
  \font\sixsy=cmsy6  \font\eightsy=cmsy8 
  \font\sixbf=cmbx6  \font\eightbf=cmbx8 
                     \font\eightit=cmti8 
                     \font\eightsl=cmsl8 
                     \font\eighttt=cmtt8 

\font\eightfrak=eufm7 at 8pt

\def\eightpoint{\def\rm{\fam0\eightrm}
 \textfont0=\eightrm \scriptfont0=\sixrm \scriptscriptfont0=\fiverm
 \textfont1=\eighti \scriptfont1=\sixi \scriptscriptfont1=\fivei
 \textfont2=\eightsy \scriptfont2=\sixsy \scriptscriptfont2=\fivesy
 \textfont3=\tenex \scriptfont3=\tenex \scriptscriptfont3=\tenex
 \textfont\itfam=\eightit \def\it{\fam\itfam\eightit}%
 \textfont\slfam=\eightsl \def\sl{\fam\slfam\eightsl}%
 \textfont\ttfam=\eighttt \def\tt{\fam\ttfam\eighttt}%
 \textfont\frakfam=\eightfrak \def\frak{\fam\frakfam\tenfrak}%
 \textfont\bffam=\eightbf \scriptfont\bffam=\sixbf
 \scriptscriptfont\bffam=\fivebf \def\bf{\fam\bffam\eightbf}%
 \normalbaselineskip=9pt
 \setbox\strutbox=\hbox{\vrule height7pt depth2pt width0pt}%
 \let\sc=\sixrm \let\big=\eightbig \normalbaselines\rm}
\catcode`@=11
\def\footnote#1{\edef\@sf{\spacefactor\the\spacefactor}#1\@sf
     \insert\footins\bgroup\eightpoint
     \interlinepenalty100 \let\par=\endgraf
     \leftskip=0pt \rightskip=0pt
     \splittopskip=10pt plus 1pt minus 1pt \floatingpenalty=20000
     \smallskip\item{#1}\bgroup\strut\aftergroup\@foot\let\next}
\skip\footins=12pt plus 2pt minus 4pt
\dimen\footins=30pc
\catcode`@=12


  \IgNoRe{PG}
  \IgNoRe{STM Assertion }
  \IgNoRe{PG}
  \IgNoRe{PG}
  \IgNoRe{STM Assertion }
  \IgNoRe{PG}
  \IgNoRe{STM Assertion }
  \IgNoRe{STM Assertion }
  \IgNoRe{EQN}
  \IgNoRe{STM Assertion }
  \IgNoRe{STM Assertion }
  \IgNoRe{PG}
  \IgNoRe{STM Assertion }
  \IgNoRe{STM Assertion }
  \IgNoRe{EQN}
  \IgNoRe{STM Assertion }
  \IgNoRe{STM Assertion }
  \IgNoRe{STM Assertion }
  \IgNoRe{STM Assertion }
  \IgNoRe{STM Assertion }
  \IgNoRe{STM Assertion }
  \IgNoRe{PG}
  \IgNoRe{STM Assertion }
  \IgNoRe{STM Assertion }
  \IgNoRe{STM Assertion }
  \IgNoRe{STM Assertion }
  \IgNoRe{STM Assertion }
  \IgNoRe{STM Assertion }
  \IgNoRe{STM Assertion }
  \IgNoRe{STM Assertion }
  \IgNoRe{STM Assertion }
  \IgNoRe{STM Assertion }
  \IgNoRe{STM Assertion }
  \IgNoRe{STM Assertion }
  \IgNoRe{PG}
  \IgNoRe{EQN}
  \IgNoRe{STM Assertion }
  \IgNoRe{STM Assertion }
  \IgNoRe{STM Assertion }
  \IgNoRe{PG}
  \IgNoRe{STM Assertion }
  \IgNoRe{STM Assertion }
  \IgNoRe{STM Assertion }
  \IgNoRe{STM Assertion }
  \IgNoRe{EQN}
  \IgNoRe{EQN}
  \IgNoRe{STM Assertion }
  \IgNoRe{PG}
  \IgNoRe{PG}
  \IgNoRe{STM Assertion }
  \IgNoRe{EQN}
  \IgNoRe{STM Assertion }
  \IgNoRe{STM Assertion }
  \IgNoRe{STM Assertion }
  \IgNoRe{EQN}
  \IgNoRe{STM Assertion }
  \IgNoRe{STM Assertion }
  \IgNoRe{STM Assertion }
  \IgNoRe{PG}
  \IgNoRe{STM Assertion }
  \IgNoRe{STM Assertion }
  \IgNoRe{PG}
  \IgNoRe{STM Assertion }
  \IgNoRe{STM Assertion }
  \IgNoRe{STM Assertion }
  \IgNoRe{PG}
  \IgNoRe{STM Assertion }
  \IgNoRe{STM Assertion }
  \IgNoRe{STM Assertion }
  \IgNoRe{STM Assertion }
  \IgNoRe{STM Assertion }
  \IgNoRe{STM Assertion }
  \IgNoRe{STM Assertion }
 \def\propBII{\frefwarning A.2} \IgNoRe{STM Assertion }
  \IgNoRe{PG}
  \IgNoRe{STM Assertion }
 \def\lemBIV{\frefwarning A.4} \IgNoRe{STM Assertion }
  \IgNoRe{STM Assertion }
  \IgNoRe{STM Assertion }
  \IgNoRe{STM Assertion }
  \IgNoRe{STM Assertion }
  \IgNoRe{STM Assertion }
  \IgNoRe{STM Assertion }
  \IgNoRe{PG}
  \IgNoRe{STM Assertion }
  \IgNoRe{STM Assertion }
  \IgNoRe{PG}
  \IgNoRe{PG}
  \IgNoRe{STM Assertion }
  \IgNoRe{STM Assertion }
  \IgNoRe{PG}
  \IgNoRe{PG}
  \IgNoRe{STM Assertion }
  \IgNoRe{STM Assertion }
  \IgNoRe{EQN}
  \IgNoRe{STM Assertion }
  \IgNoRe{PG}
  \IgNoRe{STM Assertion }
 \def\remtheotheo{\frefwarning VI.7} \IgNoRe{STM Assertion }
  \IgNoRe{STM Assertion }
  \IgNoRe{PG}
  \IgNoRe{STM Assertion }
  \IgNoRe{STM Assertion }
  \IgNoRe{STM Assertion }
  \IgNoRe{EQN}
  \IgNoRe{STM Assertion }
  \IgNoRe{PG}
  \IgNoRe{EQN}
  \IgNoRe{STM Assertion }
  \IgNoRe{STM Assertion }
  \IgNoRe{STM Assertion }
  \IgNoRe{STM Assertion }
  \IgNoRe{EQN}
  \IgNoRe{EQN}
  \IgNoRe{PG}
  \IgNoRe{PG}
  \IgNoRe{STM Assertion }
  \IgNoRe{STM Assertion }
  \IgNoRe{STM Assertion }
  \IgNoRe{STM Assertion }
  \IgNoRe{STM Assertion }
  \IgNoRe{STM Assertion }
  \IgNoRe{STM Assertion }
  \IgNoRe{STM Assertion }
  \IgNoRe{PG}
  \IgNoRe{STM Assertion }
  \IgNoRe{STM Assertion }
  \IgNoRe{STM Assertion }
  \IgNoRe{STM Assertion }
  \IgNoRe{STM Assertion }
  \IgNoRe{STM Assertion }
  \IgNoRe{STM Assertion }
  \IgNoRe{STM Assertion }
  \IgNoRe{STM Assertion }
  \IgNoRe{STM Assertion }
  \IgNoRe{PG}
  \IgNoRe{STM Assertion }
  \IgNoRe{STM Assertion }
  \IgNoRe{STM Assertion }
  \IgNoRe{PG}
  \IgNoRe{STM Assertion }
  \IgNoRe{STM Assertion }
  \IgNoRe{STM Assertion }
  \IgNoRe{STM Assertion }
  \IgNoRe{PG}
  \IgNoRe{STM Assertion }
  \IgNoRe{STM Assertion }
  \IgNoRe{PG}
  \IgNoRe{STM Assertion }
  \IgNoRe{PG}
  \IgNoRe{STM Assertion }
  \IgNoRe{PG}
  \IgNoRe{STM Assertion }
  \IgNoRe{STM Assertion }
  \IgNoRe{STM Assertion }
  \IgNoRe{STM Assertion }
  \IgNoRe{PG}
  \IgNoRe{PG}
  \IgNoRe{STM Assertion }
  \IgNoRe{STM Assertion }
  \IgNoRe{EQN}
  \IgNoRe{EQN}
  \IgNoRe{STM Assertion }
  \IgNoRe{STM Assertion }
  \IgNoRe{STM Assertion }
  \IgNoRe{STM Assertion }
  \IgNoRe{PG}
  \IgNoRe{STM Assertion }
  \IgNoRe{STM Assertion }
  \IgNoRe{STM Assertion }
  \IgNoRe{STM Assertion }
  \IgNoRe{STM Assertion }
  \IgNoRe{STM Assertion }
  \IgNoRe{STM Assertion }
  \IgNoRe{PG}
  \IgNoRe{STM Assertion }
  \IgNoRe{EQN}
  \IgNoRe{EQN}
  \IgNoRe{PG}
  \IgNoRe{STM Assertion }
  \IgNoRe{EQN}
  \IgNoRe{STM Assertion }
  \IgNoRe{STM Assertion }
  \IgNoRe{STM Assertion }
  \IgNoRe{PG}
  \IgNoRe{STM Assertion }
  \IgNoRe{EQN}
  \IgNoRe{STM Assertion }
  \IgNoRe{PG}
  \IgNoRe{PG}


  \IgNoRe{PG}
  \IgNoRe{PG}
 \def\lemLADprimitivemanfred{\frefwarning I.1} \IgNoRe{STM Assertion }
  \IgNoRe{EQN}
  \IgNoRe{STM Assertion }
  \IgNoRe{PG}
  \IgNoRe{STM Assertion }
  \IgNoRe{EQN}
  \IgNoRe{STM Assertion }
  \IgNoRe{EQN}
  \IgNoRe{STM Assertion }
  \IgNoRe{STM Assertion }
  \IgNoRe{STM Assertion }
  \IgNoRe{STM Assertion }
  \IgNoRe{PG}
  \IgNoRe{STM Assertion }
  \IgNoRe{STM Assertion }
  \IgNoRe{STM Assertion }
  \IgNoRe{STM Assertion }
  \IgNoRe{PG}
  \IgNoRe{STM Assertion }
  \IgNoRe{STM Assertion }
  \IgNoRe{EQN}
  \IgNoRe{STM Assertion }
  \IgNoRe{STM Assertion }
  \IgNoRe{STM Assertion }
  \IgNoRe{PG}
  \IgNoRe{PG}
  \IgNoRe{STM Assertion }
  \IgNoRe{EQN}
  \IgNoRe{STM Assertion }
  \IgNoRe{STM Assertion }
  \IgNoRe{PG}
  \IgNoRe{STM Assertion }
  \IgNoRe{STM Assertion }

  \IgNoRe{STM Assertion }
  \IgNoRe{STM Assertion }
  \IgNoRe{PG}
  \IgNoRe{PG}
  \IgNoRe{STM Assertion }
  \IgNoRe{STM Assertion }
  \IgNoRe{STM Assertion }
  \IgNoRe{STM Assertion }
  \IgNoRe{STM Assertion }
  \IgNoRe{STM Assertion }
  \IgNoRe{PG}
  \IgNoRe{STM Assertion }
  \IgNoRe{STM Assertion }
  \IgNoRe{STM Assertion }
  \IgNoRe{STM Assertion }
  \IgNoRe{STM Assertion }
  \IgNoRe{STM Assertion }
  \IgNoRe{EQN}
  \IgNoRe{PG}
  \IgNoRe{STM Assertion }
  \IgNoRe{STM Assertion }
  \IgNoRe{STM Assertion }
  \IgNoRe{EQN}
  \IgNoRe{STM Assertion }
  \IgNoRe{STM Assertion }
  \IgNoRe{STM Assertion }
  \IgNoRe{PG}
  \IgNoRe{STM Assertion }
  \IgNoRe{STM Assertion }
  \IgNoRe{STM Assertion }
  \IgNoRe{STM Assertion }
  \IgNoRe{EQN}
  \IgNoRe{EQN}
  \IgNoRe{EQN}
  \IgNoRe{EQN}
  \IgNoRe{STM Assertion }
  \IgNoRe{STM Assertion }
  \IgNoRe{EQN}
  \IgNoRe{STM Assertion }
  \IgNoRe{EQN}
  \IgNoRe{EQN}
  \IgNoRe{EQN}
  \IgNoRe{EQN}
  \IgNoRe{EQN}
  \IgNoRe{PG}
  \IgNoRe{STM Assertion }
  \IgNoRe{STM Assertion }
  \IgNoRe{EQN}

  \IgNoRe{EQN}
  \IgNoRe{EQN}
  \IgNoRe{PG}
  \IgNoRe{EQN}
  \IgNoRe{EQN}
  \IgNoRe{STM Assertion }
  \IgNoRe{STM Assertion }
  \IgNoRe{EQN}
  \IgNoRe{EQN}
  \IgNoRe{EQN}
  \IgNoRe{EQN}
  \IgNoRe{STM Assertion }
  \IgNoRe{STM Assertion }
  \IgNoRe{STM Assertion }
  \IgNoRe{STM Assertion }
  \IgNoRe{STM Assertion }
  \IgNoRe{STM Assertion }
  \IgNoRe{STM Assertion }
  \IgNoRe{STM Assertion }
  \IgNoRe{STM Assertion }
  \IgNoRe{STM Assertion }
  \IgNoRe{STM Assertion }
  \IgNoRe{STM Assertion }
  \IgNoRe{EQN}
  \IgNoRe{EQN}
  \IgNoRe{EQN}
  \IgNoRe{EQN}
  \IgNoRe{STM Assertion }
  \IgNoRe{EQN}
  \IgNoRe{STM Assertion }
  \IgNoRe{EQN}
  \IgNoRe{EQN}
  \IgNoRe{EQN}
  \IgNoRe{EQN}
  \IgNoRe{EQN}
  \IgNoRe{STM Assertion }
  \IgNoRe{STM Assertion }
  \IgNoRe{STM Assertion }
  \IgNoRe{STM Assertion }
  \IgNoRe{STM Assertion }
  \IgNoRe{EQN}
  \IgNoRe{EQN}
  \IgNoRe{STM Assertion }
  \IgNoRe{STM Assertion }
  \IgNoRe{EQN}
  \IgNoRe{EQN}
  \IgNoRe{STM Assertion }
  \IgNoRe{EQN}
  \IgNoRe{STM Assertion }
  \IgNoRe{STM Assertion }
  \IgNoRe{EQN}
  \IgNoRe{STM Assertion }
  \IgNoRe{EQN}
  \IgNoRe{EQN}
  \IgNoRe{EQN}
  \IgNoRe{EQN}
  \IgNoRe{STM Assertion }
  \IgNoRe{PG}
  \IgNoRe{STM Assertion }
  \IgNoRe{STM Assertion }
  \IgNoRe{PG}
  \IgNoRe{STM Assertion }

  \IgNoRe{PG}
  \IgNoRe{STM Assertion }
  \IgNoRe{EQN}
  \IgNoRe{EQN}
  \IgNoRe{EQN}
  \IgNoRe{STM Assertion }
  \IgNoRe{STM Assertion }
  \IgNoRe{STM Assertion }
  \IgNoRe{STM Assertion }
  \IgNoRe{EQN}
  \IgNoRe{STM Assertion }
  \IgNoRe{EQN}
  \IgNoRe{EQN}
  \IgNoRe{EQN}
  \IgNoRe{STM Assertion }
  \IgNoRe{STM Assertion }
  \IgNoRe{STM Assertion }
  \IgNoRe{EQN}

  \IgNoRe{STM Assertion }
  \IgNoRe{PG}
  \IgNoRe{STM Assertion }
  \IgNoRe{EQN}
  \IgNoRe{EQN}
  \IgNoRe{STM Assertion }
  \IgNoRe{EQN}

  \IgNoRe{STM Assertion }
  \IgNoRe{EQN}
  \IgNoRe{EQN}
  \IgNoRe{PG}
  \IgNoRe{EQN}
  \IgNoRe{EQN}
  \IgNoRe{EQN}
  \IgNoRe{EQN}
  \IgNoRe{EQN}
  \IgNoRe{STM Assertion }
  \IgNoRe{EQN}
  \IgNoRe{EQN}
  \IgNoRe{EQN}
  \IgNoRe{EQN}
  \IgNoRe{EQN}
  \IgNoRe{STM Assertion }
  \IgNoRe{EQN}
  \IgNoRe{EQN}
  \IgNoRe{EQN}
  \IgNoRe{EQN}
  \IgNoRe{EQN}
  \IgNoRe{EQN}
  \IgNoRe{EQN}
  \IgNoRe{EQN}
  \IgNoRe{STM Assertion }

  \IgNoRe{STM Assertion }
  \IgNoRe{PG}
  \IgNoRe{EQN}
  \IgNoRe{EQN}
  \IgNoRe{STM Assertion }
  \IgNoRe{EQN}
  \IgNoRe{EQN}
  \IgNoRe{STM Assertion }
  \IgNoRe{EQN}
  \IgNoRe{STM Assertion }
  \IgNoRe{EQN}
  \IgNoRe{EQN}
  \IgNoRe{PG}
  \IgNoRe{PG}


  \IgNoRe{PG}
  \IgNoRe{EQN}
  \IgNoRe{STM Assertion }
  \IgNoRe{PG}
  \IgNoRe{STM Assertion }
  \IgNoRe{STM Assertion }
  \IgNoRe{STM Assertion }
  \IgNoRe{STM Assertion }
  \IgNoRe{STM Assertion }
  \IgNoRe{STM Assertion }
  \IgNoRe{EQN}
  \IgNoRe{STM Assertion }
  \IgNoRe{STM Assertion }
  \IgNoRe{STM Assertion }
  \IgNoRe{STM Assertion }
  \IgNoRe{STM Assertion }
  \IgNoRe{STM Assertion }
  \IgNoRe{STM Assertion }
  \IgNoRe{STM Assertion }
  \IgNoRe{PG}
  \IgNoRe{STM Assertion }
  \IgNoRe{STM Assertion }
  \IgNoRe{STM Assertion }
  \IgNoRe{STM Assertion }
  \IgNoRe{STM Assertion }
  \IgNoRe{STM Assertion }
  \IgNoRe{STM Assertion }
  \IgNoRe{STM Assertion }
  \IgNoRe{EQN}
  \IgNoRe{PG}
  \IgNoRe{PG}
  \IgNoRe{STM Assertion }
  \IgNoRe{STM Assertion }
  \IgNoRe{STM Assertion }
  \IgNoRe{STM Assertion }
  \IgNoRe{STM Assertion }
  \IgNoRe{STM Assertion }
  \IgNoRe{PG}
  \IgNoRe{EQN}
  \IgNoRe{EQN}
  \IgNoRe{EQN}
  \IgNoRe{EQN}
  \IgNoRe{EQN}
  \IgNoRe{EQN}
  \IgNoRe{STM Assertion }
  \IgNoRe{STM Assertion }
  \IgNoRe{STM Assertion }
  \IgNoRe{STM Assertion }
  \IgNoRe{PG}
  \IgNoRe{STM Assertion }
  \IgNoRe{EQN}
  \IgNoRe{STM Assertion }
  \IgNoRe{STM Assertion }
  \IgNoRe{STM Assertion }
  \IgNoRe{STM Assertion }
  \IgNoRe{PG}
  \IgNoRe{STM Assertion }
  \IgNoRe{STM Assertion }
  \IgNoRe{STM Assertion }
  \IgNoRe{STM Assertion }
  \IgNoRe{PG}
  \IgNoRe{EQN}
  \IgNoRe{EQN}
  \IgNoRe{PG}
  \IgNoRe{STM Assertion }
  \IgNoRe{STM Assertion }
 \def\lemOStworengrpmaps{\frefwarning VII.3} \IgNoRe{STM Assertion }
  \IgNoRe{EQN}
  \IgNoRe{PG}
  \IgNoRe{STM Assertion }
  \IgNoRe{STM Assertion }
  \IgNoRe{EQN}
  \IgNoRe{STM Assertion }
  \IgNoRe{STM Assertion }
  \IgNoRe{STM Assertion }
  \IgNoRe{STM Assertion }
  \IgNoRe{PG}
  \IgNoRe{STM Assertion }
  \IgNoRe{STM Assertion }
  \IgNoRe{STM Assertion }
  \IgNoRe{STM Assertion }
  \IgNoRe{STM Assertion }
  \IgNoRe{STM Assertion }
  \IgNoRe{STM Assertion }
  \IgNoRe{STM Assertion }
  \IgNoRe{PG}
  \IgNoRe{STM Assertion }
  \IgNoRe{STM Assertion }
  \IgNoRe{STM Assertion }
  \IgNoRe{STM Assertion }
  \IgNoRe{STM Assertion }
  \IgNoRe{EQN}
  \IgNoRe{STM Assertion }
  \IgNoRe{STM Assertion }
  \IgNoRe{PG}
  \IgNoRe{STM Assertion }
  \IgNoRe{STM Assertion }
  \IgNoRe{STM Assertion }
  \IgNoRe{STM Assertion }
  \IgNoRe{STM Assertion }
  \IgNoRe{STM Assertion }
  \IgNoRe{STM Assertion }
  \IgNoRe{STM Assertion }
  \IgNoRe{STM Assertion }
  \IgNoRe{EQN}
  \IgNoRe{STM Assertion }
  \IgNoRe{STM Assertion }
  \IgNoRe{PG}
  \IgNoRe{STM Assertion }
  \IgNoRe{STM Assertion }
  \IgNoRe{STM Assertion }
 \def\remOSrengrppreserves{\frefwarning B.5} \IgNoRe{STM Assertion }
  \IgNoRe{STM Assertion }
  \IgNoRe{STM Assertion }
 \def\lemOSappGrassII{\frefwarning C.2} \IgNoRe{STM Assertion }
  \IgNoRe{PG}
  \IgNoRe{STM Assertion }
  \IgNoRe{PG}
  \IgNoRe{PG}
  \IgNoRe{STM Assertion }
  \IgNoRe{STM Assertion }
  \IgNoRe{PG}
 \def\lemOSsectpartunit{\frefwarning XII.3} \IgNoRe{STM Assertion }
  \IgNoRe{STM Assertion }
  \IgNoRe{STM Assertion }
  \IgNoRe{STM Assertion }
  \IgNoRe{STM Assertion }
  \IgNoRe{STM Assertion }
  \IgNoRe{STM Assertion }
  \IgNoRe{STM Assertion }
  \IgNoRe{STM Assertion }
  \IgNoRe{STM Assertion }
  \IgNoRe{STM Assertion }
  \IgNoRe{STM Assertion }
  \IgNoRe{STM Assertion }
  \IgNoRe{STM Assertion }
  \IgNoRe{STM Assertion }
  \IgNoRe{STM Assertion }
  \IgNoRe{EQN}
  \IgNoRe{STM Assertion }
 \def\propOSGenDecay{\frefwarning XIII.1} \IgNoRe{STM Assertion }
  \IgNoRe{PG}
 \def\lemOSsectorderiv{\frefwarning XIII.2} \IgNoRe{STM Assertion }
  \IgNoRe{EQN}
 \def\eqnOSpartunit{\frefwarning XIII.2} \IgNoRe{EQN}
  \IgNoRe{STM Assertion }
  \IgNoRe{EQN}
  \IgNoRe{EQN}
  \IgNoRe{STM Assertion }
  \IgNoRe{EQN}
  \IgNoRe{STM Assertion }
  \IgNoRe{EQN}
  \IgNoRe{STM Assertion }
  \IgNoRe{STM Assertion }
  \IgNoRe{STM Assertion }
  \IgNoRe{STM Assertion }
  \IgNoRe{PG}
  \IgNoRe{STM Assertion }
  \IgNoRe{STM Assertion }
  \IgNoRe{STM Assertion }
  \IgNoRe{STM Assertion }
  \IgNoRe{EQN}
  \IgNoRe{EQN}
  \IgNoRe{STM Assertion }
  \IgNoRe{STM Assertion }
  \IgNoRe{PG}
  \IgNoRe{STM Assertion }
  \IgNoRe{STM Assertion }
  \IgNoRe{STM Assertion }
  \IgNoRe{EQN}
  \IgNoRe{STM Assertion }
  \IgNoRe{EQN}
  \IgNoRe{STM Assertion }
  \IgNoRe{STM Assertion }
  \IgNoRe{EQN}
  \IgNoRe{STM Assertion }
  \IgNoRe{EQN}
  \IgNoRe{EQN}
  \IgNoRe{EQN}
  \IgNoRe{EQN}
  \IgNoRe{STM Assertion }
  \IgNoRe{STM Assertion }
  \IgNoRe{STM Assertion }
  \IgNoRe{STM Assertion }
  \IgNoRe{PG}
  \IgNoRe{STM Assertion }
  \IgNoRe{STM Assertion }
  \IgNoRe{STM Assertion }
  \IgNoRe{STM Assertion }
  \IgNoRe{STM Assertion }
  \IgNoRe{STM Assertion }
  \IgNoRe{STM Assertion }
  \IgNoRe{STM Assertion }
  \IgNoRe{STM Assertion }
  \IgNoRe{STM Assertion }
  \IgNoRe{EQN}
  \IgNoRe{EQN}
  \IgNoRe{STM Assertion }
  \IgNoRe{PG}
  \IgNoRe{STM Assertion }
  \IgNoRe{STM Assertion }
  \IgNoRe{EQN}
  \IgNoRe{STM Assertion }
  \IgNoRe{STM Assertion }
  \IgNoRe{EQN}
  \IgNoRe{EQN}
  \IgNoRe{EQN}
  \IgNoRe{EQN}
  \IgNoRe{EQN}
  \IgNoRe{STM Assertion }
  \IgNoRe{STM Assertion }
  \IgNoRe{STM Assertion }
  \IgNoRe{EQN}
  \IgNoRe{EQN}
  \IgNoRe{EQN}
  \IgNoRe{EQN}
  \IgNoRe{EQN}
  \IgNoRe{EQN}
  \IgNoRe{STM Assertion }
  \IgNoRe{STM Assertion }
  \IgNoRe{STM Assertion }
  \IgNoRe{PG}
  \IgNoRe{STM Assertion }
  \IgNoRe{STM Assertion }
  \IgNoRe{EQN}
  \IgNoRe{EQN}
  \IgNoRe{STM Assertion }
  \IgNoRe{EQN}
  \IgNoRe{EQN}
  \IgNoRe{STM Assertion }
  \IgNoRe{EQN}
  \IgNoRe{EQN}
  \IgNoRe{STM Assertion }
  \IgNoRe{STM Assertion }
  \IgNoRe{PG}
  \IgNoRe{STM Assertion }
  \IgNoRe{STM Assertion }
  \IgNoRe{STM Assertion }
  \IgNoRe{PG}
 \def\remModII{\frefwarning XVIII.4} \IgNoRe{STM Assertion }
 \def\propOSthreetoonenorm{\frefwarning XIX.1} \IgNoRe{STM Assertion }
  \IgNoRe{STM Assertion }
  \IgNoRe{STM Assertion }
  \IgNoRe{PG}
 \def\propOSresectorI{\frefwarning XIX.4} \IgNoRe{STM Assertion }
  \IgNoRe{STM Assertion }
  \IgNoRe{STM Assertion }
  \IgNoRe{STM Assertion }
  \IgNoRe{STM Assertion }
  \IgNoRe{STM Assertion }
  \IgNoRe{STM Assertion }
  \IgNoRe{STM Assertion }
  \IgNoRe{STM Assertion }
  \IgNoRe{STM Assertion }
  \IgNoRe{STM Assertion }
  \IgNoRe{STM Assertion }
  \IgNoRe{STM Assertion }
  \IgNoRe{STM Assertion }
  \IgNoRe{PG}
  \IgNoRe{STM Assertion }
  \IgNoRe{PG}
  \IgNoRe{STM Assertion }
  \IgNoRe{STM Assertion }
  \IgNoRe{PG}
  \IgNoRe{STM Assertion }
  \IgNoRe{STM Assertion }
  \IgNoRe{EQN}
  \IgNoRe{PG}
  \IgNoRe{EQN}
  \IgNoRe{STM Assertion }
  \IgNoRe{STM Assertion }
  \IgNoRe{PG}
  \IgNoRe{EQN}
  \IgNoRe{EQN}
  \IgNoRe{EQN}
  \IgNoRe{EQN}
  \IgNoRe{EQN}
  \IgNoRe{STM Assertion }
  \IgNoRe{STM Assertion }
  \IgNoRe{EQN}
  \IgNoRe{STM Assertion }
  \IgNoRe{PG}
  \IgNoRe{PG}
  \IgNoRe{PG}
  \IgNoRe{STM Assertion }
  \IgNoRe{EQN}
  \IgNoRe{STM Assertion }
  \IgNoRe{PG}
  \IgNoRe{STM Assertion }
  \IgNoRe{EQN}
  \IgNoRe{STM Assertion }
  \IgNoRe{STM Assertion }
  \IgNoRe{PG}
  \IgNoRe{EQN}
  \IgNoRe{EQN}
  \IgNoRe{EQN}
  \IgNoRe{STM Assertion }
  \IgNoRe{STM Assertion }
  \IgNoRe{STM Assertion }
  \IgNoRe{EQN}
  \IgNoRe{STM Assertion }
  \IgNoRe{STM Assertion }
  \IgNoRe{STM Assertion }
  \IgNoRe{STM Assertion }
  \IgNoRe{STM Assertion }
  \IgNoRe{STM Assertion }
  \IgNoRe{PG}
  \IgNoRe{STM Assertion }
  \IgNoRe{STM Assertion }
  \IgNoRe{STM Assertion }
  \IgNoRe{STM Assertion }
  \IgNoRe{STM Assertion }
  \IgNoRe{STM Assertion }
  \IgNoRe{STM Assertion }
  \IgNoRe{PG}
  \IgNoRe{PG}


\newcount\CHAPNO
\newcount\APPNO
\CHAPNO=0
\APPNO=1
\def\advCHAPNO{\advance\CHAPNO by 1}
\def\advAPPNO{\advance\APPNO by 1}

\def\caproman#1{\ifcase#1\or I\or II\or III\or IV\or V\or VI\or VII\or
VIII\or IX\or X\or XI\or XII\or XIII\or XIV\or XV\or XVI\or XVII\or XVIII\or
XIX\or XX\or XXI\or XXII\or XXIII\or XXIV\or XXV\or XXVI\or XXVII\or XXVIII\or XXIX\or XXX\or XXXI\or XXXII\or XXXIII\or XXXIV\or XXXV\or XXXVI\or XXXVII\or XXXVIII\or XXXIX\fi}%

\def\capletter#1{\ifcase#1\or A\or B\or C\or D\or E\or F\or G\or
H\or I\or J\or K\or L\or M\or N\or O\or P\or Q\or R\or
S\or T\or U\or V\or W\or X\or Y\or Z\fi}%

\newcount\cHintroI \cHintroI=\CHAPNO \advCHAPNO 
                              \edef\CHintroI{\caproman\CHAPNO}
\newcount\cHnorms  \cHnorms=\CHAPNO \advCHAPNO 
                              
\newcount\cHproprengrp \cHproprengrp=\CHAPNO \advCHAPNO 
                              
\newcount\cHcovbounds  \cHcovbounds=\CHAPNO \advCHAPNO 
                              
\newcount\cHinsulator \cHinsulator=\CHAPNO \advCHAPNO

 \advAPPNO

\newcount\cHintroII \cHintroII=\CHAPNO \advCHAPNO 
                              \edef\CHintroII{\caproman\CHAPNO}
\newcount\cHamputate \cHamputate=\CHAPNO \advCHAPNO
                              
\newcount\cHscales \cHscales=\CHAPNO \advCHAPNO
                              \edef\CHscales{\caproman\CHAPNO}
\newcount\cHfourier \cHfourier=\CHAPNO \advCHAPNO
                              
\newcount\cHmomentum \cHmomentum=\CHAPNO \advCHAPNO

\edef\APappSymmetries{\capletter\APPNO} \advAPPNO
 \advAPPNO

\newcount\cHintroIII \cHintroIII=\CHAPNO \advCHAPNO
                              \edef\CHintroIII{\caproman\CHAPNO}
\newcount\cHsectors \cHsectors=\CHAPNO \advCHAPNO
                              
\newcount\cHsecpropbounds \cHsecpropbounds=\CHAPNO \advCHAPNO
                              
\newcount\cHladdersNotn  \cHladdersNotn=\CHAPNO \advCHAPNO
                              
\newcount\cHestren  \cHestren=\CHAPNO \advCHAPNO
                              
\newcount\cHsecmomnorm \cHsecmomnorm=\CHAPNO \advCHAPNO
                              
\newcount\cHmomestren \cHmomestren=\CHAPNO \advCHAPNO

 \advAPPNO

\newcount\cHintroIV  \cHintroIV=\CHAPNO \advCHAPNO
                              
\newcount\cHcomparison   \cHcomparison=\CHAPNO \advCHAPNO
                              
\newcount\cHsumsmom  \cHsumsmom=\CHAPNO \advCHAPNO
                              
\newcount\cHsectorsmom   \cHsectorsmom=\CHAPNO \advCHAPNO
                              
\newcount\cHppladsect    \cHppladsect=\CHAPNO \advCHAPNO

 \advAPPNO


 \def\defNPCTMSpace{\frefwarning I.1} \IgNoRe{STM Assertion }
 \def\pgNPI{\frefwarning 1} \IgNoRe{PG}
 \def\eqnNPreal{\frefwarning I.1} \IgNoRe{EQN}
 \def\eqnNPphexchange{\frefwarning I.2} \IgNoRe{EQN}
 \def\eqnNPinteraction{\frefwarning I.3} \IgNoRe{EQN}
 \def\eqnNPforgenfn{\frefwarning I.4} \IgNoRe{EQN}
 \def\eqnNPphijpsi{\frefwarning I.5} \IgNoRe{EQN}
 \def\defNPscales{\frefwarning I.2} \IgNoRe{STM Assertion }
 \def\remNPlargej{\frefwarning I.3} \IgNoRe{STM Assertion }
 \def\theoremNPmainthI{\frefwarning I.4} \IgNoRe{STM Assertion }
 \def\theoremNPmainthII{\frefwarning I.5} \IgNoRe{STM Assertion }
 \def\remNPmainthII{\frefwarning I.6} \IgNoRe{STM Assertion }
 \def\theoremNPmainthIII{\frefwarning I.7} \IgNoRe{STM Assertion }
 \def\remNPmainthIII{\frefwarning I.8} \IgNoRe{STM Assertion }
 \def\remNPmainthI{\frefwarning I.9} \IgNoRe{STM Assertion }
 \def\defNPstrongasymm{\frefwarning I.10} \IgNoRe{STM Assertion }
 \def\remModII{\frefwarning I.11} \IgNoRe{STM Assertion }
 \def\hypNPdisprel{\frefwarning I.12} \IgNoRe{STM Assertion }
 \def\pgNPII{\frefwarning 11} \IgNoRe{PG}
 \def\pgNPIIa{\frefwarning 11} \IgNoRe{PG}
 \def\pgNPIIb{\frefwarning 12} \IgNoRe{PG}
 \def\pgNPIIc{\frefwarning 12} \IgNoRe{PG}
 \def\eqnOVlinfty{\frefwarning II.1} \IgNoRe{EQN}
 \def\eqnOVshellvol{\frefwarning II.2} \IgNoRe{EQN}
 \def\eqnOVvertexbnd{\frefwarning II.3} \IgNoRe{EQN}
 \def\eqnOVtripleintersection{\frefwarning II.4} \IgNoRe{EQN}
 \def\pgNPIId{\frefwarning 14} \IgNoRe{PG}
 \def\pgNPIIe{\frefwarning 16} \IgNoRe{PG}
 \def\pgNPIIf{\frefwarning 17} \IgNoRe{PG}
 \def\eqnOVabvertexnorm{\frefwarning II.5} \IgNoRe{EQN}
 \def\pgNPIIg{\frefwarning 18} \IgNoRe{PG}
 \def\pgNPIIh{\frefwarning 20} \IgNoRe{PG}
 \def\eqnOVtriponesi{\frefwarning II.6} \IgNoRe{EQN}
 \def\eqnOVsectcontrI{\frefwarning II.7} \IgNoRe{EQN}
 \def\eqnOVsectcontrII{\frefwarning II.8} \IgNoRe{EQN}
 \def\eqnOVsectcontrIII{\frefwarning II.9} \IgNoRe{EQN}
 \def\eqnOVsectcontrbnd{\frefwarning II.10} \IgNoRe{EQN}
 \def\eqnOVsectorvertexreq{\frefwarning II.11} \IgNoRe{EQN}
 \def\eqnOVsecmap{\frefwarning II.12} \IgNoRe{EQN}
 \def\eqnOVsecnmbr{\frefwarning II.13} \IgNoRe{EQN}
 \def\eqnOVtripthreesi{\frefwarning II.14} \IgNoRe{EQN}
 \def\eqnOVsectorvertexthreereq{\frefwarning II.15} \IgNoRe{EQN}
 \def\pgNPIIi{\frefwarning 25} \IgNoRe{PG}
 \def\pgNPIIj{\frefwarning 30} \IgNoRe{PG}
 \def\eqnNPcovFT{\frefwarning III.1} \IgNoRe{EQN}
 \def\eqnNPantisymmCov{\frefwarning III.2} \IgNoRe{EQN}
 \def\eqnNPjdef{\frefwarning III.3} \IgNoRe{EQN}
 \def\defNPrengroupmap{\frefwarning III.1} \IgNoRe{STM Assertion }
 \def\pgNPIII{\frefwarning 32} \IgNoRe{PG}
 \def\eqnNPsemigrp{\frefwarning III.4} \IgNoRe{EQN}
 \def\eqnNPtworengrp{\frefwarning III.5} \IgNoRe{EQN}
 \def\eqnNPrginduct{\frefwarning III.6} \IgNoRe{EQN}
 \def\eqnNPrginductpsi{\frefwarning III.7} \IgNoRe{EQN}
 \def\defNPformalfinalCTmSpace{\frefwarning III.2} \IgNoRe{STM Assertion }
 \def\defNPformalscaleCTmSpace{\frefwarning III.3} \IgNoRe{STM Assertion }
 \def\defNPeffinttriple{\frefwarning III.4} \IgNoRe{STM Assertion }
 \def\defNPformalCovariances{\frefwarning III.5} \IgNoRe{STM Assertion }
 \def\defNPinoutmap{\frefwarning III.6} \IgNoRe{STM Assertion }
 \def\remNPinoutmap{\frefwarning III.7} \IgNoRe{STM Assertion }
 \def\defNPinputData{\frefwarning III.8} \IgNoRe{STM Assertion }
 \def\defNPoutputData{\frefwarning III.9} \IgNoRe{STM Assertion }
 \def\lemNPinOut{\frefwarning III.10} \IgNoRe{STM Assertion }
 \def\eqnNPoutin{\frefwarning III.8} \IgNoRe{EQN}
 \def\eqnNPinouttriple{\frefwarning III.9} \IgNoRe{EQN}
 \def\eqnNPgenfnrengrp{\frefwarning III.10} \IgNoRe{EQN}
 \def\eqnNPKuprime{\frefwarning III.11} \IgNoRe{EQN}
 \def\lemNPformalselfconsistent{\frefwarning III.11} \IgNoRe{STM Assertion }
 \def\eqnNPformalselfconsistent{\frefwarning III.12} \IgNoRe{EQN}
 \def\propNPoutIn{\frefwarning III.12} \IgNoRe{STM Assertion }
 \def\eqnOutII{\frefwarning III.13} \IgNoRe{EQN}
 \def\exNPsectorizebound{\frefwarning A.1} \IgNoRe{STM Assertion }
 \def\pgNPA{\frefwarning 44} \IgNoRe{PG}
 \def\eqnSMPlone{\frefwarning A.1} \IgNoRe{EQN}
 \def\exNPloopsector{\frefwarning A.2} \IgNoRe{STM Assertion }
 \def\exNPsectorizescalechange{\frefwarning A.3} \IgNoRe{STM Assertion }
 \def\eqnSMPresector{\frefwarning A.2} \IgNoRe{EQN}
 \def\pgNPIref{\frefwarning 50} \IgNoRe{PG}
  \IgNoRe{PG}
  \IgNoRe{STM Assertion }
 \def\defNPFancynormdomain{\frefwarning V.2} \IgNoRe{STM Assertion }
  \IgNoRe{PG}
  \IgNoRe{STM Assertion }
  \IgNoRe{STM Assertion }
  \IgNoRe{STM Assertion }
  \IgNoRe{STM Assertion }
  \IgNoRe{STM Assertion }
  \IgNoRe{STM Assertion }
  \IgNoRe{STM Assertion }
 \def\defNPfourtrans{\frefwarning VI.1} \IgNoRe{STM Assertion }
  \IgNoRe{PG}
 \def\defNPsectors{\frefwarning VI.2} \IgNoRe{STM Assertion }
  \IgNoRe{STM Assertion }
  \IgNoRe{STM Assertion }
 \def\defNPtens{\frefwarning VI.5} \IgNoRe{STM Assertion }
  \IgNoRe{EQN}
  \IgNoRe{STM Assertion }
 \def\defNPsectGrnorm{\frefwarning VI.7} \IgNoRe{STM Assertion }
  \IgNoRe{STM Assertion }
  \IgNoRe{EQN}
  \IgNoRe{STM Assertion }
  \IgNoRe{STM Assertion }
  \IgNoRe{STM Assertion }
  \IgNoRe{STM Assertion }
  \IgNoRe{PG}
  \IgNoRe{STM Assertion }
  \IgNoRe{STM Assertion }
  \IgNoRe{STM Assertion }
  \IgNoRe{STM Assertion }
  \IgNoRe{STM Assertion }
  \IgNoRe{STM Assertion }
 \def\defcompLadder{\frefwarning VII.7} \IgNoRe{STM Assertion }
 \def\theoremcompLadder{\frefwarning VII.8} \IgNoRe{STM Assertion }
  \IgNoRe{STM Assertion }
  \IgNoRe{EQN}
  \IgNoRe{EQN}
  \IgNoRe{EQN}
  \IgNoRe{STM Assertion }
  \IgNoRe{PG}
  \IgNoRe{STM Assertion }
  \IgNoRe{STM Assertion }
 \def\theoremNPinduction{\frefwarning VIII.5} \IgNoRe{STM Assertion }
  \IgNoRe{EQN}
  \IgNoRe{EQN}
  \IgNoRe{EQN}
  \IgNoRe{STM Assertion }
  \IgNoRe{STM Assertion }
  \IgNoRe{EQN}
  \IgNoRe{EQN}
  \IgNoRe{EQN}
  \IgNoRe{STM Assertion }
  \IgNoRe{PG}
  \IgNoRe{PG}
  \IgNoRe{STM Assertion }
  \IgNoRe{STM Assertion }
  \IgNoRe{PG}
  \IgNoRe{STM Assertion }
  \IgNoRe{STM Assertion }
  \IgNoRe{EQN}
  \IgNoRe{EQN}
  \IgNoRe{EQN}
  \IgNoRe{EQN}
  \IgNoRe{EQN}
  \IgNoRe{EQN}
  \IgNoRe{EQN}
  \IgNoRe{EQN}
  \IgNoRe{EQN}
  \IgNoRe{EQN}
  \IgNoRe{EQN}
  \IgNoRe{EQN}
  \IgNoRe{EQN}
  \IgNoRe{EQN}
  \IgNoRe{STM Assertion }
  \IgNoRe{EQN}
  \IgNoRe{PG}
  \IgNoRe{EQN}
  \IgNoRe{STM Assertion }
  \IgNoRe{EQN}
  \IgNoRe{STM Assertion }
  \IgNoRe{STM Assertion }
  \IgNoRe{EQN}
  \IgNoRe{EQN}
  \IgNoRe{EQN}
  \IgNoRe{STM Assertion }
  \IgNoRe{EQN}
  \IgNoRe{EQN}
  \IgNoRe{EQN}
  \IgNoRe{EQN}
  \IgNoRe{EQN}
  \IgNoRe{EQN}
  \IgNoRe{EQN}
  \IgNoRe{EQN}
  \IgNoRe{EQN}
  \IgNoRe{EQN}
  \IgNoRe{EQN}
  \IgNoRe{EQN}
  \IgNoRe{EQN}
  \IgNoRe{EQN}
  \IgNoRe{EQN}
  \IgNoRe{EQN}
  \IgNoRe{EQN}
  \IgNoRe{EQN}
  \IgNoRe{EQN}
  \IgNoRe{EQN}
  \IgNoRe{EQN}
  \IgNoRe{STM Assertion }
  \IgNoRe{PG}
  \IgNoRe{PG}
  \IgNoRe{STM Assertion }
  \IgNoRe{PG}
  \IgNoRe{EQN}
  \IgNoRe{EQN}
  \IgNoRe{EQN}
  \IgNoRe{EQN}
  \IgNoRe{STM Assertion }
  \IgNoRe{EQN}
  \IgNoRe{PG}
  \IgNoRe{EQN}
  \IgNoRe{EQN}
  \IgNoRe{EQN}
  \IgNoRe{EQN}
  \IgNoRe{EQN}
  \IgNoRe{EQN}
  \IgNoRe{EQN}
  \IgNoRe{EQN}
  \IgNoRe{EQN}
  \IgNoRe{EQN}
  \IgNoRe{EQN}
  \IgNoRe{EQN}
  \IgNoRe{STM Assertion }
  \IgNoRe{STM Assertion }
  \IgNoRe{EQN}
  \IgNoRe{EQN}
  \IgNoRe{PG}
  \IgNoRe{PG}
  \IgNoRe{STM Assertion }
  \IgNoRe{EQN}
  \IgNoRe{STM Assertion }
  \IgNoRe{PG}
  \IgNoRe{EQN}
  \IgNoRe{EQN}
  \IgNoRe{EQN}
  \IgNoRe{STM Assertion }
 \def\lemTNPabstractjump{\frefwarning XII.4} \IgNoRe{STM Assertion }
  \IgNoRe{EQN}
  \IgNoRe{EQN}
  \IgNoRe{EQN}
  \IgNoRe{EQN}
  \IgNoRe{EQN}
  \IgNoRe{STM Assertion }
  \IgNoRe{EQN}
  \IgNoRe{EQN}
  \IgNoRe{EQN}
  \IgNoRe{EQN}
  \IgNoRe{STM Assertion }
  \IgNoRe{STM Assertion }
  \IgNoRe{EQN}
  \IgNoRe{STM Assertion }
  \IgNoRe{STM Assertion }
  \IgNoRe{STM Assertion }
  \IgNoRe{STM Assertion }
  \IgNoRe{PG}
  \IgNoRe{STM Assertion }
  \IgNoRe{STM Assertion }
  \IgNoRe{STM Assertion }
  \IgNoRe{STM Assertion }
  \IgNoRe{STM Assertion }
  \IgNoRe{STM Assertion }
  \IgNoRe{STM Assertion }
  \IgNoRe{STM Assertion }
  \IgNoRe{STM Assertion }
  \IgNoRe{STM Assertion }
  \IgNoRe{STM Assertion }
  \IgNoRe{STM Assertion }
  \IgNoRe{STM Assertion }
  \IgNoRe{STM Assertion }
  \IgNoRe{STM Assertion }
  \IgNoRe{PG}
  \IgNoRe{STM Assertion }
  \IgNoRe{STM Assertion }
  \IgNoRe{STM Assertion }
  \IgNoRe{STM Assertion }
  \IgNoRe{STM Assertion }
  \IgNoRe{STM Assertion }
  \IgNoRe{STM Assertion }
  \IgNoRe{STM Assertion }
  \IgNoRe{STM Assertion }
  \IgNoRe{STM Assertion }
  \IgNoRe{STM Assertion }
  \IgNoRe{STM Assertion }
  \IgNoRe{EQN}
  \IgNoRe{STM Assertion }
  \IgNoRe{STM Assertion }
  \IgNoRe{STM Assertion }
  \IgNoRe{STM Assertion }
  \IgNoRe{STM Assertion }
  \IgNoRe{STM Assertion }
  \IgNoRe{EQN}
  \IgNoRe{STM Assertion }
  \IgNoRe{PG}
  \IgNoRe{PG}
  \IgNoRe{STM Assertion }
  \IgNoRe{STM Assertion }
  \IgNoRe{STM Assertion }
  \IgNoRe{EQN}
  \IgNoRe{STM Assertion }
  \IgNoRe{PG}
  \IgNoRe{EQN}
  \IgNoRe{STM Assertion }
  \IgNoRe{STM Assertion }
  \IgNoRe{EQN}
  \IgNoRe{EQN}
  \IgNoRe{EQN}
  \IgNoRe{EQN}
  \IgNoRe{EQN}
  \IgNoRe{EQN}
  \IgNoRe{EQN}
  \IgNoRe{EQN}
  \IgNoRe{EQN}
  \IgNoRe{EQN}
  \IgNoRe{EQN}
  \IgNoRe{EQN}
  \IgNoRe{EQN}
  \IgNoRe{EQN}
  \IgNoRe{EQN}
  \IgNoRe{EQN}
  \IgNoRe{EQN}
  \IgNoRe{EQN}
  \IgNoRe{EQN}
  \IgNoRe{EQN}
  \IgNoRe{STM Assertion }
  \IgNoRe{STM Assertion }
  \IgNoRe{PG}
  \IgNoRe{EQN}
  \IgNoRe{EQN}
  \IgNoRe{EQN}
  \IgNoRe{EQN}
  \IgNoRe{EQN}
  \IgNoRe{EQN}
  \IgNoRe{EQN}
  \IgNoRe{EQN}
  \IgNoRe{EQN}
  \IgNoRe{EQN}
  \IgNoRe{EQN}
  \IgNoRe{EQN}
  \IgNoRe{EQN}
  \IgNoRe{EQN}
  \IgNoRe{EQN}
  \IgNoRe{EQN}
  \IgNoRe{EQN}
  \IgNoRe{EQN}
  \IgNoRe{EQN}
  \IgNoRe{EQN}
  \IgNoRe{EQN}
  \IgNoRe{STM Assertion }
  \IgNoRe{PG}
  \IgNoRe{STM Assertion }
  \IgNoRe{STM Assertion }
  \IgNoRe{EQN}
  \IgNoRe{EQN}
  \IgNoRe{PG}
  \IgNoRe{EQN}
  \IgNoRe{EQN}
  \IgNoRe{EQN}
  \IgNoRe{EQN}
  \IgNoRe{EQN}
  \IgNoRe{EQN}
  \IgNoRe{EQN}
  \IgNoRe{EQN}
  \IgNoRe{STM Assertion }
  \IgNoRe{STM Assertion }
  \IgNoRe{STM Assertion }
  \IgNoRe{STM Assertion }
  \IgNoRe{STM Assertion }
  \IgNoRe{PG}
 \def\pgNPInot{\frefwarning 53} \IgNoRe{PG}
  \IgNoRe{PG}
  \IgNoRe{PG}


\newcount\CHAPNO
\newcount\APPNO
\CHAPNO=0
\APPNO=1
\def\advCHAPNO{\advance\CHAPNO by 1}
\def\advAPPNO{\advance\APPNO by 1}

\def\caproman#1{\ifcase#1\or I\or II\or III\or IV\or V\or VI\or VII\or
VIII\or IX\or X\or XI\or XII\or XIII\or XIV\or XV\or XVI\or XVII\or XVIII\or
XIX\or XX\or XXI\or XXII\or XXIII\or XXIV\or XXV\or XXVI\or XXVII\or XXVIII\or XXIX\or XXX\or XXXI\or XXXII\or XXXIII\or XXXIV\or XXXV\or XXXVI\or XXXVII\or XXXVIII\or XXXIX\fi}%

\def\capletter#1{\ifcase#1\or A\or B\or C\or D\or E\or F\or G\or
H\or I\or J\or K\or L\or M\or N\or O\or P\or Q\or R\or
S\or T\or U\or V\or W\or X\or Y\or Z\fi}%

\newcount\cHintroI \cHintroI=\CHAPNO \advCHAPNO 
                              \edef\CHintroI{\caproman\CHAPNO}         
\newcount\cHintroOverview  \cHintroOverview=\CHAPNO \advCHAPNO 
                              \edef\CHintroOverview{\caproman\CHAPNO}  
\newcount\cHrenmap \cHrenmap=\CHAPNO \advCHAPNO 
                              \edef\CHrenmap{\caproman\CHAPNO}         

\edef\APappModelComp{\capletter\APPNO} \advAPPNO

\newcount\cHintroII \cHintroII=\CHAPNO \advCHAPNO 
                              \edef\CHintroII{\caproman\CHAPNO}
\newcount\cHfirstscale \cHfirstscale=\CHAPNO \advCHAPNO
                              
\newcount\cHnewsectors \cHnewsectors=\CHAPNO \advCHAPNO
                              \edef\CHnewsectors{\caproman\CHAPNO}
\newcount\cHphladders \cHphladders=\CHAPNO \advCHAPNO
                              
\newcount\cHfinitescale \cHfinitescale=\CHAPNO \advCHAPNO
                              
\newcount\cHstep \cHstep=\CHAPNO \advCHAPNO
                              
\newcount\cHrecurs \cHrecurs=\CHAPNO \advCHAPNO
                              \edef\CHrecurs{\caproman\CHAPNO}
\edef\APappRewick{\capletter\APPNO} \advAPPNO

\newcount\cHintroIII \cHintroIII=\CHAPNO \advCHAPNO
                              \edef\CHintroIII{\caproman\CHAPNO}
\newcount\cHtildefinitescale \cHtildefinitescale=\CHAPNO \advCHAPNO
                              
\newcount\cHtildenewsectors \cHtildenewsectors=\CHAPNO \advCHAPNO
                              
\newcount\cHtildephladders \cHtildephladders=\CHAPNO \advCHAPNO
                              
\newcount\cHtildestep  \cHtildestep=\CHAPNO \advCHAPNO
                              \edef\CHtildestep{\caproman\CHAPNO}

\edef\APappHoelder{\capletter\APPNO} \advAPPNO
\edef\APappPhladders{\capletter\APPNO} \advAPPNO


{\nopagenumbers
\multiply\baselineskip by \spacingDenominator\divide \baselineskip by\spacingNumerator

\null\vskip3truecm

%
%
\centerline{\tafontt A Two Dimensional Fermi Liquid }
\vskip0.1in
\centerline{\tbfontt Part 1: Overview}

\vskip0.75in
\centerline{Joel Feldman{\parindent=.15in\footnote{$^{*}$}{Research supported 
in part by the
 Natural Sciences and Engineering Research Council of Canada and the Forschungsinstitut f\"ur Mathematik, ETH Z\"urich}}}
\centerline{Department of Mathematics}
\centerline{University of British Columbia}
\centerline{Vancouver, B.C. }
\centerline{CANADA\ \   V6T 1Z2}
\centerline{feldman@math.ubc.ca}
\centerline{http:/\hskip-3pt/www.math.ubc.ca/\squiggle feldman/}
\vskip0.3in
\centerline{Horst Kn\"orrer, Eugene Trubowitz}
\centerline{Mathematik}
\centerline{ETH-Zentrum}
\centerline{CH-8092 Z\"urich}
\centerline{SWITZERLAND}
\centerline{knoerrer@math.ethz.ch, trub@math.ethz.ch}
\centerline{http:/\hskip-3pt/www.math.ethz.ch/\squiggle knoerrer/}

\vskip0.75in
\noindent

{\bf Abstract.\ \ \ } 
In a series of ten papers, of which this is the first, we prove that 
the temperature zero renormalized perturbation expansions of a class 
of interacting many--fermion models in two space dimensions have 
nonzero radius of convergence. The models have ``asymmetric'' Fermi 
surfaces and short range interactions. One consequence of the convergence
of the perturbation expansions is the existence of a discontinuity in the 
particle number density at the Fermi surface. Here, we present a self 
contained formulation of our main results and give an overview of the 
methods used to prove them.

\vfill
\eject


\titleb{Table of Contents}
\halign{\hfill#\ &\hfill#\ &#\hfill&\ p\ \hfil#&\ p\ \hfil#\cr
\noalign{\vskip0.05in}
\S I&\omit Introduction                             \span&\:\pgNPI&\omit\cr
\noalign{\vskip0.05in}
\S II&\omit An Overview                             \span&\:\pgNPII\cr
&&Renormalization of the Fermi Surface and the Dispersion Relation
                                                  &\omit&\:\pgNPIIa\cr
&&Multi Scale Analysis                            &\omit&\:\pgNPIIb\cr
&&Integrating out a Scale                         &\omit&\:\pgNPIIc\cr
&&Overlapping Loops                               &\omit&\:\pgNPIId\cr
&&Particle--Particle Bubbles                      &\omit&\:\pgNPIIe\cr
&&Particle--Hole Ladders                          &\omit&\:\pgNPIIf\cr
&&Power Counting in Position Space                &\omit&\:\pgNPIIg\cr
&&Sectors                                         &\omit&\:\pgNPIIh\cr
&&Cancellation Between Diagrams                   &\omit&\:\pgNPIIi\cr
&&The Counterterm                                 &\omit&\:\pgNPIIj\cr
\noalign{\vskip0.05in}
\S III&\omit Formal Renormalization Group Maps     \span&\:\pgNPIII\cr
\noalign{\vskip0.05in}
{\bf Appendices}\span\cr
\noalign{\vskip0.05in}
\S A&\omit Model Computations                       \span&\:\pgNPA\cr
\noalign{\vskip0.05in}
 &\omit References                                    \span&\:\pgNPIref \cr
\noalign{\vskip0.05in}
 &\omit Notation                                      \span&\:\pgNPInot \cr
}
\vfill\eject
\multiply\baselineskip by \spacingNumerator
\divide \baselineskip by\spacingDenominator}
\pageno=1


\chap{Introduction}\PG\pgNPI

The standard model for a gas of weakly interacting fermions in a $d$-dimensional
crystal at low temperature is given in terms of
\item{$\bullet$} a single particle dispersion relation (shifted by the chemical
potential) $e(\k)$ on $\bbbr^d$,
\item{$\bullet$} an ultraviolet cutoff $U(\k)$ on $\bbbr^d$,
\item{$\bullet$} an interaction $V$.

\noindent
Here $\k$ is the momentum variable dual to the position variable 
$\x\in \bbbr^d$. The Fermi surface associated to the dispersion relation $e(\k)$ is by definition
$$
F=\set{\k\in\bbbr^d}{ e(\k)=0}
$$
The ultraviolet cutoff is a smooth function with compact support that fulfills
$0\le U(\k)\le 1$ for all $\k\in\bbbr^d$. We assume that it is identically one 
on a neighbourhood of the Fermi surface\footnote{$^{(1)}$}
{In particular, we assume that $F$ is compact.}. 

We use renormalization group techniques to show that, for $d=2$ and under the assumptions on the dispersion relation $e(\k)$ specified in Hypotheses \:\hypNPdisprel\ below, such a system is a Fermi
liquid whenever $V$ is small enough (the precise statement
is given in Theorem \:\theoremNPmainthII\ below). Renormalization is necessary 
since the Fermi
surfaces for the noninteracting (that is $V=0$) and interacting systems 
($V \ne 0$) do not, in general, agree. We therefore select ($V$--dependent) counterterms $\de e(\k)$,
from the space in Definition \:\defNPCTMSpace, below, in such a way that 
the Fermi surface of the model with dispersion relation $e(\k)-\de e(\k)$ and
 interaction $V$ is equal to $F$. 

\definition{\STM\defNPCTMSpace}{ 
The space of counterterms, $\cE$, consists of all functions 
$\de e(\k)$ on $\bbbr^d$ that are supported in $\set{\k\in\bbbr^d}{U(\k)=1}$ and
for which the $L^1$--norm of the Fourier transform is finite. That is 
$$
\int d^d\x\ \big| \de e^\wedge(\x) \big| <\infty
$$
where $\,\de e^\wedge(\x) = \int \sfrac{d^d\k}{(2\pi)^d}\, e^{-\imath\k\cdot\x} \,\de e(\k)\,$.
}

The temperature Green's functions at temperature zero (also known as the Euclidean Green's functions)
for this model can be described in field theoretic terms 
using the anticommuting fields
$\psi_\si(x_0,\x),\,\bar \psi_\si(x_0,\x)$, where $x_0\in \bbbr$ is the temperature (or Euclidean time) argument and $\si \in \{\uparrow,\downarrow\}$ is the spin argument. For $x=(x_0,\x,\si)$ we write 
$\psi(x)= \psi_\si(x_0,\x)$ and $\bar \psi(x)=\bar \psi_\si(x_0,\x)$. 

For a model with dispersion relation $e(\k)-\de e(\k)$ and
 interaction $V=0$, the Green's functions are the
moments of  the Grassmann Gaussian measure, 
$d\mu_{C(\de e)}$, whose covariance is the Fourier transform of
$$
C(k_0,\k;\de e) = \frac{U(\k)}{\imath k_0 - e(\k)+\de e(\k)}
$$
Precisely, for  $x=(x_0,\x,\si),\ x'=(x_0',\x',\si') 
\in \bbbr\times\bbbr^d\times\{\uparrow,\downarrow\}$
$$
C(x,x';\de e) = \int \psi(x)\bar\psi(x')\ d\mu_{C(\de e)}(\psi,\bar\psi) 
= \de_{\si,\si'} \int \frac{d^{d+1}k}{(2\pi)^{d+1}} 
e^{\imath<k,x-x'>_-}C(k;\de e)
$$
where $<k,x-x'>_- = -k_0(x_0-x'_0) + \k\cdot(\x-\x')$ for 
$k=(k_0,\k)\in\bbbr\times\bbbr^d$. To simplify notation we set
$$
C(k) = C(k;0) \qquad,\qquad C(x,x')=C(x,x';0)
$$

The interaction between the fermions is determined by the effective potential
$$
\cV(\psi,\bar\psi) = \int_{(\bbbr\times\bbbr^d\times\{\uparrow,\downarrow\})^4} 
\hskip-.7in V(x_1,x_2,x_3,x_4)\, \bar\psi(x_1)\psi(x_2)\bar\psi(x_3)\psi(x_4)\
dx_1dx_2dx_3dx_4
$$
We assume that $V$ is translation invariant and spin independent. For 
some results, we also assume that $V$ obeys
$$
V(R_0x_1,R_0x_2,R_0x_3,R_0x_4)=\overline{V(-x_1,-x_2,-x_3,-x_4)}
\EQN\eqnNPreal$$
and
$$
V(-x_2,-x_1,-x_4,-x_3)=V(x_1,x_2,x_3,x_4)
\EQN\eqnNPphexchange$$
where $R_0(x_0,\x,\si)=(-x_0,\x,\si)$ and $-(x_0,\x,\si)=(-x_0,-\x,\si)$.
We call (\eqnNPreal) ``$k_0$--reversal reality'' and  
(\eqnNPphexchange)``bar/unbar exchange invariance''. Precise definitions and 
a discussion of the properties of these symmetries are given in Appendix 
\APappSymmetries\ of [FKTo2].

In the case of a two--body interaction $v(x_0,\x)$, the interaction kernel is
$$
V({\sst (x_{1,0},\x_1,\si_1),\cdots,(x_{4,0},\x_4,\si_4)})
= -\half \de({\sst x_1,x_2}) \de({\sst x_3,x_4})\de({\sst x_{1,0}-x_{3,0}}) 
v({\sst x_{1,0}-x_{3,0},\x_1-\x_3})
\EQN\eqnNPinteraction$$
where $\de({\sst(x_0,\x,\si),(x_0',\x',\si')})
=\de({\sst x_0-x_0'})\de({\sst \x-\x'})\de_{\si,\si'}$.
If the Fourier transform, $\check v(k_0,\k)$, of the  two--body interaction 
$v(x_0,\x)$ obeys $\check v(-k_0,\k)=\overline{\check v(k_0,\k)}$ , 
then the interaction kernel $V$ has all four symmetries mentioned above.
In addition, $\cV$ always conserves particle number.

We briefly discuss the norms imposed on interaction kernels. 
For a function $f(x_1,\cdots,x_n)$ on
$(\bbbr\times\bbbr^d\times\{\uparrow,\downarrow\})^n$ we define its
 $L_1$--$L_\infty$--norm as
$$
\tn f\tn_{1,\infty} 
=\max\limits_{1\le j_0 \le n}\ 
\sup\limits_{x_{j_0} \in \bbbr\times\bbbr^d\times\{\uparrow,\downarrow\}}\  
\int \prod\limits_{j=1,\cdots, n \atop j\ne j_0}\hskip-5pt dx_j\ \  
| f( x_1,\cdots,x_n) | 
$$
A multiindex is an element 
$\de=(\de_0,\de_1,\cdots,\de_d) \in \bbbn_0 \times \bbbn_0^d$.
The length of a multiindex $\de=(\de_0,\de_1,\cdots\de_d)$ is 
$|\de| =\de_0+\de_1+\cdots+\de_d$ and its factorial is
$\de!=\de_0!\de_1!\cdots\de_d!$.  For two multiindices $\de,\de'$ we say that
$\de \le \de'$ if $\de_i \le \de'_i$ for $i=0,1,\cdots,d$.
The spatial part of the multiindex $\de=(\de_0,\de_1,\cdots\de_d)$ is 
$\bde=(\de_1,\cdots\de_d)\in\bbbn_0^d$. It has length
 $|\bde| =\de_1+\cdots+\de_d$.
For a  multiindex $\de$ and 
$x=(x_0,\x,\si),\,x'=(x'_0,\x',\si') \in \bbbr\times\bbbr^d\times 
\{ \uparrow,\,\downarrow\}$ set
$$
(x-x')^\de = (x_0-x_0')^{\de_0}\,(\x_1-\x'_1)^{\de_1}\cdots
(\x_d-\x_d')^{\de_d}
$$
We fix $r_0,r\ge 6$ for the numbers of temporal and spatial momentum 
derivatives that we will control. The norm imposed on an interaction 
kernel will be 
$$
\max_{\de_{i,j}\in \bbbn_0 \times \bbbn_0^d \atop{ {\rm for\ }1\le i<j \le 4
\atop \Si |\bde_{i,j}| \le r,\  \Si |\de_{i,j;0}| \le r_0 }     }
\TTN \smprod_{1\le i<j \le 4} \sfrac{1}{\de_{i,j}!}
(x_i-x_j)^{\de_{i,j}} V(x_1,x_2,x_3,x_4)\TTN_{1,\infty} 
$$
When $V$ is of the form (\eqnNPinteraction),
$$
\max_{\de_{i,j}\in \bbbn_0 \times \bbbn_0^d \atop{ {\rm for\ }1\le i<j \le 4
\atop \Si |\bde_{i,j}| \le r,\  \Si |\de_{i,j;0}| \le r_0 }     }
\TTN \smprod_{1\le i<j \le 4} \sfrac{1}{\de_{i,j}!}
(x_i-x_j)^{\de_{i,j}} V(x_1,x_2,x_3,x_4)\TTN_{1,\infty} 
=\max_{\bde\in\bbbn_0^d\atop |\bde|\le r}\sfrac{1}{\bde!}
\int \big|\x^\bde v(\x)\big|\ d\x
$$

\vskip 1cm
Formally, the generating function for the connected Green's functions is
$$
\cG(\phi,\bar\phi;\de e) =  \log\sfrac{1}{Z} 
\int  e^{\phi J\psi}\,  e^{\cV(\psi,\bar \psi)}\,
e^{-\cK(\psi,\bar\psi;\de e)}\,d\mu_C(\psi,\bar\psi)
\EQN\eqnNPforgenfn$$
where the source term is 
$$
\phi J\psi 
=  \int dx \ \bar\phi(x)\psi(x)+\bar\psi(x)\phi(x)
\EQN\eqnNPphijpsi$$
The counterterm is implemented in 
$$
\cK(\psi,\bar\psi;\de e) = \half \smsum_{\si\in\{\uparrow,\downarrow\}}
\int d\tau d^d\x d^d\y\  \de e^\wedge(\x-\y)\, \bar\psi_\si(\tau,\x) \psi_\si(\tau,\y)
$$
and
$\ Z=\int e^{\cV(\psi,\bar\psi)}\,e^{-\cK(\psi,\bar\psi;\de e)}\,
d\mu_C(\psi,\bar\psi) \ $ is the partition function.
The fields $\phi,\bar\phi$ are called source fields and
 the fields $\psi,\bar\psi$ are called internal fields. The 
connected Green's functions themselves are determined by
$$
\cG(\phi,\bar \phi;\de e) = 
\sum_{n=1}^\infty \sfrac{1}{(n!)^2} \int\smprod_{i=1}^n dx_idy_i\ 
G_{2n}(x_1,y_1,\cdots,x_n,y_n;\de e) 
\smprod_{i=1}^n \bar\phi(x_i) \phi(y_i) 
$$
Observe that for $\de e\in \cE$, formally,
$$
\cG(\phi,\bar\phi;\de e) 
=  \log\sfrac{1}{Z'}  \int e^{\phi J\psi}\, e^{\cV(\psi,\bar \psi)}\,
d\mu_{C(\de e)}(\psi,\bar\psi)
$$
where
$ \ Z'=\int e^{\cV(\psi,\bar\psi)} d\mu_{C(\de e)}(\psi,\bar\psi) \ $.
 See [FKTo2, Lemma \lemOSappGrassII].

\vskip .3cm
We show in this paper that, for $d=2$ and under the Hypotheses \:\hypNPdisprel\ 
on $e(\k)$, there exists, for every sufficiently small interaction,
a counterterm $\de e\in \cE$ such that connected Green's functions 
$G_{2n}(\,\cdot\,;\de e)$ exist and have Fermi surface $F$. This statement
needs to be made precise, because the functional integrals in the definition of
$\cG$ are not, a priori, well defined due to the singularities of 
the propagator. This problem is dealt with by a multiscale analysis. 

We introduce scales by slicing momentum space into shells around the Fermi surface. We choose a  ``scale parameter'' $M>1$ and a function 
$\nu\in C^\infty_0([\sfrac{1}{M},\,2M])$ that 
takes values in $[0,1]$, is identically 1 on $[\sfrac{2}{M},M]$ and obeys
$$
\sum_{j=0}^\infty \nu\big(M^{2j}x\big) = 1
$$
for $0<x<1$ (see also [FKTo2, \S\CHscales]). The function $\nu$
may be constructed by choosing a function
$\varphi\in C_0^\infty\big((-2,2)\big)$ that is identically one on $[-1,1]$
and setting $\nu(x)=\varphi(x/M)~-~\varphi(Mx)$ for $x>0$ and zero otherwise.

\goodbreak
\definition{\STM\defNPscales}{
\Item i) 
For $j\ge 1$, the $j^{\rm th}$ scale function on $\bbbr\times\bbbr^d$ is defined as
$$ 
\nu^{(j)}(k)=\nu\left(M^{2j}(k_0^2+e(\k)^2)\right) 
$$
By construction, $\nu^{(j)}$ is identically one on
$$
\set{k=(k_0,\k) \in \bbbr\times\bbbr^d}
{\sqrt{\sfrac{2}{M}}\, \sfrac{1}{M^j}\le |ik_0-e(\k)|\le \sqrt{M} \sfrac{1}{M^j} }
$$
The support of $\nu^{(j)}$ is called the $j^{\rm th}$ shell. By construction, it is contained in
$$
\set{k\in \bbbr\times\bbbr^d}
{\sfrac{1}{\sqrt{M}}\, \sfrac{1}{M^j}\le |ik_0-e(\k)|\le \sqrt{2M} \sfrac{1}{M^j} }
$$
The momentum $k$ is said to be of scale $j$ if $k$ lies in the $j^{\rm th}$ shell.
\Item ii) 
For real $j\ge 1$, set
$$
\nu^{(\ge j)}(k)=\varphi\big(M^{2j-1}(k_0^2+e(\k)^2)\big)
$$
By construction, $\nu^{(\ge j)}$ is identically 1 on 
$$
\set{k \in \bbbr\times\bbbr^d} {|ik_0-e(\k)|\le \sqrt{M} \sfrac{1}{M^j} }
$$ 
Observe that if $j$ is an integer, then for $|ik_0-e(\k)|>0$
$$
\nu^{(\ge j)}(k)=\smsum_{i\ge j}\nu^{(i)}(k)
$$
The support of $\nu^{(\ge j)}$ is called the $j^{\rm th}$ neighbourhood of 
the Fermi surface. By construction, it  is contained in
$$
\set{k\in \bbbr\times\bbbr^d}{|ik_0-e(\k)|\le \sqrt{2M} \sfrac{1}{M^j} }
$$
The support of $\varphi\big(M^{2j-2}(k_0^2+e(\k)^2)\big)$ is called the 
$j^{\rm th}$ extended neighbourhood. It  is contained in
$$
\set{k\in \bbbr\times\bbbr^d}{|ik_0-e(\k)|\le \sqrt{2}M\sfrac{1}{M^{j}} }
$$
\Item iii) 
For real $\il\ge 1$, the covariance with infrared cutoff at scale $\il$ and 
counterterm $\de e$ is
$$\eqalign{
C^{\IR(\il)}(k_0,\k;\de e) 
&= \frac{U(\k)-\nu^{(\ge\il)}(k)}{\imath k_0 - e(\k)
                +\de e(\k)[1-\nu^{(\ge\il)}(k)]}\cr
\noalign{\vskip0.05in}
C^{\IR(\il)}(x,x';\de e) &= \de_{\si,\si'} \int \frac{d^{d+1}k}{(2\pi)^{d+1}} 
e^{\imath<k,x-x'>_-}C^{\IR(\il)}(k;\de e)\cr
}$$
Also write
$$
C^{\IR(\il)}(k) = C^{\IR(\il)}(k;0) \qquad,\qquad C^{\IR(\il)}(x,x')=C^{\IR(\il)}(x,x';0)
$$
The Grassmann Gaussian measure $d\mu_{C^{\IR(\il)}(\de e)}$ is characterized by
$$\eqalign{
 \int \psi(x)\bar\psi(x')\, d\mu_{C^{\IR(\il)}(\de e)}(\psi,\bar\psi) 
&=C^{\IR(\il)}(x,x';\de e) \cr
 \int \psi(x)\psi(x')\, d\mu_{C^{\IR(\il)}(\de e)}(\psi,\bar\psi) &= 
 \int \bar\psi(x)\bar\psi(x')\, d\mu_{C^{\IR(\il)}(\de e)}(\psi,\bar\psi) =0
}$$
}

\remark{\STM\remNPlargej}{
As the scale parameter  $M>1$, the shells near the Fermi curve have $\il$ near 
$+\infty$, and the neighbourhoods shrink as $\il \rightarrow \infty$. Also, 
$$
\lim_{\il\rightarrow \infty} C^{\IR(\il)}(k) = C(k)
$$
pointwise.
}

Even for the cutoff, and hence bounded, covariance $C^{\IR(\il)}$, 
it is not a priori clear that the generating functional
$$\eqalign{
\cG_\il(\phi,\bar\phi;\cV,\de e) =  \log\sfrac{1}{Z} 
\int e^{\phi J\psi}\, e^{\cV(\psi,\bar \psi)}\,
&d\mu_{C^{\IR(\il)}(\de e)}(\psi,\bar\psi)\cr
&\hbox{where}\quad
Z=\int e^{\la\cV(\psi,\bar\psi)}\,d\mu_{C^{\IR(\il)}(\de e)}(\psi,\bar\psi)
}$$
or the corresponding connected Green's functions,
$G_{2n;\il}(x_1,y_1,\cdots,x_n,y_n)$,  defined by
$$
\cG_\il(\phi,\bar \phi;\cV,\de e) = 
\sum_{n=1}^\infty \sfrac{1}{(n!)^2} \int\smprod_{i=1}^n dx_idy_i\ 
G_{2n;\il}(x_1,y_1,\cdots,x_n,y_n) 
\smprod_{i=1}^n \bar\phi(x_i) \phi(y_i) 
$$
make sense for a reasonable set of $V$'s and $\de e$'s. 
On the other hand, it is easy
to see, using graphs, that each term in the formal Taylor expansion 
of the Grassmann function\footnote{$^{(2)}$}{We shall, in Definition \defNPsectGrnorm,
introduce a norm on the Grassmann algebra generated by $\phi$ and $\bar\phi$.
All of our generating functionals will in fact have finite norm.} $\cG_\il(\phi,\bar\phi;\cV,\de e(V))$ in powers of $V$ is well--defined for a 
large class of $V$'s and $\de e(V)$'s.  
The Taylor expansion of 
$\int e^{\phi J\psi}\, e^{\cV(\psi,\bar \psi)}\,
d\mu_{C^{\IR(\il)}}(\psi,\bar\psi)$
is $\sum_{n=1}^\infty\cG_{\il,n}(\cV,\cdots,\cV)$ where the 
$n^{\rm th}$ term is the multilinear form
$$
\cG_{\il,n}(\cV_1,\cdots,\cV_n)= \sfrac{1}{n!}
\int e^{\phi J\psi} \cV_1(\psi,\bar \psi)
\cdots\cV_n(\psi,\bar \psi)\ d\mu_{C^{\IR(\il)}}(\psi,\bar\psi)
$$
restricted to the diagonal. Explicit evaluation of the Grassmann 
integral expresses $\cG_{\il,n}$ as the sum of all graphs with vertices 
$\cV_1,\ \cdots,\ \cV_n$ and lines $C^{\IR(\il)}$. 
The (formal) Taylor coefficient 
$\sfrac{d\hfill}{dt_1}\cdots\sfrac{d\hfill}{dt_n}
\cG_\il(\phi,\bar\phi;t_1\cV_1+\cdots +t_n\cV_n,0)\Big|_{t_1=\cdots=t_n=0}$ 
of $\cG_\il(\phi,\bar\phi;\cV,0)$ is similar, but with only 
connected graphs. Choosing $\de e$ to be an appropriate function of $V$
produces renormalized connected graphs\footnote{$^{(3)}$}{Under the hypotheses of Theorem \:\theoremNPmainthI, below, when $\il$ is finite, both the propagator $C^{\IR(\il)}$ and the vertex $V$ are continuous in 
momentum space. The values of all connected graphs, whether renormalized or not,
are well--defined. However renormalization is essential for the limit $\il\rightarrow\infty$.}.
We prove here that, under suitable hypotheses, for each $\il$,
the renormalized formal Taylor series for 
$\cG_\il\big(\phi,\bar \phi;\cV,\de e(V)\big)$ 
converges to an analytic\footnote{$^{(4)}$}{For an elementary discussion of analytic maps between Banach spaces see, for example, Appendix A of [PT].}
function of $V$ with a radius of convergence that is independent of $\il$.
We further show that the limit as $\il\rightarrow\infty$ exists.

\theorem{\STM\theoremNPmainthI}{
Assume that $d=2$ and that $e(\k)$ fulfills the Hypotheses \:\hypNPdisprel\ 
below. There is an open ball, centered on the origin, in the Banach 
space of  translation invariant and spin independent interaction  kernels $V$
with norm
$$
\max_{\de_{i,j}\in \bbbn_0 \times \bbbn_0^d \atop{ {\rm for\ }1\le i<j \le 4
\atop \Si |\bde_{i,j}| \le r,\  \Si |\de_{i,j;0}| \le r_0 }     }
\tn \smprod_{1\le i<j \le 4} \sfrac{1}{\de_{i,j}!}
(x_i-x_j)^{\de_{i,j}} V(x_1,x_2,x_3,x_4) \tn_{1,\infty}
$$
and an analytic counterterm function $V\mapsto\de e(V) \in \cE$ on the 
ball, that vanishes for $V=0$, such that the following holds:

\noindent
For any real $\il\ge 1$, the formal Taylor series
$$
\cG_\il(\phi,\bar\phi) =  \log\sfrac{1}{Z} 
\int e^{\phi J\psi}\, e^{\la\cV(\psi,\bar \psi)}\,
d\mu_{C^{\IR(\il)}(\de e(V))}(\psi,\bar\psi)
$$
converges to an analytic function on the ball. 
As $\il\rightarrow\infty$, the Green's functions 
$ G_{2n;\il} $ converge uniformly, in $x_1,\ \cdots,\ y_n$ and $V$,
 to a translation invariant, spin independent, particle number 
conserving function $ G_{2n}$ on 
$\big( \bbbr\times \bbbr^d \times \{\uparrow,\downarrow\}\big)^{2n}$
that is analytic in $V$.

\noindent
If, in addition, $V$ is $k_0$--reversal real, as in (\eqnNPreal), 
then $\de e(\k;V)$ is real for all $\k$.

}

\noindent
The proof of Theorem  \theoremNPmainthI\ follows the statement of 
Theorem \:\theoremNPinduction\ in [FKTf2].

\theorem{\STM\theoremNPmainthII}{
Under the hypotheses of Theorem \theoremNPmainthI\ and the assumption that 
$V$ obeys the symmetries (\eqnNPreal) and (\eqnNPphexchange), 
the Fourier transform
$$\eqalign{
\check G_{2}(k_0,\k)  
&= \int dx_0 d^d\x\ e^{\imath(-k_0x_0 +\k\cdot\x)}\,
    G_{2} \big((0,0,\uparrow), (x_0,\x,\uparrow) \big) \cr
&=  \int dx_0 d^d\x\ e^{\imath(-k_0x_0 +\k\cdot\x)}\,
    G_{2}\big((0,0,\downarrow), (x_0,\x,\downarrow) \big) \cr
}$$
of the two--point function exists and is continuous, except on the Fermi 
surface (precisely, except when $k_0=0$ and $e(\k)=0$). Define
$$
n(\k)=\lim_{\tau\rightarrow 0+}\int\sfrac{dk_0}{2\pi}\ 
         e^{\imath k_0\tau}\ \check G_{2}(k_0,\k)
$$
 Then $n(\k)$ is continuous except on the Fermi surface $F$. 
If $\bar\k\in F$, then $\lim\limits_{\k\rightarrow\bar\k\atop e(\k)>0}n(\k)$ and 
$\lim\limits_{\k\rightarrow\bar\k\atop e(\k)<0}n(\k)$ exist and obey
$$
\lim_{\k\rightarrow\bar\k\atop e(\k)<0}n(\k)
-\lim_{\k\rightarrow\bar\k\atop e(\k)>0}n(\k)
>\half
$$

}

\noindent
The proof of Theorem  \theoremNPmainthII\ follows  
Lemma \:\lemTNPabstractjump \ in [FKTf3].

\remark{\STM\remNPmainthII}{
The quantity $n(\k)$ is known as the momentum distribution function. The
jump discontinuity in $n(\k)$ at the Fermi surface exhibited in Theorem
\theoremNPmainthII\ is generally viewed as the most basic characteristic 
of a Fermi liquid [MCD, \S4.1]. The number $\half$ in the bound of \theoremNPmainthII\ 
has no special significance. It may be replaced by any number strictly
smaller than one, provided the interaction is made sufficiently weak.

}

\theorem{\STM\theoremNPmainthIII}{
Let
$$\eqalign{
\check G_{4;\si_1,\si_2,\si_3,\si_4}(k_1,k_2,k_3)  
&= \int G_{4} \big(x_1,x_2,x_3,(0,0,\si_4) \big) \ 
\smprod_{\ell=1}^3 e^{-(-1)^\ell\imath<k_\ell,x_\ell>_-}\,dx_{0,\ell} d^d{\x_\ell}
\cr
}$$
be the Fourier transform of the four--point function and
$$
\check G^A_{4;\si_1,\si_2,\si_3,\si_4}(k_1,k_2,k_3)
=\check G_{4;\si_1,\si_2,\si_3,\si_4}(k_1,k_2,k_3)
\smprod_{\ell=1}^4\sfrac{1}{\check G_2(k_\ell)}\Big|_{k_4=k_1-k_2+k_3}
$$
its amputation by the physical propagator.
Under the hypotheses of Theorem \theoremNPmainthII, 
$\check G^A_{4}$ is continuous on
$$
\set{(k_1,k_2,k_3)}{k_1\ne k_2,\ k_2\ne k_3,\ 
U(\k_1)=U(\k_2)=U(\k_3)=U(\k_1-\k_2+\k_3)=1}
$$ 
Furthermore, $\check G^A_{4}$ has a decomposition
$$\eqalign{
\check G^A_{4;\si_1,\si_2,\si_3,\si_4}(k_1,k_2,k_3)
&=N_{\si_1,\si_2,\si_3,\si_4}(k_1,k_2,k_3)\cr
&\hskip.25in+\half L_{\si_1,\si_2,\si_3,\si_4}\big(\sfrac{k_1+k_2}{2},\sfrac{k_3+k_4}{2},k_2-k_1\big)\Big|_{k_4=k_1-k_2+k_3}\cr
&\hskip.25in-\half L_{\si_1,\si_2,\si_3,\si_4}\big(\sfrac{k_3+k_2}{2},\sfrac{k_1+k_4}{2},k_2-k_3\big)\Big|_{k_4=k_1-k_2+k_3}
}$$
with
{\parindent=0.25in
\item{$\circ$}$N_{\si_1,\si_2,\si_3,\si_4}$ continuous
\item{$\circ$}$L_{\si_1,\si_2,\si_3,\si_4}(q_1,q_2,t)$ continuous
                   except at $t=0$
\item{$\circ$}
$L_{\si_1,\si_2,\si_3,\si_4}(q_1,q_2,(0,\t))$ having an extension
to $\t=0$  which is continuous in $(q_1,q_2,\t)$
\item{$\circ$}
$L_{\si_1,\si_2,\si_3,\si_4}(q_1,q_2,(t_0,\0))$ having an extension
to $t_0=0$  which is continuous in $(q_1,q_2,t_0)$

}
}
\noindent
The proof of Theorem  \theoremNPmainthIII\ is at the end of \S\CHtildestep\ 
in [FKTf3].
\remark{\STM\remNPmainthIII}{
$L_{\si_1,\si_2,\si_3,\si_4}(q_1,q_2,t)$ consists of particle--hole
ladder contributions of the form
$$
\figput{phladder2}
$$
}

\remark{\STM\remNPmainthI}{
Theorem \theoremNPmainthI\ defines  $G_{2n;\il}(V)$ through its
renormalized perturbation expansion. Alternatively, one could introduce 
an additional finite volume cutoff by replacing the space--time $\bbbr\times\bbbr^d$ with $(\bbbr/L\bbbz)\times(\bbbr^d/L\bbbz^d)$. 
The associated Green's functions, $G_{2n;\il, L}(V)$,
are trivially  meromorphic functions of $V$ that are analytic at $V=0$. 
The methods of these papers could be used to show that the domain of 
analyticity of $G_{2n;\il, L}(V)$ includes the ball of Theorem 
\theoremNPmainthI\ and that, 
as $L\rightarrow\infty$, $G_{2n;\il, L}$ converges uniformly to $G_{2n;\il}$.
We do not do so.

We also do not deal with the question of independence of the renormalization
prescription. Suppose that $e(\k)$, $\de e(\k;V)$
and $e'(\k)$, $\de e'(\k;V)$ satisfy the appropriate regularity conditions
and that $e(\k)-\de e(\k;V)=e'(\k)-\de e'(\k;V)$, for some specific $V$. 
Observe that $G_{2n;\il}$ depends on $e(\k)$ and $\de e(\k;V)$ only through 
the combinations $e(\k)-\de e(\k;V)[1-\nu^{(\ge\il)}(k)]$ and 
$\nu^{(\ge\il)}(k)=\varphi\big(M^{2j-1}(k_0^2+e(\k)^2)\big)$. 
Hence, it is likely that the limiting Green's functions constructed using 
$e(\k)$, $\de e(\k;V)$ coincide
with those constructed using  $e'(\k)$, $\de e'(\k;V)$, for the specific 
$V$, but we do not attempt to prove so here.

}

We now state the hypotheses on the dispersion relation $e(\k)$ used in
Theorems \theoremNPmainthI, \theoremNPmainthII\ and \theoremNPmainthIII. 
First, we assume that the dispersion relation $e(\k)$ is $r+6 $ times 
continuously differentiable. Furthermore, we assume that the Fermi curve
$$
F=\set{\k\in\bbbr^2}{ e(\k)=0}
$$
is a strictly convex, smooth connected curve with curvature bounded away from zero. Since $F$ is strictly convex, for each point $\k\in F$ there is a unique
point $a(\k)\in F$ different from $\k$ such that the tangent lines to $F$ at $\k$ and $a(\k)$ are parallel. $a(\k)$ is called the antipode of $\k$. Choose an orientation for $F$.

\definition{\STM\defNPstrongasymm}{ 
\Item i)
Let $\k\in F$, $\tanV$ the oriented unit tangent vector to $F$ at $\k$
and $\normV$ the inward pointing unit normal vector to $F$ at $\k$. Then
there is a function $\varphi_\k(s)$, defined on a neighbourhood of $0$
in $\bbbr$, such that $\, s\mapsto \k+ s\tanV+\varphi_\k(s)\normV\ $ is an
oriented parametrization of $F$ near $\k$.
\Item ii)
We say that $F$ is strongly asymmetric if there is $n_0\in\bbbn$, with
$n_0\le r$, such
that for each $\k\in F$ there exists an $n\le n_0$ with
$$
\varphi_\k^{(n)}(0)\ne \varphi_{a(\k)}^{(n)}(0)
$$
}

\remark{\STM\remModII}{

\noindent
i) By construction, $\varphi_\k(0)=\dot\varphi_\k(0)=0$ and
$\ddot\varphi_\k(0)$ is the curvature of $F$ at $\k$.
\Item ii)
If $F$ is symmetric about a point $\p\in\bbbr^2$, that is
$\,F=\{\,2\p-\k\,\big|\,\k\in F\,\}$,
then $\varphi_\k=\varphi_{a(\k)}$ for all $\k\in F$.
Symmetry of the Fermi curve about a point promotes the formation of 
Cooper pairs and the phase transition to a superconducting state. Theorem  
\theoremNPmainthI\ shows that --- at temperature zero ---
 this is the only instability in a broad class of 
short range many fermion models, at least when $d=2$. Sufficiently high
temperature also blocks the Cooper instability and leads to Fermi liquid
behaviour, even for a round Fermi surface. This was shown in [DR1, DR2]
using the criterion proposed in [S].

When $d=1$, fermionic many--body models exhibit Luttinger liquid rather than Fermi liquid behaviour. See Chapter 11 of [BG] and the references therein. 
We would expect that results like Theorems \theoremNPmainthI\ and 
\theoremNPmainthII\ also hold for $d=3$. There has been some progress in this
direction [MR,DMR].  
\Item iii)
In [FKTa] we show that independent fermions in a suitably chosen periodic
electromagnetic background field have a dispersion relation whose associated
Fermi curve, for suitably chosen chemical potential, is smooth, 
strictly convex, strongly asymmetric and has nonzero curvature everywhere.
}

\noindent
\hypothesis{\STM\hypNPdisprel (on the dispersion relation):}
{We assume that $e(\k)$ is $r+6$ times continuously differentiable with 
$r\ge 6$, that the Fermi curve $F$ is a strictly convex, smooth,  
strongly asymmetric, connected curve whose curvature is bounded away from
zero and that $\nabla e(\k)$ does not vanish on $F$.}

This paper is divided into three parts. This first part, which consists
of Sections \CHintroI\ through \CHrenmap\ and Appendix \APappModelComp, contains an overview of the
main ideas involved in the construction (\S\CHintroOverview) and
the algebraic structure of the construction (\S\CHrenmap). Part 2,
[FKTf2], contains Sections \CHintroII\ through \CHrecurs\ and 
Appendix \APappRewick\ 
and is concerned with the proof of convergence of the expansion. 
Part 3, [FKTf3], contains  Sections \CHintroIII\ through \CHtildestep\ and 
Appendices \APappHoelder\ and \APappPhladders\
and is concerned with the proof of the existence of the Fermi surface. 
Cumulative notation tables are provided at the end of each part.
The construction described in these papers was outlined in
[FKLT1, FKLT2, FKLT3].

\vfill\eject
\chap{An Overview}\PG\pgNPII
In this Section, we describe the main difficulties in the proof of Theorems 
\theoremNPmainthI\ and \theoremNPmainthII\  and outline our strategy 
to overcome them. For simplicity, we omit spins in this discussion. 
The notation in this section is close, but not 
always identical to that used in the rest of this paper. As the main theorems
are for $d=2$ space dimensions, we describe all constructions only for $d=2$, 
even those that can be extended to other dimensions.

\Item{1.} { \it Renormalization of the Fermi Surface and the Dispersion Relation}\PG\pgNPIIa
\vskip .0in
The Fermi surface of a fermionic many particle system is the locus in momentum
space where the two point function $\check G_2(0,\k)$ has a discontinuity -- if
such a discontinuity occurs at all.\footnote{$^{(1)}$}{For example, in a superconductor there is no such discontinuity.} For a system of non interacting 
fermions, the Fermi surface coincides with the zero--set of the 
dispersion relation. In a metal or a crystal, the dispersion relation is 
a datum derived from first principles; it is determined by the associated 
periodic Schr\"odinger operator. 
On the other hand, the Fermi surface of the system of
interacting fermions is accessible to measurement. See, for example,
[AM]. 

\noindent 
As already mentioned in \S\CHintroI, the Fermi surface of a system of
interacting fermions is, in general, different from that of the 
system of non interacting fermions with the same  dispersion relation. 
This shift in the Fermi surface is responsible 
for the divergence of many coefficients in naive perturbation expansions. It
is controlled by renormalizing the dispersion relation. Theorems 
\theoremNPmainthI\ and \theoremNPmainthII\ state
that, given a function $e(\k)$ and an interaction $V$ fulfilling all of
the Theorems' hypotheses, there is a function $\de e(\k)$, called the
``counterterm", such that system with dispersion relation $e(\k)-\de e(\k)$
and interaction $V$ has a Fermi surface and that Fermi surface is precisely
$F=\{\k\big| \ e(\k)=0\ \}$.

\noindent
We pointed out in the previous paragraph that the data derived from first
principles are the dispersion relation and the interaction $V$. Therefore it
is desirable to prove that every reasonable function $e'(\k)$ is of the form
$e(\k)-\de e(\k)$ as above. This could be done by proving the invertibility 
of the map $e(\k) \mapsto e(\k)-\de e(\k)$ in an appropriate function space. 
To all orders in perturbation theory, this has been achieved in [FST4]. 
The bounds of this paper are not yet strong enough to prove the 
corresponding result non perturbatively.

\noindent
Even for $C^\infty$ functions $e(\k)$ and $V$, it is not known how smooth the
counterterm $\de e(\k)$ is. In this paper, we show that $\de e$ 
is $C^{\ep}$.
In [FST3] it is shown that $\de e$ is $C^{2+\ep}$ to all orders in perturbation 
theory. Later in this overview (subsection 10) we shall point out where 
this lack of smoothness in the counterterm creates difficulties for 
the construction.

\vskip .5cm
\goodbreak
\Item{2.} { \it Multi Scale Analysis}\PG\pgNPIIb
\vskip .0in
We cannot treat the functional integral (\eqnNPforgenfn) defining the 
formal Green's functions in one piece, because the propagator is singular. 
Similarly it is probably impossible to determine the counterterm $\de e(\k)$ 
in one step. Therefore we introduce scales adjusted to the size of the 
propagator in momentum space and to the infrared cut off propagators 
$C^{\IR(j)}(k_0,\k;\de e)$ of Definition \eqnNPforgenfn, 
construct an appropriate counterterm for each scale $j$ and take the limit 
$j  \rightarrow \infty$. The
limit is controlled by comparing, for each $j$, the model with covariance 
$C^{\IR(j+1)}(k_0,\k;\de e)$ to that with covariance  
$C^{\IR(j)}(k_0,\k;\de e)$. This comparison amounts to
``integrating out scale $j$". 
We give an introduction to ``integrating out a scale'' in the next subsection.
In \S\CHrenmap, we describe, formally but in more detail, how the limit $j\rightarrow \infty$ is taken.

\vskip .5cm
\Item{3.} { \it Integrating out a Scale}\PG\pgNPIIc
\vskip .0in
The discussion of the previous subsection shows that the essential estimates in
our construction concern the effect of integrating out one scale as
above. To simplify the discussion, we consider the case in which
 the covariance $C^{\IR(j)}(k;\de e)$  is replaced by $C^{\IR(j)}(k;0)$.
 That is,  for the moment, we ignore the effect of the counterterm. 
Since $C^{\IR(j+1)}(k;0)=C^{\IR(j)}(k;0)+C^{(j)}(k)$ with
$ C^{(j)}(k) = \sfrac{\nu^{(j)}(k)}{\imath k_0 - e(\k)}$,
$$\eqalign{
\cG_{j+1}(\phi,\bar\phi) 
&= \log \sfrac{1}{Z'}\int \int e^{\phi J(\psi+\ze)+\cV(\psi+\ze, \bar\psi+\bar\ze)}
   d\mu_{C^{\IR(j)}}(\ze,\bar\ze)d\mu_{C^{(j)}}(\psi,\bar\psi)\cr
&= \log \sfrac{1}{Z}\int e^{\phi J\psi+\cW(\phi,\bar\phi,\psi,\bar\psi)}
  d\mu_{C^{(j)}}(\psi,\bar\psi)\cr
}$$
where 
$$
\cW(\phi,\bar\phi,\psi,\bar\psi) 
= \log \sfrac{1}{Z_j}  \int e^{\phi J\ze+\cV(\psi+\ze, \bar\psi+\bar\ze)}
   d\mu_{C^{\IR(j)}}(\ze,\bar\ze)
$$ 
is the effective interaction at scale $j$. The partition functions $Z$, $Z'$ and $Z_j$ are chosen so that $\cG_{j+1}(0,0)=\cW(0,0,0,0)=0$.
To iterate this construction, we need an effective interaction 
$$
\cW'(\phi,\bar\phi,\psi,\bar\psi) 
= \log \sfrac{1}{Z_{j+1}}  \int e^{\phi J\ze+\cV(\psi+\ze, \bar\psi+\bar\ze)}
   d\mu_{C^{\IR(j+1)}}(\ze,\bar\ze)
$$
at scale $j+1$. So the problem is to estimate
$$
\cW'(\phi,\bar\phi,\psi,\bar\psi) 
= \log \sfrac{1}{Z} \int e^{\phi J\ze+\cW(\phi,\bar\phi,\psi+\ze, \bar\psi+\bar\ze)}
   d\mu_{C^{(j)}}(\ze,\bar\ze)
$$
in terms of estimates on $\cW$.  The main difficulties already 
occur when $\phi=\bar\phi =0$, so we concentrate on this special
case. Write
$$\eqalign{
&\cW(0,0,\psi, \bar\psi) \cr
& = \smsum_{n\ge 0} \int {\sst dp_1\cdots dp_n\,dq_1\cdots dq_n} 
   \,w_{2n}{\sst (p_1,\cdots, p_n,\,q_1,\cdots, q_n)}\, 
    \de{\sst (p_1+\cdots+ p_n -q_1-\cdots- q_n)}
     \bar\psi{\sst (p_1) \cdots }\bar\psi({\sst p_n)} \,
     \psi{\sst (q_1) \cdots} \psi({\sst q_n)} \cr
}$$
and
$$\eqalign{
&\cW'(0,0,\psi, \bar\psi) \cr
& = \smsum_{n\ge 0} \int {\sst dp_1\cdots dp_n\,dq_1\cdots dq_n} 
   \,w'_{2n}{\sst (p_1,\cdots, p_n,\,q_1,\cdots, q_n)}\, 
    \de{\sst (p_1+\cdots+ p_n -q_1-\cdots- q_n)}
     \bar\psi{\sst (p_1) \cdots} \bar\psi{\sst (p_n)} \,
     \psi{\sst (q_1) \cdots} \psi{\sst (q_n)} \cr
}$$
where $\psi(k)$ and $\bar\psi(k)$ are the Fourier transforms of $\psi(x)$ and
$\bar\psi(x)$ respectively\footnote{$^{(2)}$}{Precise Fourier transform 
conventions are formulated in \S\CHnewsectors.}.

\noindent
Then $w_{2n}'$ can be written as a sum of values of connected directed 
graphs with
vertices $w_2, w_4, \cdots$ and propagator $C^{(j)}(k)$. See [FW, Chapter 3]. 
Naive power counting just uses that
$$
\|C^{(j)}(k)\|_\infty = \sup_k |C^{(j)}(k)|  \ \ {\rm is\ of\ order\ } M^j
\EQN\eqnOVlinfty$$
and
$$\eqalign{
{\rm volume\ of\ the\ }j^{\rm th}\ {\rm shell\  is\ of\ order\ } \sfrac{1}{M^{2j}} \cr
}\EQN\eqnOVshellvol$$
This is because the $j^{\rm th}$ shell has width of order $\sfrac{1}{M^j}$ in 
the $k_0$ direction and in the $\k$--direction transversal to $F$ and has 
circumference, in the direction along $F$, of order one. If one assumes 
that 
$$
\|w_{2n}\|_\infty \ \ {\rm is\ of\ order\ } M^{j(n-2)}\ \ {\rm for\ all\ }n 
\EQN\eqnOVvertexbnd$$ 
then every graph contributing to $w_{2n}'$ is again of order 
$$
M^{j\Si_i(n_i-2)}M^{j\Si_i(2n_i-2n)/2}M^{-2j[\Si_i(2n_i-2n)/2-\Si_i 1+1]}
=M^{j(n-2)}
$$ 
The three factors come from the suprema of the vertex functions 
$w_{2n_i}$, the suprema of the $\sfrac{\Si_i 2n_i-2n}{2}$ propagators and the 
volume of the domain of integration respectively. For
$n\ge 3$ the Condition (\eqnOVvertexbnd) grows nicely in $j$. 
Four legged vertices ($n=2$) are marginal --- the estimate (\eqnOVvertexbnd) 
alone would lead to divergences in powers of $j$ when the sum over 
$j$ is performed. We exploit ``overlapping loops" and special estimates on 
ladders to derive better bounds on the four point functions. 
The counterterm is built up 
from contributions at each scale chosen so that (\eqnOVvertexbnd) holds 
for n=1, too. Overlapping loops will be
discussed in subsection 4, ladders in subsections 5 and 6, and the 
construction of the counterterm in subsection 10.

\noindent
It is well known that the sum of the norms of all graphs diverges. 
In this paper we use cancellations between different
graphs to derive convergent bounds. There are several schemes to implement such
cancellations, all variants of the basic scheme of [C]. 
In all of these schemes the cancellation is only seen when one writes the model 
in position space variables. This is a major technical difficulty, since 
the geometry of the Fermi surface and special effects like ``overlapping loops" 
and ladder estimates are naturally formulated in momentum space. We use 
the cancellation scheme developed in [FMRT, FKTcf, FKTr1, FKTr2].
It is sufficiently close to the graphical picture to allow us to implement
overlapping loops in collections of diagrams within which cancellations take
place.

\vskip .5cm
\Item{4.} { \it Overlapping Loops}\PG\pgNPIId
\vskip .0in
We first describe the effect of overlapping loops in an example. Consider the
diagram
$$
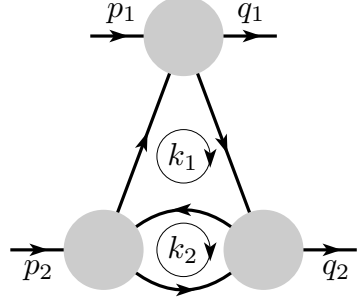
\figput{dbubble8}
$$
contributing to $w_4'(p_1,p_2,q_1,q_2)$. If $w_4=1$, its value is 
$$
\Ga(p_1,p_2,q_1,q_2)= \int  dk_1\,dk_2
\ C^{(j)}(k_1+q_1-p_1)\,C^{(j)}(k_1)\,C^{(j)}(k_1-k_2)\,C^{(j)}(q_2-k_2)
$$
Naive power counting gives
$$
\|\Ga\|_\infty \le \|C^{(j)}\|_\infty^4\cdot
  ( \hbox{volume of the $j^{\rm th}$ shell})^2
= O(1)
$$
However, taking into account that $C^{(j)}$ is supported on the $j^{\rm th}$ 
shell,
$$
\|\Ga\|_\infty \le \|C^{(j)}\|_\infty^4 \cdot
  ( {\rm volume\ of\ } T)
$$
where
$$
T= \{(k_1,k_2)\,\big|\, k_1,k_1-k_2,q_2-k_2 \in j^{\rm th}
   \ {\rm neighbourhood} \}
\EQN\eqnOVtripleintersection$$
Let $S$ be the set of all $k_2$ for which $q_2-k_2$ lies in the $j^{\rm th}$
neighbourhood and  $S'$ be the set of all $k_2$ for which 
$|k_2| \ge \sfrac{1}{\sqrt{M^j}}$. Then $T \subset T_1\cup T_2$ where
$$\eqalign{
T_1&= \{(k_1,k_2)\,\big|\, k_1,k_1-k_2 \in j^{\rm th}
   \ {\rm neighbourhood},\ k_2\in S\cap S' \} \cr
T_2&=\{(k_1,k_2)\,\big|\, k_1 \in j^{\rm th}
   \ {\rm neighbourhood},\ k_2\in S\setminus S' \}
}$$
For $k_2 \in S'$ the volume of the 
set
$$
\{k_1\,\big|\, k_1,k_1-k_2 \in j^{\rm th}\ {\rm neighbourhood} \}
= \big(j^{\rm th}\ {\rm neighbourhood} \big) 
\cap \big(k_2 + j^{\rm th}\ {\rm neighbourhood}\big)
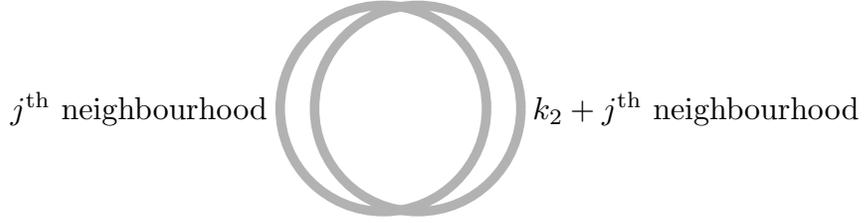
$$
\centerline{\figput{voleffectOverlap}}

\noindent
is of order $\sfrac{1}{M^{2j}}\cdot \sfrac{1}{\sqrt{M^j}}$, since this set has 
width of order $\sfrac{1}{M^j}$ in $k_0$ direction and in the $\k$
direction transversal to $F$ and width at most of order $\sfrac{1}{\sqrt{M^j}}$ in
the direction along $F$. Here we use that the Fermi surface
$F$ has curvature bounded away from zero. Therefore the volume of
$T_1$ is of order
$$
\sfrac{1}{M^{5j/2}}\cdot ({\rm volume\ of\ } S) 
= \sfrac{1}{M^{5j/2}}\cdot O\big(\sfrac{1}{M^{2j}}\big)
=O\big(\sfrac{1}{M^{9j/2}}\big)
$$
Similarly the volume of $T_2$ is bounded by
$$
({\rm volume\ of\ }j^{\rm th}\ {\rm neighbourhood}) 
\cdot ({\rm volume\ of\ }S\setminus S') 
= O(\sfrac{1}{M^{2j}})\cdot\sfrac{1}{M^{5j/2}} 
= O(\sfrac{1}{M^{9j/2}})
$$
Therefore, using (\eqnOVlinfty), $\|\Ga\|_\infty = O(\sfrac{1}{\sqrt{M^j}})$.
The volume estimate on $T$ derived above is not optimal. By [FST2, Theorem
1.1], the volume of $T$ is $O\big(\sfrac{j}{M^j}\big)$.

Similar improvements are possible for all diagrams with ``overlapping loops", i.e.
with two different simple loops which have at least one line in common. If
$w_2=0$, the only four legged diagrams without overlapping loops and tadpoles
are particle--particle ladders
$$
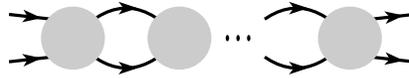
\figplace{CmpLadder3}{-.1 in}{0 in}
$$
and particle--hole ladders
$$
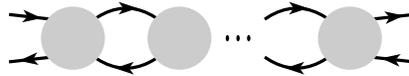
\figplace{CmpLadder2}{-.1 in}{0 in}
$$
See [FST1, \S2.4].
We shall Wick order the effective interactions with respect to the future
covariance in order to exclude tadpoles. The condition that $w_2=0$ is
achieved by moving the two legged part of the effective interaction into
 the covariance.

\noindent
We have already mentioned that the effect of overlapping loops and special ladder estimates are
used to get convergent bounds on the four point functions. The effect of
overlapping loops has to be combined with the cancellation scheme between diagrams
mentioned at the end of subsection 3 and discussed in subsection 9 below. Thus the
cancellation scheme has to be sensitive enough to detect simultaneous 
overlapping loops in all mutually cancelling diagrams and to isolate
ladder diagrams. Furthermore the geometric estimates on $T$ above have
to be exploitable in a position space setting. This is done using sectors.
See subsection 8.

\goodbreak\vskip .5cm
\Item{5.} { \it Particle--Particle Bubbles}\PG\pgNPIIe
\vskip .0in
The strong asymmetry condition of Definition \defNPstrongasymm\ 
is used to get improved power counting on particle--particle ladders.
We describe the effect in the example of the particle--particle bubble
$$
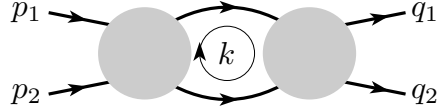
\figput{ppbubble2}
$$
again assuming that $w_4=1$. The value of this graph is
$$
\int dk \,C^{(j)}(t-k)\,C^{(j)}(k)
$$
where $t =p_1+p_2=q_1+q_2$ is the transfer momentum. Naive power counting again
gives that this value is $O(1)$. On the other hand, taking the support 
condition of $C^{(j)}$ into account, the value of this graph is bounded by
$$
\|C^{(j)}\|_\infty^2 \cdot
  ( {\rm volume\ of\ } \{k\,\big|\,k \ {\rm and}\ t-k \ {\rm lie\ in\ the \ }
      j^{\rm th}\ {\rm neighbourhood} \}
$$
By the strong asymmetry condition of Definition \defNPstrongasymm, 
$F$ and the shifted reflected Fermi surface
$t-F=\{t-k\,\big|\,k\in F \}$ have tangency of order at most $n_0$. From 
this one deduces that 
$$
\{k\,\big|\,k,\ t-k \in
      j^{\rm th}\ {\rm neighbourhood} \}
= \big(j^{\rm th}\ {\rm neighbourhood} \big) \cap
\big( t- (j^{\rm th}\ {\rm neighbourhood}) \big) 
$$
has width of order $\sfrac{1}{M^j}$ in $k_0$ direction and in the $\k$
direction transversal to $F$ and width at most of order $\sfrac{1}{M^{j/n_0}}$%
\vadjust{\centerline{\figput{voleffectR}}} 
in the direction along $F$. Therefore its volume is of order 
$\sfrac{1}{M^{2j}}\sfrac{1}{M^{j/n_0}}$ and the value of the particle--particle
bubble is of order $\sfrac{1}{M^{j/n_0}}$.

\noindent
As in the case of overlapping loops, this estimate is based on a volume estimate
in momentum space. Again sectors will used to implement this in position space
variables.

\vskip .5cm
\Item{6.} { \it Particle--Hole Ladders}\PG\pgNPIIf
\vskip .0in
Our estimate on particle--hole ladders is not based on a geometric argument as 
in the case of particle--particle ladders or overlapping loops, but on cancellations
between scales. The limit as $j\rightarrow\infty$ of the particle--hole bubble $B_j(p_1,p_2,q_1,q_2)$
$$
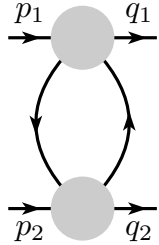
\figput{phbubble}
$$
with $w_4=1$ and propagator $\smsum_{i\le j} C^{(i)}$  has a
discontinuity in the transfer momentum $t=p_1-q_1$ at $t=0$, but is continuous for $t\ne 0$ and smooth in a neighbourhood of the origin in  $\{(t_0,\t)\,\big|\,t_0=0\}$. 
See the introduction to [FKTl], and 
in particular Lemma \lemLADprimitivemanfred\ there. The proof of this 
Lemma is based on integration by parts and thus cancellation between scales.

\noindent
In the multi scale analysis, we combine all the contributions of particle--hole
ladders created at scales $\le j$, and give a uniform estimate on the result. A
vertex of a particle--hole ladder at scale $j$ may be a particle--hole ladder
created at a previous scale $i<j$, as in the diagram
$$
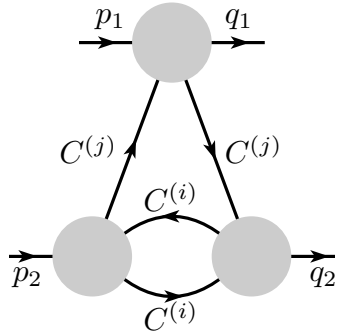
\figput{dbubble9}
$$
The iteration of these effects (and of the Wick ordering with respect to future
covariances) leads to the concept of iterated particle--hole ladders of 
Definition \:\defcompLadder.
Uniform estimates on these iterated particle--hole ladders are stated in 
Theorem \:\theoremcompLadder\ and proven in [FKTl].
They have to be in position space, as they have to be combined with the other
estimates derived from the ``cancellation scheme" mentioned in subsection 3.
This is technically difficult, because it amounts to taking Fourier 
transforms of quantities whose limit, as $j\rightarrow \infty$, is 
discontinuous.

\vskip .5cm
\Item{7.} { \it Power Counting in Position Space}\PG\pgNPIIg
\vskip .0in

Proving power counting bounds on graphs is usually split into
two steps. To illustrate the two steps, we consider the situation that we 
have two vertices $\varphi_1$ and $ \varphi_2$ with $n_1$ and $n_2$ legs, 
respectively. For simplicity we ignore orientation of the lines. We 
assume we form a diagram $\Ga$ by connecting $\varphi_1$ and $ \varphi_2$ 
by $1\le r \le \min\{n_1,n_2\}$ lines
$$
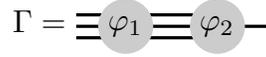
\figput{pcGraph1}
$$
This can be done in two steps: First connect $\varphi_1$ and $ \varphi_2$ by one
line and call the result $\Ga'$.
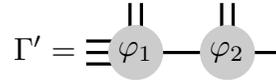
$$
\figput{pcGraph2}
$$
Secondly, pairwise contract $2r-2$ legs of $\Ga'$ to form $r-1$ lines.
$$
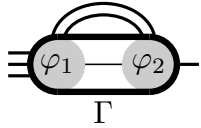
\figput{pcGraph3}
$$
The first estimate is to bound the norm of $\Ga'$ in terms of a constant
times the product of the norms of $\varphi_1$ and $ \varphi_2$. We call such a
constant a ``contraction bound" $c$. If one uses $\|\cdot\|_\infty$ norms in
momentum space, then $\|C^{(j)}(k)\|_\infty$ is a contraction bound. 

\noindent
A tadpole bound is a number $\tb$ with the following property. Let $\varphi$ be
any graph with at least two legs, and $\varphi'$ the graph obtained from $\varphi$ by connecting two legs to form
a line.
$$
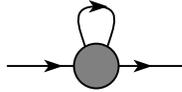
\figput{selfWick}
$$
Then the norm of $\varphi'$ is bounded by $\tb$ times the norm of $\varphi$. 
If one uses $\|\cdot\|_\infty$ norms in
momentum space, then $\|C^{(j)}(k)\|_1$ is a tadpole bound. 

\noindent
Applying one contraction bound and $r-1$ tadpole bounds, one sees that the norm of $\Ga$ is bounded by $c\,\tb^{r-1}$ times the product of the norms of $\varphi_1$ and $ \varphi_2$. By (\eqnOVlinfty) and (\eqnOVshellvol), for the $\|\cdot\|_\infty$ norms in momentum space, the contraction bound is of order $M^j$, while the tadpole bound is of order $\sfrac{1}{M^j}$. The power counting 
for the $\|\ \cdot\ \|_\infty$ norm in momentum space described in subsection 3 can be generalized to the abstract setting of contraction and tadpole bounds:
If one assumes that 
$$
\hbox{the norm of $w_{2n}$ is of order } 
\sfrac{1}{c\tb^{n-1}}\ \ {\rm for\ all\ }n  
\EQN\eqnOVabvertexnorm$$ 
then every graph contributing to $w_{2n}'$ is again of order 
$\sfrac{1}{c\tb^{n-1}}$. For example, if such a graph $\Ga$ has two vertices, $w_{2n_1}$ and $w_{2n_2}$ then there are $r=n_1+n_2-n$ connecting lines and 
the norm of $\Ga$ is bounded by
$$\eqalign{
c\tb^{n_1+n_2-n-1}\big(\hbox{norm of $w_{2n_1}$}\big)
\big(\hbox{norm of $w_{2n_2}$}\big)
&= O\big(c\tb^{n_1+n_2-n-1}\sfrac{1}{c\tb^{n_1-1}}\sfrac{1}{c\tb^{n_2-1}}\big)\cr
&=O\big(\sfrac{1}{c\tb^{n-1}}\big)
}$$
A general graph may be bounded by building it up one vertex at a time.

\noindent
As pointed out in subsection 3, we need to use position space variables. A
natural position space norm for translation invariant vertices
$\varphi(x_1,\cdots,x_n)$, which mimics the $\|\ \cdot\ \|_\infty$ norm
in momentum space, is 
$$
\tn\varphi\tn_{1,\infty} = \max\limits_{1\le p \le n}\ 
\sup\limits_{x_{p}}\  
\int \prod\limits_{j=1,\cdots, n \atop j\ne p} dx_j\, | \varphi(x_1,\cdots,x_n) |
$$
It is easily seen that the $L^1$ norm, $\|C^{(j)}(x)\|_1$, of the Fourier transform of $C^{(j)}(k)$ is a contraction bound for this norm and that 
the $L^\infty$ norm of $C^{(j)}$ in position space,
$\|C^{(j)}(x)\|_\infty$, is a tadpole bound. Clearly, 
$\ \|C^{(j)}(x)\|_\infty \le  \|C^{(j)}(k)\|_1\,$, so that we again have a
tadpole bound of order $\sfrac{1}{M^j}$. A naive computation, given in
the next paragraph, gives a bound on $\|C^{(j)}(x)\|_1$ that is of order $M^{2j}$.
A more refined argument, sketched in the next but one paragraph, gives a 
(realistic --- see Example \exNPsectorizebound) bound of order $M^{3j/2}$. 
In any event, 
$M^{3j/2}\gg \|C^{(j)}(k)\|_\infty$ and naive power counting in 
position space does not coincide with power counting in momentum space. 
Substituting $c=O\big(M^{3j/2}\big)$ and $\tb=O\big(\sfrac{1}{M^j}\big)$ into 
(\eqnOVabvertexnorm) yields the requirement that $\tn\hat w_{2n}\tn_{1,\infty}$ 
be order $M^{j(n-{5\over 2})}$. In particular the norm of the four point function would have to decrease like $\sfrac{1}{\sqrt{M^j}}$ as $j$ increased. This is absurd, since the original interaction $V$ is, at each scale, the dominant part of the four point function. 
Again, we use sectors to cope with this problem.

We first sketch the standard calculation that gives the naive bound
on $\|C^{(j)}(x)\|_1$. For a multi index $\de=(\de_0,\de_1,\de_2)$ of non negative
integers write $|\de| = \de_0+\de_1+\de_2$ and 
$x^\de=x_0^{\de_0} x_1^{\de_1} x_2^{\de_2}$. Then, integrating by parts
$|\de|$ times,
$$
\big( \sfrac{x}{M^j} \big)^\de |C^{(j)}(x)| 
\le \sfrac{1}{M^{j|\de|}} \|\sfrac{\partial^{|\de|}\ }{\partial
k^\de}C^{(j)}(k)\|_1 =O(\sfrac{1}{M^j})
$$
since the support of $\sfrac{\partial^{|\de|}\ }
{\partial k^\de}C^{(j)}(k)$ has volume of
order $\sfrac{1}{M^{2j}}$ and 
$\|\sfrac{\partial^{|\de|}\ }{\partial k^\de}C^{(j)}(k)\|_\infty$ is of order
$M^{j(|\de|+1)}$. Therefore
$$
\Big(1+ \big(\sfrac{x_0}{M^j} \big)^2 \Big) 
\Big(1+ \big(\sfrac{x_1}{M^j} \big)^2 \Big)
\Big(1+ \big(\sfrac{x_2}{M^j} \big)^2 \Big)
|C^{(j)}(x)| = O(\sfrac{1}{M^j})
$$
Dividing by 
$\!\!\smprod\limits_{\nu=0,1,2}\!\!\Big(1+ \big(\sfrac{x_\nu}{M^j} \big)^2 \Big) $ and
integrating over $\bbbr^3$ gives the bound $\|C^{(j)}(x)\|_1 = O(M^{2j})$.

As it motivates the construction of sectors, we indicate how the more refined
bound on $\|C^{(j)}(x)\|_1$ is derived. Assume first that $F$ has a straight
segment of length $\fl$ on the $k_2$ axis\footnote{$^{(3)}$}{This of course
contradicts our hypotheses that $F$ is strictly convex. We will remove this
assumption immediately.} and that $e(k)=k_1$ in a neighbourhood of this segment.
Choose a cutoff function
$\chi(k_2)$ that is identically one on most of the segment, zero outside the
the segment and for which $|\sfrac{\partial^n\ }{\partial k_2^n}\chi(k_2)|$ 
is of order $\sfrac{1}{\fl^n}$. Set $C_s^{(j)}(k)= \chi(k_2) C^{(j)}(k)$. 
An argument similar to that of the previous paragraph shows that
$$
\Big(1+ \big(\sfrac{x_0}{M^j} \big)^2 \Big) 
\Big(1+ \big(\sfrac{x_1}{M^j} \big)^2 \Big)
\Big(1+ (\fl x_2 )^2 \Big)
|C_s^{(j)}(x)| = O(\sfrac{\fl}{M^j})
$$
and therefore that $\|C_s^{(j)}(x)\|_1 = O(M^{j})$. The same argument also 
works for a realistic Fermi surface, if one cuts out a ``sector" of length
$ \fl \le \sfrac{1}{\sqrt{M^j}} $ as indicated in the figure below.
$$
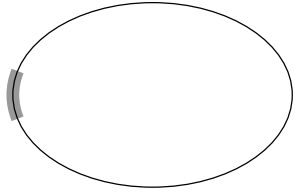
\figput{oneSector}
$$
The precise computation is in [FKTo3, Proposition \propOSGenDecay\ 
and Lemma \lemOSsectorderiv]. If the sector is too long, the curvature of the
Fermi surface causes deterioration of the bounds on the derivatives parallel to 
the Fermi surface. One can divide up the $j^{\rm th}$ neighbourhood into
$O(\sfrac{1}{\fl})$ sectors and use a partition of unity by functions 
like $\chi$, to see that
$$
\|C^{(j)}(x)\|_1 \le ({\rm number\ of\ sectors})\cdot O(M^{j}) =
O(\sfrac{1}{\fl}M^j)
$$
If one chooses $\fl = \sfrac{1}{\sqrt{M^j}}$, one gets the bound 
$\|C^{(j)}(x)\|_1 = O(M^{3j/2})$.

\vskip .5cm
\goodbreak
\Item{8.} { \it Sectors}\PG\pgNPIIh
\vskip .0in
We cover the $j^{\rm th}$ neighbourhood by slightly overlapping sectors of length $ \fl \le \sfrac{1}{\sqrt{M^j}} $ as indicated in the figure below.
$$
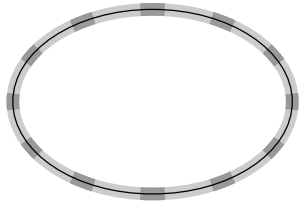
\figput{allSectors}
$$ 
The set $\Si$ of sectors is called a sectorization of scale $j$ and length $\fl$. Furthermore we select a partition of unity 
$\{ \chi_s(k)\,\big|\, s\in\Si \}$ subordinate to the sectorization, such that each $\chi_s$ has properties analogous to those of the function $\chi$ in the last subsection\footnote{$^{(4)}$}{For precise Definitions of sectors, sectorizations, and the partition of unity see Definition
\defNPsectors\ through Definition \defNPtens.}. 
Recall that we want to integrate out scale $j$ and that 
$$
\cW'(0,0,\psi, \bar\psi) 
= \sfrac{1}{Z} \int e^{\cW(0,0,\psi+\ze, \bar\psi+\bar\ze)}
   d\mu_{C^{(j)}}(\ze,\bar\ze)
$$
Decompose the kernel $w_{2n}$ of the part of $\cW$ that is homogeneous of degree
$2n$ in the fields 
$$
w_{2n}{\sst (p_1,\cdots, p_n,\,q_1,\cdots, q_n)}
=\sum_{s_1,\cdots,s_{2n}\in\Si}
\om_{2n}({\sst (p_1,s_1),\cdots, (p_n,s_n),\,(q_1,s_{n+1}),\cdots, (q_n,s_{2n})})
$$
for $p_i,q_i$ in the $j^{\rm th}$ neighbourhood. Here, $\om_{2n}$ is a function
that vanishes unless $p_i$ lies in the sector $s_i$ and $q_i$ lies in the sector $s_{n+i}$. $\om_{2n}$ is called a $\Si$--sectorized representative of $w_{2n}$.

\noindent
A $\Si$--sectorized representative $\om'_{2n}$ of $w_{2n}'$ can than be written
as a sum of values of connected directed graphs with
vertices $\om_2, \om_4, \cdots$ and propagator $C^{(j)}(k)$, where the momentum integrals in the graphs also include sector sums. The main norm
we use for the Fourier transforms 
$\,\hat\om_{2n}({\sst (x_1,s_1),\cdots, (x_{2n},s_{2n})})\,$
of $\om_{2n}$ is
$$
\tn \hat\om_{2n} \tn_{1,\Si} =  
\max_{1\le i_0\le 2n}\ \ 
 \max_{s_{i_0} \in \Si} \ \ 
 \sum_{s_i \in \Si \ {\rm for} \atop i \ne i_0} 
 \tn \hat\om_{2n}({\sst (x_1,s_1),\cdots,(x_{2n},s_{2n})} ) \tn_{1,\infty}
\EQN\eqnOVtriponesi$$
That is, we first fix one sector and then take the sum over all other sectors of the $\tn\cdot\tn_{1,\infty}$ norms of the functions
$ \hat\om_{2n}({\sst (\cdot,s_1),\cdots, (\cdot,s_{2n})})$.
With respect to this norm, we have a contraction bound of order $M^j$
and a tadpole bound of order $\sfrac{\fl}{M^j}$. 

\noindent
We first indicate how the contraction bound is derived. Let 
$\varphi_1({\sst (x_1,s_1),\cdots, (x_n,s_n)})$ and 
$\varphi_2({\sst (x_1,s_1),\cdots, (x_m,s_m)})$ be vertices, and 
$$\eqalign{
\Ga'&({\sst (x_1,s_1),\cdots, (x_{n-1},s_{n-1}),\,(x'_2,s'_2),\cdots, (x'_m,s'_m) }) \cr
 &\hskip 1.5cm = \smsum_{s_n,s_1'\in\Si} \int {\sst dx_n\,dx_1'}\
\varphi_1({\sst (x_1,s_1),\cdots, (x_n,s_n)})\,
C^{(j)}{\sst (x_n-x_1')}\,
\varphi_2({\sst (x_1',s_1'),\cdots, (x_m',s_m')}) \cr
}$$
be the graph constructed by connecting $\varphi_1$ and $\varphi_2$ with
one line.  Write 
$\ 
C^{(j)}(k) = \smsum\limits_{s\in\Si} C^{(j)}_s(k) 
\ $
where $C^{(j)}_s(k) =\chi_s(k)\,C^{(j)}(k)$. Let $C^{(j)}_s(x) $ be the Fourier transform of $C^{(j)}_s(k) $. Then
$$\eqalign{
\Ga'= \smsum_{s_n,s,s_1'\in\Si} \int {\sst dx_n\,dx_1'}\
\varphi_1({\sst (x_1,s_1),\cdots, (x_n,s_n)})\,
C^{(j)}_s{\sst (x_n-x_1')}\,
\varphi_2({\sst (x_1',s_1'),\cdots, (x_m',s_m')})
}\EQN\eqnOVsectcontrI$$
By conservation of momentum, the integral in (\eqnOVsectcontrI) vanishes if
$s_n\cap s\cap  s_1' = \emptyset$. For fixed sectors $s_1,\cdots,s_n,\,s_1',\cdots,s_m',\,s$, by double convolution
$$\eqalign{
&\TTN  \int {\sst dx_n\,dx_1'}\
\varphi_1({\sst (x_1,s_1),\cdots, (x_n,s_n)})\,
C^{(j)}_s{\sst (x_n-x_1')}\,
\varphi_2({\sst (x_1',s_1'),\cdots, (x_m',s_m')}) \TTN_{1,\infty} \cr
&\hskip 3cm 
\le \tn \varphi_1({\sst (\cdot,s_1),\cdots, (\cdot,s_n)}) \tn_{1,\infty}
\|C^{(j)}_s{\sst (x)}\|_1
\tn \varphi_2({\sst (\cdot,s_1'),\cdots, (\cdot,s_m')}) \tn_{1,\infty}
}\EQN\eqnOVsectcontrII$$
We consider the contribution to $\tn\Ga'\tn_{1,\Si}$ having the sector $s_1$
fixed. By (\eqnOVsectcontrI), (\eqnOVsectcontrII) and conservation of momentum
it is bounded by
$$
\sum_{s_2,\cdots,s_{n-1}\in\Si \atop s_2',\cdots,s_m'\in\Si}
\sum_{s_n,s,s_1'\in\Si \atop s_n\cap s\cap s_1' \ne \emptyset}
\tn \varphi_1({\sst (\cdot,s_1),\cdots, (\cdot,s_n)}) \tn_{1,\infty}
\|C^{(j)}_s{\sst (x)}\|_1
\tn \varphi_2({\sst (\cdot,s_1'),\cdots, (\cdot,s_m')}) \tn_{1,\infty}
\EQN\eqnOVsectcontrIII 
$$
Observe that, for given $s_n$, there are at most three sectors $s$ and at most three sectors $s_1'$ with 
$s_n\cap s \ne \emptyset,\ s_n\cap s_1' \ne \emptyset$. 
Therefore (\eqnOVsectcontrIII ) is bounded by
$$\eqalign{
9 \max_{s,s_1' \in \Si} & \
\sum_{s_2,\cdots,s_{n-1},s_n\in\Si \atop s_2',\cdots,s_m'\in\Si}
\tn \varphi_1({\sst (\cdot,s_1),\cdots, (\cdot,s_n)}) \tn_{1,\infty}
\|C^{(j)}_s{\sst (x)}\|_1
\tn \varphi_2({\sst \cdot,s_1'),\cdots, \cdot,s_m')}) \tn_{1,\infty} \cr
& \hskip 7cm\le 9 \ \tn\varphi_1\tn_{1,\Si} \tn\varphi_2\tn_{1,\Si}\ 
\max_{s\in \Si} \|C^{(j)}_s(x)\|_1
}$$
Fixing $s_i$ with $2\le i\le n-1$ or $s'_i$ with $2\le i\le m$ 
leads to the same bound, so that
$$
\tn\Ga'\tn_{1,\Si}
\le
 9 \ \tn\varphi_1\tn_{1,\Si} \tn\varphi_2\tn_{1,\Si}\ 
\max_{s\in \Si} \|C^{(j)}_s(x)\|_1
\EQN\eqnOVsectcontrbnd$$
As at the end of the previous subsection, one sees that 
$\ \max\limits_{s\in \Si} \|C^{(j)}_s(x)\|_1 = O(M^j)\ $. This gives the contraction bound of order $M^j$.

\noindent
To derive the tadpole bound, let $\varphi({\sst (x_1,s_1),\cdots, (x_n,s_n)})$
be a vertex and 
$$\eqalign{
\Ga&({\sst (x_1,s_1),\cdots, (x_{n-2},s_{n-2})}) \cr
 &\hskip 1.5cm =\smsum_{s_{n-1},s_n\in\Si} \int {\sst dx_{n-1}\,dx_n}\
\varphi({\sst (x_1,s_1),\cdots, (x_{n-2},s_{n-2}), (x_{n-1},s_{n-1}), (x_n,s_n)})\,C^{(j)}{\sst (x_{n-1}-x_n)}
}$$
be obtained by joining the last two legs of $\varphi$ to form a tadpole.
As above, by conservation of momentum, for each choice of sectors $s_1,\cdots,s_{n-2}$
$$\eqalign{
& \tn \Ga({\sst (\cdot,s_1),\cdots, (\cdot,s_{n-2})}) \tn_{1,\infty}\cr
&\hskip 1.5cm
 \le \sum_{s_{n-1},s,s_n \in \Si\atop s_{n-1}\cap s\cap s_n \ne \emptyset}
\TTN  \int {\sst dx_{n-1}\,dx_n}\
\varphi({\sst \cdot,s_1),\cdots, (\cdot,s_{n-2}), (x_{n-1},s_{n-1}), (x_n,s_n)})\,C^{(j)}_s{\sst (x_{n-1}-x_n)} \TTN_{1,\infty} \cr
&\hskip 1.5cm
 \le 3\sum_{s_{n-1},s_n \in \Si}
\tn   \varphi({\sst (\cdot,s_1),\cdots, (\cdot,s_{n-2}), (\cdot,s_{n-1}), (\cdot,s_n)})\tn_{1,\infty} \,\ 
\max_{s\in \Si}\|C^{(j)}_s(x) \|_\infty\cr
}$$ 
as, for a given sector $s_{n-1}$, there are at most three sectors $s$ for which $s\cap s_{n-1} \ne \emptyset$. Consequently
$$
\tn \Ga \tn_{1,\Si} \le 3 \,\tn\varphi  \tn_{1,\Si} \ \max_{s\in \Si}\|C^{(j)}_s(x) \|_\infty 
$$
As in the previous subsection,
 $\ \|C^{(j)}_s(x) \|_\infty = O\big(\sfrac{\fl}{M^j}\big)\ $ and the tadpole bound of order $\sfrac{\fl}{M^j}$ follows.

Substituting $c=O\big(M^j\big)$ and $\tb=O\big(\sfrac{\fl}{M^j}\big)$ into 
(\eqnOVabvertexnorm) yields the requirement that 
$$
\tn\hat\om_{2n}\tn_{1,\Si} \ \ {\rm is\ of\ order\ } 
\sfrac{M^{j(n-2)}}{\fl^{n-1}}\ \ {\rm for\ all\ }n 
\EQN\eqnOVsectorvertexreq$$
In contrast to the norm $\tn\ \cdot\ \tn_{1,\infty}$  of subsection 7, this norm is compatible with change of scale. In \S\CHnewsectors, we choose, for each scale $i$, a sectorization $\Si_i$ of length $\fl_i$ where $\fl_i$ goes to zero   as some power of $\sfrac{1}{M^i}$\footnote{$^{(5)}$}{This is forced on us by the condition $\fl_i\le\sfrac{1}{\sqrt{M^i}}$ which is imposed to ensure that the curvature of the sector does not affect Fourier transform estimates.}.
In going from scale $j$ to scale $j+1$, we construct from a $\Si_j$--sectorized 
representative $\om'_{2n}$ of $w'_{2n}$ a $\Si_{j+1}$--sectorized 
representative $\om''_{2n}$ of $w'_{2n}$ that fulfills 
(\eqnOVsectorvertexreq) with $j$ replaced by $j+1$ and $\fl$ replaced by 
$\fl_{j+1}$. It is constructed using a partition of unity subordinate to $\Si_{j+1}$. See Example \exNPsectorizescalechange.

\noindent
To give an idea of the underlying mechanism, we show that the problem with the $\tn\ \cdot\ \tn_{1,\infty}$ norm of the four point function described
in subsection 7 does not occur for the $\tn\ \cdot\ \tn_{1,\Si_j}$ norm. 
To do so, we need a $\Si=\Si_j$--sectorized representative for the original 
interaction kernel $V(p_1,p_2,q_1,q_2)$ whose 
$\tn \cdot\tn_{1,\Si}$ norm is of order $\sfrac{1}{\fl}$. A natural such
representative is
$$
v\big((p_1,s_1), (p_2,s_2),\,(q_1,s_3), (q_2,s_4)\big) 
= \chi_{s_1}(p_1)\,\chi_{s_2}(p_2)\,\chi_{s_3}(q_1)\,\chi_{s_4}(q_2)\,
V(p_1,p_2,q_1,q_2)
$$
By conservation of momentum, the Fourier transform 
$\hat v\big((\cdot,s_1), (\cdot,s_2),\,(\cdot,s_3), (\cdot,s_4)\big) $ vanishes
if $(s_1+s_2)\cap (s_3+s_4) =\emptyset$. Here, 
$\,s_1+s_2=\{p_1+p_2\,\big|\,p_1\in s_1,\ p_2\in s_2\,\}\,$. 
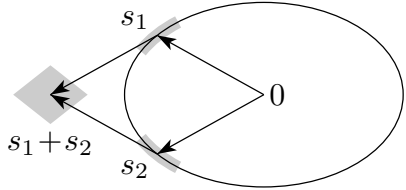
$$
\figput{twoSectors}
$$
One sees, by the same method that yielded $\| C_s(x)\|_1 = O(M^j)$, 
that the Fourier transform of each $\chi_s(k)$ fulfills 
$\| \hat\chi_s(x)\|_1 = O(1)$. Therefore, for each choice of sectors
$s_1,s_2,s_3,s_4$, 
$$
\tn \hat v\big((\cdot,s_1), (\cdot,s_2),\,(\cdot,s_3), (\cdot,s_4)\big)
\tn_{1,\infty} = O\big( \tn \hat V \tn_{1,\infty} \big)
$$
Therefore, $\tn \hat v \tn_{1,\Si}$ is of the order
$$
\max_{s_1\in \Si} \,\sharp \set{(s_2,s_3,s_4) \in \Si^3}
{(s_1+s_2)\cap (s_3+s_4)\ne\emptyset}
\EQN\eqnOVsecmap$$
To estimate this number, fix a sector $s_1$. Observe that the map
$$ 
F\times F \rightarrow\ \bbbr^2,\ (k_1,k_2) \mapsto k_1+k_2
\EQN\eqnOVsecnmbr$$
is locally invertible at every $(k_1,k_2)\in F\times F$ for which the tangent
vector to $F$ at $k_1$ is not parallel to the tangent vector to $F$ at $k_2$.
 From this one concludes that for 
a general choice of $s_2$ (such that $s_1+s_2$ is not close to $2k$ or 
$k+{\rm antipode}(k)$ for all $k\in F$), there are only
$O(1)$ choices of sectors $s_3,s_4$ such that 
$\,(s_1+s_2)\cap (s_3+s_4)\ne\emptyset \,$. Since there are
$O\big(\sfrac{1}{\fl}\big)$ sectors, the contribution to (\eqnOVsecmap) from all general
sectors $s_2$ is $O\big(\sfrac{1}{\fl}\big)$. One can see that the few other
sectors $s_2$ also only contribute $O\big(\sfrac{1}{\fl}\big)$. Thus, indeed
$\tn \hat v \tn_{1,\Si}=O\big(\sfrac{1}{\fl}\big)$. The precise argument is 
given in [FKTo4, Proposition \propOSthreetoonenorm].

The estimates on the contraction and tadpole bounds and about change of sectorization in this subsection all are specific to two space dimensions. 
In three
space dimensions, similar arguments would give that
 \item{} a contraction bound for $C^{(j)}$ is of order $M^j$
 \item{} a tadpole bound for $C^{(j)}$ is of order $\sfrac{\fl^2}{M^j}$
 \item{} $\tn \hat v \tn_{1,\Si}=O\big(\sfrac{1}{\fl^3}\big)$, since the map 
(\eqnOVsecnmbr) has one dimensional fibers in this case.

\noindent
Thus, for the case of three space dimensions, the power counting suggested by
sectors is not compatible with change of sectorization. This is the reason
why, in this paper, we restrict ourselves to two space dimensions%
\footnote{$^{(6)}$}{Progress in the use of sectorization in three 
space dimensions has been made in [MR, DMR].}.

The advantage of sectors is that they allow exploitation of conservation of momentum, while working in position space. One example is the estimate  
on $\tn \hat v\tn_{1,\Si}$ derived two paragraphs ago. Another important 
example is that the improvements due to overlapping loops and 
particle--particle bubbles described in subsections 4 and 5 can be implemented 
using sectors. To this end we use a variant of the norm (\eqnOVtriponesi), with 
three sectors held fixed,
$$
\tn \hat\om_{2n} \tn_{3,\Si} =  
\max_{1\le i_1<i_2<i_3\le 2n}\ \ 
 \max_{s_{i_1},s_{i_2},s_{i_3} \in \Si} \ \ 
 \sum_{s_i \in \Si \ {\rm for} \atop i \ne i_1,i_2,i_3} 
 \tn \hat\om_{2n}({\sst (x_1,s_1),\cdots,(x_{2n},s_{2n})} ) \tn_{1,\infty}
\EQN\eqnOVtripthreesi$$
Its power counting is better by one factor of $\fl$ than the 
power counting of the $\tn \ \cdot\  \tn_{1,\Si}$ norm. That is, we expect
$$
\tn\hat\om_{2n}\tn_{3,\Si} \ \ {\rm is\ of\ order\ } 
\big(\sfrac{M^j}{\fl}\big)^{n-2}\ \ {\rm for\ all\ }n 
\EQN\eqnOVsectorvertexthreereq$$
and in particular that $\tn\hat\om_4\tn_{3,\Si}$ is of order 
one\footnote{$^{(7)}$}{The argument above that the sectorized representative $v$ of the original interaction fulfills $\tn \hat v \tn_{1,\Si}
=O\left(\sfrac{1}{\fl}\right)$ also shows that $\tn \hat v \tn_{3,\Si}
=O(1)$.}. This norm is particularly useful for the four point function.
Since a given momentum can lie in at most two sectors,
$$
\| \om_4(p_1,p_2,q_1,q_2) \|_\infty\le 2^3 \tn\hat\om_4\tn_{3,\Si}
$$
In Example \:\exNPloopsector, we show how (\eqnOVsectorvertexreq) 
and (\eqnOVsectorvertexthreereq)  can be used to get improved power counting 
for the $\tn \ \cdot\  \tn_{1,\Si}$ norm of the diagram discussed in 
subsection 4.

\vskip .5cm
\Item{9.} { \it Cancellation Between Diagrams}\PG\pgNPIIi
\vskip .0in
In subsections 7 and 8, we described how the power counting bound on a diagram $\Ga$
formed by connecting two vertices $\varphi_1$ and $ \varphi_2$ by $r\ge 1$
lines\footnote{$^{(8)}$}{Again, for simplicity, we ignore orientation of 
the lines.} is obtained. First we connect $\varphi_1$ and $ \varphi_2$ by one
line and call the resulting graph $\Ga'$ and apply a contraction bound. Then we
form the remaining $r-1$ lines, each time applying a tadpole bound. Iterating
this procedure, adding vertex after vertex, leads to the power counting for
arbitrary diagrams.

\noindent
Since we are dealing with fermions, we may assume in our discussion that all
vertex functions are antisymmetric. Let $n_1\ge r$ and $n_2\ge r$ be the 
number of legs of $\varphi_1$ and $ \varphi_2$, respectively. Assume that
$\max\{n_1,n_2\}>r$. Denote by $\cal G$ the set of all
diagrams obtained from joining $\varphi_1$ and $ \varphi_2$ by $r$ lines. 
$\cal G$ has cardinality $\ {n_1\choose r}\,{n_2\choose r}\,r!\ $. If 
one bounds each individual diagram by power counting, and sums over 
all diagrams in $\cal G$, this large number of diagrams leads to 
divergences (due to the factor $r!$).

We first describe how one finds cancellations between diagrams of $\cG$, then
describe a blocking of diagrams that allows one to find similar cancellations 
for arbitrary numbers of vertices, and then show how, using this 
blocking, one may simultaneously exploit both these cancellations 
and overlapping loops.

The first step in constructing the diagrams of $\cal G$ is again to choose one 
leg of $\varphi_1$ and one leg of $ \varphi_2$, form a line between these two 
legs, call the resulting graph $\Ga'$ and estimate the norm of $\Ga'$ in terms 
of the norms of $\varphi_1$ and $ \varphi_2$ using a contraction bound. The 
second step is to choose $(r-1)$ additional legs of $\varphi_1$ and 
$(r-1)$ additional legs of $\varphi_2$.
$$
\figput{pcGraph4}
$$
The third step is to form all possible connections between the $(r-1)$ legs of
$\varphi_1$ and the $(r-1)$ legs of $\varphi_2$ chosen in the second step. 
There are $n_1n_2$ choices in the first step, 
$\ {n_1-1\choose r-1}\,{n_2-1\choose r-1}\,$ choices in the second step and 
$(r-1)!$ choices in the third step. Each
diagram of $\cal G$ is obtained $r$ times, since with the first step we
distinguish one of the $r$ lines. Observe that
$$
\sfrac{1}{r}\,n_1n_2\,{\tst{n_1-1\choose r-1}}\,
{\tst{n_2-1\choose r-1}}\,(r-1)! 
= {\tst{n_1\choose r}}\,{\tst{n_2\choose r}}\,r!
$$
There are cancellations amongst the diagrams formed in the third step described
above. For simplicity, assume that the $(r-1)$ legs of $\varphi_1$ chosen in the
second step are labeled by $2,\cdots,r$ and are all incoming, and that 
the $(r-1)$ legs of $\varphi_1$ are labeled by $n_2-r+2,\cdots,n_2$ and 
are all outgoing. Then the sum of all diagrams formed in the third step is
$$
\int\! {\sst dx_2\cdots dx_r\,dx'_{n_2-r+2}\cdots dx'_{n_2}}\
  \Ga'{\sst (x_2,\cdots, x_r,\,\cdots\,x'_{n_2-r+2},\cdots x'_{n_2})}
   \int\! \bar\psi{\sst (x_2)}\cdots \bar\psi{\sst (x_r)}\,
     \psi{\sst (x'_{n_2-r+2})} \cdots \psi{\sst (x'_{n_2})}\ d\mu_{C^{(j)}}
$$
Its $\|\cdot\|_{1,\infty}$ norm is bounded by
$$
\|\Ga'\|_{1,\infty}\ \sup_{x_2,\cdots,x_r \atop x'_{n_2-r+2},\cdots x'_{n_2}}\
\Big| \int \bar\psi{\sst (x_2)}\cdots \bar\psi{\sst (x_r)}\,
     \psi{\sst (x'_{n_2-r+2})} \cdots \psi{\sst (x'_{n_2})}\ d\mu_{C^{(j)}}
   \Big|
$$
The magnitude of the functional integral is
$$
\Big|\int \bar\psi{\sst (x_2)}\cdots \bar\psi{\sst (x_r)}\,
     \psi{\sst (x'_{n_2-r+2})} \cdots \psi{\sst (x'_{n_2})}\ d\mu_{C^{(j)}}\Big|
= \Big|\det \big[ C^{(j)}{\sst (x_\mu-x'_{n_2-r+\nu})}\big]_{\mu,\nu=2,\cdots,r}\Big|
$$
Observe that the $\mu$--$\nu$ matrix entry
$$\eqalign{
C^{(j)}(x_\mu-x'_{n_2-r+\nu}) 
&= \int dk\  e^{\imath <k,x_\mu-x'_{n_2-r+\nu}>_-}C^{(j)}(k) \cr
&= \<  e^{\imath <k,x_\mu>_-} \sqrt{ |C^{(j)}(k)|}\ ,
 \ e^{\imath <k,x'_{n_2-r+\nu}>_-} \sfrac{C^{(j)}(k)}{\sqrt{ |C^{(j)}(k)|}} \ \>_{L^2}
}$$
(we are deliberately ignoring some unimportant factors of $2\pi$) 
is the $L^2$ inner product of the vector
$v_\mu(k) = e^{\imath <k,x_\mu>_-} \sqrt{ |C^{(j)}(k)|}$ and the vector
$v'_\nu(k) = e^{\imath <k,x'_{n_2-r+\nu}>_-} \sfrac{C^{(j)}(k)}{\sqrt{ |C^{(j)}(k)|}}$. The vectors $v_\mu, v'_\nu$ all have $L^2$ norm
$\sqrt{\| C^{(j)}(k) \|_1}$. Therefore, by Gram's bound on determinants, 
$$
\Big| \int \psi{\sst (x_2)}\cdots  \cdots \bar\psi{\sst (x'_{n_2})}\ 
               d\mu_{C^{(j)}} \Big|
 \le \| C^{(j)}(k) \|_1^{r-1}
$$
Thus the bound on the sum of all diagrams formed at the third step is 
$$
\tn\Ga'\tn_{1,\infty}\,\| C^{(j)}(k) \|_1^{r-1} \le 
c\,\| C^{(j)}(k) \|_1^{r-1}\,\tn\varphi_1\tn_{1,\infty}\,\tn\varphi_2\tn_{1,\infty}
$$ 
where $c$ is a contraction bound. Consequently, the 
$\tn\cdot\tn_{1,\infty}$ norms
of the sum of all graphs in $\cal G$ is bounded by 
$$
\sfrac{1}{r}\,n_1n_2\,{\tst{n_1-1\choose r-1}}\,{\tst{n_2-1\choose r-1}}\,
\|C^{(j)}(x)\|_1\,\|C^{(j)}(k)\|_1^{r-1}\,
\tn \varphi_1\tn_{1,\infty}\,\tn\varphi_2\tn_{1,\infty}
$$
since, as seen in subsection 7, $\|C^{(j)}(x)\|_1$ is a contraction bound.
This is  smaller than the bound
$$
(\sharp {\cal G})\,
\|C^{(j)}(x)\|_1\,\|C^{(j)}(k)\|_1^{r-1}\,
\tn \varphi_1\tn_{1,\infty}\,\tn\varphi_2\tn_{1,\infty}
$$
derived by summing the graphwise bounds, by a factor of $\sfrac{1}{(r-1)!}$.

\noindent
In general, we say that $\ib$ is an integral bound, if the following holds: Let
$\varphi$ be any antisymmetric vertex, $2r\le n$, and $S(x_{2r+1},\cdots,x_n)$ 
be the sum of the values of all diagrams obtained from $\varphi$ by 
joining each of the legs labeled $1,\cdots,r $ to one and only one of the
legs labeled $r+1,\cdots,2r$.
$$
\figput{canGraph1}
$$
Then the norm of $S$ is bounded by $\ib^{2r}$ times the norm of $\varphi$. 
The argument in the previous paragraph shows that $C^{(j)}$ has an 
integral bound with respect to the $\tn \cdot\tn_{1,\infty}$ norm that is 
of order $\sqrt{\|C^{(j)}(k)\|_1}=O(\sfrac{1}{\sqrt{M^j}})$. Similarly one 
sees that $C^{(j)}$ has an integral bound, with respect to the $\tn \cdot\tn_{1,\Si}$ norm,  that is of order $O\Big(\sqrt{\sfrac{\fl}{M^j}}\,\Big)$. 
Thus, in both cases, the integral bound is
of the order of the square root of the tadpole bound found before.

The discussion above shows how to get cancellations between diagrams that have only
two vertices. The combinatorics for treating diagrams of arbitrary size was
developed in  [FMRT, FKTcf, FKTr1, FKTr2]. We sketch it here. 
As mentioned in  subsection 4, we Wick order with respect to future 
covariances in order to avoid tadpoles. Thus, we write
$$\eqalign{
\cW(0,0,\psi,\bar\psi) &= \lw U(\psi,\bar\psi)\rw_{C^{(\ge j)}} \cr
\cW'(0,0,\psi,\bar\psi) &= \lw U'(\psi,\bar\psi)\rw_{C^{(\ge j+1)}} \cr
}$$
where $C^{(\ge j)}= \smsum\limits_{i\ge j} C^{(i)}$. Then the kernels of $U'$ 
are sums of diagrams whose vertices are kernels of $U$ and which have two kinds 
of lines. The first arises from integrating with respect to $d\mu_{C^{(j)}}$
and has propagator $C^{(j)}$. The second arises in Wick ordering $\cW'$
and has propagator $C^{(\ge j+1)}$. The subgraph of each diagram obtained
by deleting the $C^{(\ge j+1)}$ lines must be connected and tadpole--free. 
For clarity, we ignore the Wick lines. That is, we discuss a simplified
situation in which $\cW$ is Wick ordered with respect to  $C^{(j)}$ and
$\cW'$ is not Wick ordered at all\footnote{$^{(9)}$}{General Wick ordering
is treated in [FKTr2].}, so that we write
$$\eqalign{
\cW(0,0,\psi,\bar\psi) &= \lw\widetilde\cW(\psi,\bar\psi)\rw_{C^{( j)}} \cr
\widetilde\cW(\psi,\bar\psi)& = \smsum_{n\ge 0} \int {\sst dp_1 \cdots dq_n} 
   \,\tilde w_{2n}{\sst (p_1,\cdots, q_n)}\, 
    \de{\sst (p_1+\cdots+ p_n -q_1-\cdots- q_n)}
     \bar\psi{\sst (p_1)} \cdots \bar\psi{\sst (p_n)} \,
     \psi{\sst (q_1)} \cdots \psi{\sst (q_n)} \cr
}$$
Then
$$
w'_{2n} =\sum_\ell \frac{1}{\ell!}\ 
{{\rm connected\ diagrams\ without\ tadpoles\ with\ }
\choose  2n\ {\rm external\ legs\ and\ } \ell {\rm \ vertices\ from\ } \tilde w_2,
\tilde w_4,\cdots}
$$
All diagrams are labeled. A rooted diagram is a diagram with one distinguished
vertex, called the root. Clearly,
$$ 
w'_{2n} =\sum_\ell \frac{1}{\ell!}\frac{1}{\ell}\ 
{{\rm connected,\ rooted,\ tadpole\hbox{--}free\ diagrams\ with\ }
\choose  2n\ {\rm external\ legs\ and\ } \ell {\rm \ vertices\ from\ } \tilde w_2,
\tilde w_4,\cdots}
$$
The distance between two vertices in a diagram is the minimal number of 
lines needed to form a path connecting these two vertices. In a rooted 
diagram, the $r^{\rm th}$ ring is defined as the set of vertices of 
distance $r$ from the root. Thus, the ${\rm zero}^{\rm th}$ ring is 
the root itself, and 
the first ring consists of all vertices that are directly connected to the 
root by a line. Observe that the full subgraph formed by the union of the 
first $r$ rings of a diagram $G$ is again a connected diagram $G_r$. Each 
leg emanating from a vertex of the $r^{\rm th}$ ring that is not part
of a line in $G_r$ (that is, each external leg of $G_r$) is either an
external leg for the whole diagram or is 
connected to a vertex of the  $(r+1)^{\rm st}$ ring. Observe that, for 
each graph, there is an $r_0$ such that the $r^{\rm th}$ ring is empty for
all $r\ge r_0$.
$$
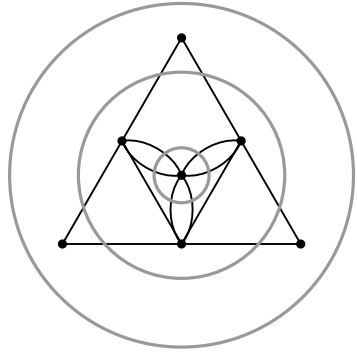
\figput{canGraph2}
$$
For simplicity, we only discuss the case that $n=0$, 
i.e. that there are no external legs.
Given a rooted diagram without external legs, we have the 
following  combinatorial data:
\item{$\bullet$}
$\ell_r = \sharp\ ({\rm vertices\ of\ the\ } r^{\rm th}\ {\rm ring})$
\item{$\bullet$} For $i=1,\cdots,\ell_r$, let 
$\left\{ \matrix{ \ga_{i,r}^-\cr\noalign{\vskip.02in} \ga_{i,r}^0\cr \noalign{\vskip.02in} \ga_{i,r}^+\cr}\right\}$ 
be the number of legs of the $i^{\rm th}$ vertex in the $r^{\rm th}$ ring
that are connected to a vertex in ring number
$\left\{ \matrix{ r-1\cr\noalign{\vskip-.04in} r\cr\noalign{\vskip-.04in} r+1\cr}\right\}$. 

\noindent
By the definition of ``ring'',
 the $i^{\rm th}$ vertex in the $r^{\rm th}$ ring has
$\ga_{i,r}^- + \ga_{i,r}^0 + \ga_{i,r}^+$ legs and
$\ga_{i,r}^- \ge 1$ for all $1\le r\le r_0$.
The sequences $\vec \ell = (\ell_1, \ell_2, \cdots)$ and 
$\vec \ga = ( \ga_{i,r}^-, \ga_{i,r}^0, \ga_{i,r}^+)_{r=1,2,\cdots \atop i=1,\cdots \ell_r}$
are called the combinatorial data of the rooted graph. 

We only exploit cancellations between connected rooted graphs with 
the same combinatorial 
data. So fix some combinatorial data  $\vec \ell, \vec \ga $ and let ${\cal G}$ 
be the set of all diagrams with these combinatorial data. Denote by 
$\tilde G_r$ the sum of the values of all subgraphs $G_r$ of graphs 
$G\in\cal G$. View it as a single vertex. It has 
$$
\smsum_{i=1}^{\ell_r} \ga_{i,r}^+ = \smsum_{i=1}^{\ell_{r+1}} \ga_{i,r+1}^-
$$
external legs. We describe how  $\tilde G_r$ is formed from $\tilde G_{r-1}$ 
and how the norm of $\tilde G_r$ is bounded in terms of the norm of 
$\tilde G_{r-1}$:
Connect each of the $\ell_r$ vertices of the $r^{\rm th}$ ring to 
$\tilde G_{r-1}$ by one line and apply contraction bounds. 
Then apply an integral bound for the 
$\sum_{i=1}^{\ell_r}(\ga_{i,r}^--1)$ remaining connections between 
$\tilde G_{r-1}$ and the $r^{\rm th}$ ring. 
$$
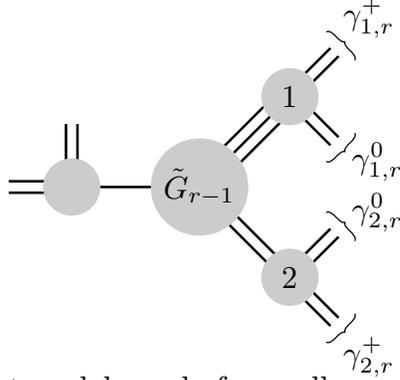
\figput{canGraph3}
$$
Then, using a variant of the integral bound, form all connections between 
the $\ga_{1,r}^0$ lines coming from the first vertex, the $\ga_{2,r}^0$ lines 
coming from the second vertex, $\cdots$ and the $\ga_{\ell_r,r}^0$ 
lines coming from the last vertex, always avoiding tadpoles. Repeat this procedure for all $r$ for
which $\ell_r\ne 0$. The bounds obtained in this way are summable over all
combinatorial data.

Observe that ladders
$$
\figput{canGraph4}
$$ 
have very special combinatorial data: Each $\ell_r$ is either zero, one or two,
$\ga_{i,r}^-=2$, $\ga_{i,r}^0=0$ and $\ga_{i,r}^+$ is either zero or two. In the
example 
$$
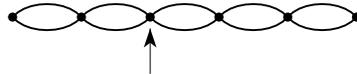
\figput{canGraph5}
$$
we have $\ell_1=\ell_2=2, \ell_3=1$, $\ga_{1,1}^+=\ga_{2,1}^+=\ga_{1,2}^+=2$,
$\ga_{2,2}^+=\ga_{1,3}^+=0$. Not all diagrams with such combinatorial data are
ladders, but there are so few of them that the non--ladder diagrams with these
combinatorial data can be bounded individually without generating divergences. 

In many cases, the combinatorial data of a diagram alone allow the detection of an overlapping loop. The two basic cases are:
\item{(i)} If, for some $r\ge 1$ and $1\le i\le\ell_r$, we have $\ga_{i,r}^-+\ga_{i,r}^0\ge 3$ then there is an overlapping loop, as indicated in the following figures. Here $v_i$ denotes the $i^{\rm th}$ vertex of the $r^{\rm th}$ ring.

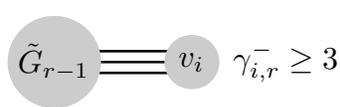
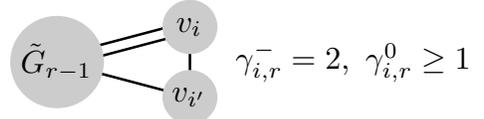
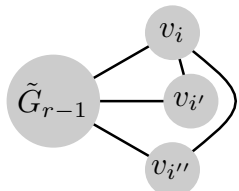
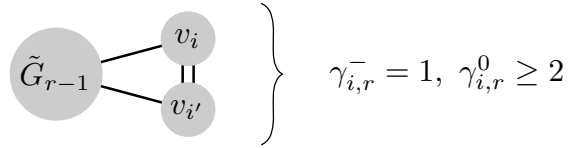
\centerline{\figput{canGraph6}\hskip1in\figput{canGraph7}}

\centerline{\figput{canGraph8}\hskip.4in\figput{canGraph9}}

\item{(ii)} If $\ell_r=2$, $\ell_{r+1}=1$ and $\ga_{1,r}^-+\ga_{1,r}^0\ge 2$,
$\ga_{1,r}^+\ge 1$, $\ga_{2,r}^+\ge 1$ then there is an overlapping loop as seen in the following figures.

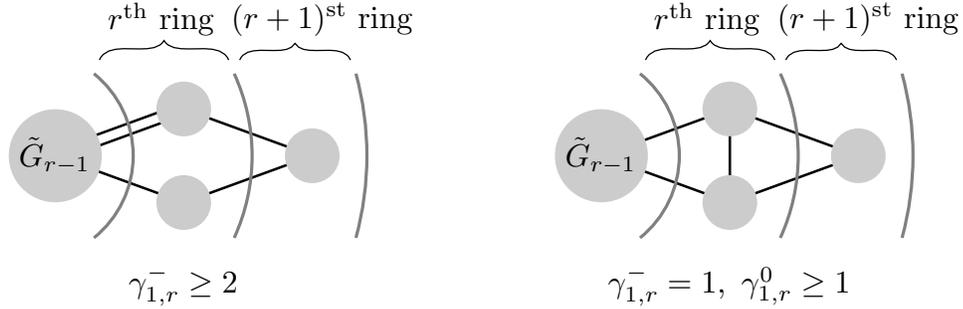
\centerline{\figput{canGraph10}\hskip.5in\figput{canGraph11}}

\noindent
These overlapping loops can be used to generate improved estimates by the techniques mentioned at the end of subsection 8 without seriously affecting 
the cancellations described above.

\noindent If one also takes external legs into account and if $w_2=0$, then 
cases (i) and (ii) are enough to identify at least one overlapping loop
in each four--legged diagram that does not have the combinatorial data of a ladder diagram. See \S VII of [FKTr2].

\vskip .5cm
\goodbreak
\Item{10.} { \it The Counterterm}\PG\pgNPIIj
\vskip .0in
The counterterm $\de e$ is constructed so that the proper self energy is
bounded by $\half|ik_0-e(\k)|$. That is, let 
$\check G_{2,j}(p)\,\de(p-q)$ be the Fourier transform of the two point Green's
function $G_{2,j}(x,y)$ constructed in Theorem \theoremNPmainthI, and 
define $\Si_j(k)$ by
$$
\check G_{2,j}(k) 
= \sfrac{U(k) - \nu^{(\ge j)}(k)}{\imath k_0 -e(\k) - \Si_j(k)}
$$
Then $|\Si_j(k)|\le \half|ik_0-e(\k)|$.

\noindent
To achieve this, we specify, for each scale $j$, a space $\cK_j$ of 
``allowed future counterterm contributions for scales after  $j$". It consists 
of functions $K(\k)$ of the vector part of $k=(k_0,\k)$ only. These functions 
are required to be bounded by a small constant times 
$\sfrac{\fl_{j+1}}{M^{j+1}}$. The numerator reflects overlapping loop 
volume improvement. We construct a map $\de e_j$ from $\cK_j$ to the 
space $\cE$ of counterterms such that, if one writes
$$
 \check G_{2,j}\big(k;\de e_j(K)\big)
= \sfrac{U(k) - \nu^{(\ge j)}(k)}{\imath k_0 -e(\k) - \Si_j(k;K)}
$$
then $\Si_j\big((0,\k);K\big)=K(\k)$ in the $j^{\rm th}$ neighbourhood. 
Think of $\de e_j(\k;K)$ as being of the form $\de \tilde e_j(\k;K)+K(\k)$
with $\de \tilde e_j(\k;K)$ implementing cancellations that have already
been identified and $K(\k)$ reserved to implement as yet unidentified 
cancellations. Since $\Si_j(k;K)$ equals $\de e_j(\k;K)$ plus higher order
contributions, it is easy to solve $\Si_j\big((0,\k);K\big)=K(\k)$ for
$\de \tilde e_j(\k;K)$. The algebra of this (recursive)
construction is presented in detail in \S\CHrenmap. We prove that $\de e =\lim\limits_{j\rightarrow \infty}\de e_j(0)$ exists and
has the required properties.

At each scale, the properties of the counterterms are used to obtain the 
bound for the $w_2(p,q)$ of order $\sfrac{1}{M^j}$ necessary for 
power counting. The choice indicated above guarantees that such a bound 
holds for momenta $k=q-p$ for which the ``temperature part" $k_0$ vanishes. 
To get this bound for arbitrary $k=(k_0,\k)$ in the $j^{\rm th}$ neighbourhood,
in particular when $|k_0|=O\big(\sfrac{1}{M^j}\big)$,  we show that the $k_0$ derivative
of this function is of order one. For this, we have to control derivatives (in
momentum space) of all the data in the renormalization construction. 
Control of derivatives in momentum space is also needed to provide decay
in position space. This is used, for example, in proving contraction bounds
through $L^1$ norms in position space.
 We pointed out in the subsection 1 that derivatives of $\de e_j(0)$
might blow up as $j\rightarrow \infty$. Therefore we have to pay special attention
to the behaviour of derivatives. This is the reason why we introduce, in
Definition \defNPFancynormdomain, the ``norm domain" that gives a convenient notation for bounds on
derivatives of functions.

\vfill\eject

\chap{Formal Renormalization Group Maps}\PG\pgNPIII

To simplify notation involving the fields, we define, for $\xi=(x_0,\x,\si,a)=(x,a) \in \bbbr \times \bbbr^d \times \{\uparrow,\downarrow\} \times \{0,1\}$, the internal fields
$$
\psi(\xi) = \cases{\psi(x) = \psi_\si(x_0,\x) & if $a=0$\cr
                   \bar\psi(x) = \bar\psi_\si(x_0,\x)& if $a=1$\cr
}$$ 
Similarly, we define for an external variable  
$\eta=(y_0,\y,\tau,b)=(y,b) \in \bbbr \times \bbbr^d \times \{\uparrow,\downarrow\} \times \{0,1\}$, the source fields
$$
\phi(\eta) = \cases{\phi(y) = \phi_\si(y_0,\y) & if $b=0$\cr
                   \bar\phi(y) = \bar\phi_\si(y_0,\y) & if $b=1$\cr
}$$ 
$\cB = \bbbr \times \bbbr^d \times \{\uparrow,\downarrow\}\times\{0,1\}$ is called the ``base space'' parameterizing the fields. 
An antisymmetric function $S(\xi,\xi')$ on $\cB\times\cB$ is called a covariance and determines a Grassmann Gaussian measure by
$$
\int \psi(\xi)\psi(\xi') \,d\mu_S(\psi) = S(\xi,\xi')
$$
A function $S(k)$ on momentum space, $\bbbr \times \bbbr^d$,
defines a function $S(x,x')$ on position space 
$\big(\bbbr \times \bbbr^d \times \{\uparrow,\downarrow\}\big)^2$ by
$$
S(x,x') = \de_{\si,\si'} \int \frac{d^{d+1}k}{(2\pi)^{d+1}} 
e^{\imath<k,x-x'>_-}S(k)
\EQN\eqnNPcovFT$$
Any function $S(x,x')$ on position space, $\big(\bbbr \times \bbbr^d \times \{\uparrow,\downarrow\}\big)^2$
defines a unique antisymmetric function $S(\xi,\xi')$ on $\cB\times\cB$ by
$$
S({\sst (x,a),(x',a')}) = \cases{ S(x,x') & if $a=0,\,a'=1$\cr
                                  -S(x',x) & if $a=1,\,a'=0$\cr
                                    0      & if $a=a'$\cr
}
\EQN\eqnNPantisymmCov$$
We denote the associated Grassmann Gaussian measure again by $d\mu_{S}$.

With the notation introduced above the source term of (\eqnNPphijpsi) is
$$
\phi J\psi 
=  \int d\xi\,d\xi' \ \phi(\xi)J(\xi,\xi')\psi(\xi')
=\psi J\phi
$$
where the operator $J$ has kernel 
$$
J\big((x_0,\x,\si,a),(x'_0,\x',\si',a')\big)
= \de(x_0-x'_0)\de(\x-\x')\de_{\si,\si'}
    \cases{1& if $a=1,\ a'=0$\cr
          -1& if $a=0,\ a'=1$\cr
           0& otherwise}
\EQN\eqnNPjdef$$

\definition{\STM\defNPrengroupmap (Renormalization Group Maps)}{
Let $S$ be a covariance and $\cW(\phi,\psi)$ a Grassmann function for which 
$Z=\int  e^{\cW(0,\ze)}\,d\mu_{S}(\ze)\ne 0$. 
We set
$$\eqalign{
\Om_S(\cW)(\phi,\psi)
& = \log\sfrac{1}{Z} 
    \int  e^{\cW(\phi,\psi+\ze)}\,d\mu_{S}(\ze)  \cr
\tilde \Om_S(\cW)(\phi,\psi)
& = \log\sfrac{1}{Z} 
    \int e^{\phi J\ze}\, e^{\cW(\phi,\psi+\ze)}\,d\mu_{S}(\ze)  \cr
}$$

}

\noindent
$\Om_S$ and $\tilde\Om_S$ map Grassmann functions in the variables $\phi,\psi,$
to Grassmann functions in the same variables. 
They obey the semigroup property
$$\eqalign{
\Om_{S_1+S_2}=\Om_{S_1}\circ\Om_{S_2}
\qquad,\qquad 
\tilde\Om_{S_1+S_2}=\tilde\Om_{S_1}\circ\tilde\Om_{S_2} 
}\EQN\eqnNPsemigrp$$
By Lemma \lemOStworengrpmaps\ of [FKTo2], they are related by 
$$\eqalign{
\tilde \Om_S(\cW)(\phi,\psi) 
&=\sfrac{1}{ 2}\phi JSJ \phi+\Om_S(\cW)(\phi,\psi+SJ\phi) \cr
}\EQN\eqnNPtworengrp$$
where, for any covariance $S$, 
$\phi S\phi=\int d\xi_1d\xi_2\ \phi(\xi_1)\, S(\xi_1,\xi_2)\,\phi(\xi_2)$.
Clearly
$$\deqalign{
\cG_\il(\phi;\de e)
&=\tilde\Om_{C^{\IR(\il)}(\de e)}(\widetilde \cV)(\phi,0)\qquad&{\rm with}\quad
\widetilde \cV(\phi,\psi)=\cV(\psi) \cr 
}$$
Observe that  
$$
C^{\IR(\il)}(k;0)=C^{\IR(1)}(k;0)+C^{(1)}(k)+\cdots+C^{(j-2)}(k)
+C^{(j-1)}(k)
$$
where
$$
C^{(i)}(k_0,\k) = \sfrac{\nu^{(i)}(k)}{\imath k_0 - e(\k)}
$$
Therefore, by induction and the semigroup property,
$$
\cG_\il(\phi;0) 
=\tilde\Om_{C^{(j-1)}}\circ\tilde\Om_{C^{(j-2)}}\circ\ \cdots\ \circ
\tilde\Om_{C^{(1)}}\circ\tilde\Om_{C^{\IR(1)}(0)} (\widetilde \cV)(\phi,0)
\EQN\eqnNPrginduct$$
Since $C^{(i)}(k)$ is supported on the $i^{\rm th}$ shell only, (\eqnNPrginduct)
would provide a convenient framework for a multiscale analysis 
of the unrenormalized generating functional $\cG_\il(\phi;0)$, by recursively controlling the effective interactions
$$\eqalign{
\cG_i(\phi,\psi;0)
&=\tilde\Om_{C^{(i-1)}}\circ\ \cdots\ \circ
\tilde\Om_{C^{(1)}}\circ\tilde\Om_{C^{\IR(1)}(0)} (\widetilde \cV)(\phi,\psi)\cr
&=\tilde\Om_{C^{(i-1)}}\big(\cG_{i-1}(0)\big)(\phi,\psi)
}\EQN\eqnNPrginductpsi$$

\vskip.3cm

We now describe an analog of (\eqnNPrginduct) for the case $\de e \ne 0$. 
It incorporates a number of technical modifications needed to maintain
control over the  bounds. These modifications include the introduction of 
a scale--dependent Wick ordering and a scale--dependent contribution to the counterterm, periodic shifting of a portion of the interaction into the covariance and the periodic isolation of purely $\phi$
dependent terms.
To this end, the effective interaction $\cG_i(\phi,\psi;0)$
of (\eqnNPrginductpsi) is replaced by a triple $(\cW,\cG,u)$ with 
\item{$\circ$}$\cG(\phi)$ being the purely $\phi$ dependent part of the effective interaction,
\item{$\circ$}$\cW(\phi,\psi)$ being the rest of the interaction and 
\item{$\circ$}$u$ being the kernel of a quadratic Grassmann function that has been moved from the effective interaction into the covariance  

\vskip.3cm

To help clarify the algebraic structure of this more complicated setting, 
we outline the construction in the category of formal power series,
without specifying the bounds that will ultimately be proven. To avoid formal power series in infinitely many variables, we introduce a coupling 
constant $\la$ into the interaction
$$
\cV(\psi) = \la\int_{(\bbbr\times\bbbr^d\times\{\uparrow,\downarrow\})^4} 
\hskip-.7in V(x_1,x_2,x_3,x_4)\, \bar\psi(x_1)\psi(x_2)\bar\psi(x_3)\psi(x_4)\
dx_1dx_2dx_3dx_4
$$
and deal with Grassmann algebra valued formal power series in $\la$. Correspondingly, the final counterterm, $\de e(\k)$ is made $\la$--dependent.
\definition{\STM\defNPformalfinalCTmSpace}{ 
The space of formal (final) counterterms, $\cE^\form$, consists of
the space of all formal power series $\de e(\k,\la)=\smsum\limits_{n=1}^\infty
\de e_n(\k)\la^n$ in $\la$ each of whose coefficients is supported in $\set{\k\in\bbbr^d}{U(\k)=1}$. 
}
\noindent
As indicated above the final counterterm $\de e$ is built up from contributions at each scale.
\definition{\STM\defNPformalscaleCTmSpace}{ 
The space $\fK_j^\form$ of formal (future) counterterms for scale $j$ is the 
space of all 
formal power series in $\la$ whose coefficients are antisymmetric, 
translation invariant functions of $\x,\ \x'$. The coefficient of $\la^0$
vanishes and the Fourier transform of each coefficient is
supported on ${\rm supp}\,\nu^{(\ge j+1)}\big((0,\k)\big)$.

}

\definition{\STM\defNPeffinttriple}{ A formal interaction triple at scale $j$
is a triple $(\cW,\cG,u)$ that obeys the following conditions.
\item{$\circ$} $\cW(\phi,\psi;K)$ is a  formal power series in $\la$,
whose coefficients are functions of $K\in\fK_{j}^\form$ that take values in the Grassmann algebra generated by the fields 
$\phi(\xi)$ and $\psi(\xi)$. Furthermore $\cW(\phi,0;K)=0$ 
and $\cW(\phi,\psi;K)\big|_{\la=0}=0$. 
 The coefficients of $\cW$ are  translation
invariant, spin independent and particle--number conserving.
\item{$\circ$} $\cG(\phi;K)$ is a formal power series in $\la$ whose coefficients are functions of $K\in\fK_{j}^\form$ that take values in the 
Grassmann algebra generated by the fields 
$\phi(\xi)$. The constant term $\cG(0;K)=0$. 
 The coefficients of $\cG$ are translation
invariant, spin independent and particle--number conserving.
\item{$\circ$}$u(\xi_1,\xi_2;K)$ is a formal power series in $\la$
whose coefficients are antisymmetric, spin independent, particle number
conserving, translation invariant functions of 
$\xi_1,\xi_2~\in~\cB$ and $K\in\fK_{j}^\form$. 
The Fourier transform\footnote{$^{(1)}$}
{A systematic set of Fourier transform conventions will be given in Definition
\:\defNPfourtrans. In the present context $\check u(k;K)$ is the Fourier
transform of $u\left((0,\0,\uparrow,1),(x_0,\x,\uparrow,0);K\right)$ as
in Theorem \theoremNPmainthII.}
$\check u(k;K)$ of $u$ 
obeys 
$$\eqalign{
\check u\big((0,\k);K\big) = - \check K(\k)\qquad\qquad
\check u\big(k;K\big)\big|_{\la=0} = -\check K(\k)\cr
}$$ 

}
\noindent The condition that $\cW(\phi,0;K)=0$ ensures that $\cG(\phi;K)$
contains the full pure $\phi$ part of the effective interaction.

As mentioned above, $u$ is the kernel of a quadratic Grassmann function 
that has been moved from the effective interaction into the covariance. 
Precisely,
\definition{\STM\defNPformalCovariances}{ 
\item{(i)} Let $u(\xi_1,\xi_2)$ be a formal power series in $\la$
whose coefficients are antisymmetric, spin independent, particle number
conserving, translation invariant functions of 
$\xi_1,\xi_2\in\cB$. Then
$$
C_u^{(j)}(k)=\frac{\nu^{(j)}(k)}{ik_0-e(\k)-\check u(k)}
$$

\item{(ii)} Let $u(\xi_1,\xi_2;K)$ be a formal power series in $\la$
whose coefficients are antisymmetric, spin independent, particle number
conserving, translation invariant functions of 
$\xi_1,\xi_2\in\cB$. Then, for $K\in\fK_{j}^\form$,
$$\eqalign{
C_j(u;K)(k) &= 
\sfrac{\nu^{(\ge j)}(k)}
{\imath k_0 -e(\k) -\check u(k;K) - \check  K(\k)\nu^{(\ge j+2)}(k)}\cr
D_j(u;K)(k) &= \sfrac{\nu^{(\ge j+1)}(k)}
{\imath k_0 -e(\k) -\check u(k;K)-\check  K(\k)\nu^{(\ge j+2)}(k)}\cr
}$$
}
For formal interaction triple $(\cW,\cG,u)$ at scale $j$, integrating out scale 
$j$ involves the evaluation of an integral with respect to the Gaussian measure with covariance $C_{u(K)}^{(j)}$. The effective interaction $\cW$ will be Wick ordered with respect to the covariance $C_j(u(K);K)$.
One important property of the Wick ordering covariances is that 
$D_j(u;K)=C_j(u;K)-C^{(j)}_{u(K)}$, so that
$$
\int \lw f(\psi+\ze)\rw_{C_j(u,K)}\ d\mu_{C^{(j)}_{u(K)}}(\ze)
=\lw f(\psi)\rw_{D_j(u,K)}
$$
for all Grassmann functions $f(\psi)$,
by Proposition \propBII.i and Lemma \lemBIV.ii  of [FKTr1].
This property prevents the formation of
Wick self--contractions

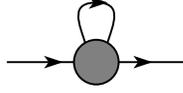
\centerline{ \figput{selfWick} }

\noindent and ensures that the
effective interaction resulting from integrating out scale 
$j$ is naturally Wick ordered with respect to 
the ``output Wick ordering covariance'' $D_j(u,K)$.
Also the Wick ordering covariances have been chosen so that, 
for $k$ of scale at least $j+3$,  
$\check u(k;K) + \check  K(\k)\nu^{(\ge j+2)}(k)=\check u(k;K)+\check  K(\k)$ vanishes for $k_0=0$. 
This property ensures that the denominator still vanishes only on the Fermi surface. 

\definition{\STM\defNPinoutmap}{
Integrating out the fields of scale $j$ is implemented
by the map $\Om_j$, 
which maps a formal interaction triple $(\cW,\cG,u)$ of scale $j$ to
the  triple  $(\cW',\cG',u)$  determined by 
$$\eqalign{
\lw\cW'(\phi,\psi;K)\rw_{\psi,D_{j}(u;K)}
&=\log\sfrac{1}{Z(\phi)}\int e^{\phi J\ze}\, 
e^{\lw \cW(\phi,\psi+\ze;K)\rw_{\psi,C_{j}(u;K)}}
\, d\mu_{C^{(j)}_{u(K)}}(\ze)\cr
\cG'(\phi) &= \cG(\phi) 
+\log\sfrac{Z(\phi)}{Z(0)}\cr
}$$
where 
$$
\log Z(\phi)=\int\Big[\log\int e^{\phi J\ze}\, 
e^{\lw \cW(\phi,\psi+\ze;K)\rw_{\psi,C_{j}(u;K)}}\, 
d\mu_{C^{(j)}_{u(K)}}(\ze)\Big]d\mu_{D_j(u;K)}(\psi)
$$
In fact $(\cW',\cG',u)$ is again a formal interaction triple of scale $j$.
That $\cW'(\phi,0;K)=0$ follows by inserting the definitions into
$$
\cW'(\phi,0;K)
=\int \lw \cW'(\phi,\psi;K)\rw_{\psi,D_j(u,K)}\ d\mu_{D_j(u,K)}(\psi)
$$
To verify $\cW'(\phi,\psi;K)\big|_{\la=0}=0$, observe that for $\la=0$
$$
\int e^{\phi J\ze}\, e^{\lw \cW(\phi,\psi+\ze;K)\rw_{\psi,C_{j}(u;K)}}\, d\mu_{C^{(j)}_{u(K)}}(\ze)
=\int e^{\phi J\ze}\,d\mu_{C^{(j)}_{u(K)}}(\ze)
=e^{{1\over 2}\phi JC^{(j)}_{u(K)}J\phi}
$$
is independent of $\psi$. To verify the various symmetries, apply remark
\remOSrengrppreserves\ of [FKTo2].

}
\remark{\STM\remNPinoutmap}{ Define, for all $1\le i\le j\le\infty$,
$$
C_u^{[i,j)}(k)=
\cases{\sfrac{\nu^{(\ge i)}(k)-\nu^{(\ge j)}(k)}
                {ik_0-e(\k)-\check u(k)[1-\nu^{(\ge j)}(k)]}& if $j<\infty$\cr
\sfrac{\nu^{(\ge i)}(k)}{ik_0-e(\k)-\check u(k)}&if $j=\infty$\cr}
$$
We also write $C_u^{(\ge i)}=C_u^{[i,\infty)}$.

Let $(\cW',\cG',u)=\Om_j(\cW,\cG,u)$.
Then, for any infrared cutoff $j+2\le\jbar\le\infty$, formally, ignoring the problems engendered by the infrared singularity,
$$\eqalign{
&\cG(\phi)+\log\frac{\int e^{\phi J\psi}\, e^{\lw \cW(\phi,\psi;K)\rw_{\psi,C_{j}(u;K)}}\, d\mu_{C^{[j,\jbar)}_u}(\psi)}
{\int e^{\lw \cW(0,\psi;K)\rw_{\psi,C_{j}(u;K)}}\, 
d\mu_{C^{[j,\jbar)}_u}(\psi)}\cr
&\hskip1in=\cG'(\phi)+\log\frac{\int e^{\phi J\psi}\, e^{\lw \cW'(\phi,\psi;K)\rw_{\psi,D_{j}(u;K)}}\, 
d\mu_{C^{[j+1,\jbar)}_u}(\psi)}
{\int e^{\lw \cW'(0,\psi;K)\rw_{\psi,D_{j}(u;K)}}\, d\mu_{C^{[j+1,\jbar)}_u}(\psi)}\cr
}$$
\prf 
Since $C^{[j,\jbar)}_u=C^{(j)}_u+C^{[j+1,\jbar)}_u$,
$$
\int f(\phi,\psi)\  d\mu_{C^{[j,\jbar)}_u}(\psi)
=\int\!\!\int f(\phi,,\psi+\ze)\  d\mu_{C^{(j)}_u}(\ze)\  d\mu_{C^{[j+1,\jbar)}_u}(\psi)
$$
by Proposition I.21 of [FKTffi]. Hence, by
Proposition \propBII.ii of [FKTr1],
$$\eqalign{
&\hskip-0.5in\log\int e^{\phi J\psi} 
e^{\lw \cW(\phi,\psi;K)\rw_{\psi,C_{j}(u;K)}}
\, d\mu_{C^{[j,\jbar)}_u}(\psi)\cr
& =\log\int\int e^{\phi J(\psi+\ze)}\,  e^{\lw \cW(\phi,\psi+\ze;K)\rw_{\psi,C_{j}(u;K)}}\, d\mu_{C^{(j)}_u}(\ze)\, d\mu_{C^{[j+1,\jbar)}_u}(\psi)\cr
& =\log\int e^{\phi J\psi}\,
 e^{\lw \cW'(\phi,\psi;K)\rw_{\psi,D_{j}(u;K)}+\log Z(\phi)}\, d\mu_{C^{[j+1,\jbar)}_u}(\psi)\cr
& =\log\int e^{\phi J\psi}\,
 e^{\lw \cW'(\phi,\psi;K)\rw_{\psi,D_{j}(u;K)}}\,d\mu_{C^{[j+1,\jbar)}_u}(\psi) +\log Z(\phi) \cr
& =\log\int e^{\phi J\psi}\,
 e^{\lw \cW'(\phi,\psi;K)\rw_{\psi,D_{j}(u;K)}}\,d\mu_{C^{[j+1,\jbar)}_u}(\psi) +\log Z(0)+\cG'(\phi)-\cG(\phi) \cr
}$$
Subtracting the same equation with $\phi=0$ gives the desired result.
\endproof

}

\vskip.3cm
When we derive bounds on the map $\Om_j$, we get improvements on the two-- and four--legged contributions to $\cW'(0,\psi)$ by exploiting overlapping loops.
See subsection 4 of \S\CHintroOverview\ and the introduction to [FKTr2]. However to ensure the presence of and to detect sufficiently many overlapping loops, we need that the two--legged part of
$\cW(0,\psi)$ vanishes. See the end of subsection 9 of \S\CHintroOverview\ and
Remark \remtheotheo\ of [FKTr2]. Therefore,
we wish that the formal interaction triple $(\cW,\cG,u)$ input to $\Om_j$
be an element of 
\definition{\STM\defNPinputData (Formal Input Data)}{ 
The space $\cD_{\rm in}^{(j,\form)}$ of formal input data consists of the set of all formal interaction triples $(\cW,\cG,u)$
at scale $j$, in the sense of Definition \defNPeffinttriple, obeying
\item{(i)} If the effective interaction $\cW(K)=\sum\limits_{m,n\ge 0} \cW_{m,n}$ with
$$\hskip-2pt
\cW_{m,n}\! =\!
\int\!\! \smprod_{i=1}^m\! d\eta_i \smprod_{\ell=1}^n\! d\xi_\ell  \ 
W_{m,n}({\sst \eta_1,\cdots,\eta_m,\xi_1,\cdots ,\xi_n})
\phi(\eta_1)\cdots\phi(\eta_m)
\psi(\xi_1)\cdots \psi(\xi_n)
$$
then $W_{0,2}=0$. 
\item{(ii)}  The coefficient of $\la^0$ in $\cG(\phi;K)$ is 
$\half\phi JC^{(< j)}_{-K}J\phi$. Here  
$C^{(< j)}_{u}=\sfrac{U(\k)-\nu^{(\ge j)}(k)}{ik_0-e(\k)-\check u(k)}$.

}

When $\Om_j$ is applied to an element of $\cD_{\rm in}^{(j,\form)}$,
the output no longer satisfies  condition (i) of Definition  \defNPinputData.
Rather, the output lies in
\definition{\STM\defNPoutputData (Formal Output Data)}{ 
The space $\cD_{\rm out}^{(j,\form)}$ of formal output data consists of the set of all formal interaction triples $(\cW,\cG,u)$
at scale $j$, in the sense of Definition
\defNPeffinttriple, for which the coefficient of $\la^0$ in $\cG(\phi;K)$ is $\half\phi JC^{(\le j)}_{-K}J\phi$.

}
\lemma{\STM\lemNPinOut}{Let $(\cW,\cG,u)\in \cD_{\rm in}^{(j,\form)}$. Then
$\Om_j(\cW,\cG,u)\in \cD_{\rm out}^{(j,\form)}$.
}
\prf Set $(\cW',\cG',u)=\Om_j(\cW,\cG,u)$. Then
$$\eqalign{
\cG'(\phi)\big|_{\la=0}&=\cG(\phi)\big|_{\la=0}
     +\log\sfrac{Z(\phi)}{Z(0)}\big|_{\la=0}\cr
&=\half\phi JC^{(< j)}_{-K}J\phi+\log\sfrac{Z(\phi)}{Z(0)}\big|_{\la=0}\cr
}$$
Since
$$
\log Z(\phi)\big|_{\la=0}
=\int\Big[\log\int e^{\ze J\phi}\, 
d\mu_{C^{(j)}_{-K}}(\ze)\Big]d\mu_{D_j(u;K)}(\psi)
=\half J\phi C^{(j)}_{-K}J\phi
$$
the result follows.
\endproof
\goodbreak
\vskip.3cm

Elements of the space $\cD_{\rm out}^{(j,\form)}$ are not of the form desired for the application of $\Om_{j+1}$, the map implementing the integration out of scale $j+1$. In particular, the two--point part of the effective interaction is nonzero and $\fK^\form_{j}$ is not the appropriate space of counterterms. Below, just before Proposition \:\propNPoutIn,
we construct maps 
$$
\cO_{j}:\cD_{\rm out}^{(j,\form)}\rightarrow \cD_{\rm in}^{(j+1,\form)}
$$ 
and, for each $(\cW,\cG,u)\in \cD_{\rm out}^{(j,\form)}$,
$$
\ren_{j,j+1}(\ \cdot\ ,\cW,u):\fK^\form_{j+1}\rightarrow\fK^\form_{j}
$$
with the following property:
If $(\cW,\cG,u)\in \cD_{\rm out}^{(j,\form)}$ and $(\cW',\cG',u')=\cO_{j}(\cW,\cG,u)$ and if $K'\in\fK^\form_{j+1}$ and 
$K=\ren_{j,j+1}(K',\cW,u)$, then, for any infrared cutoff $j+1\le\jbar\le\infty$, formally, ignoring the problems 
engendered by the infrared singularity,
$$\eqalign{
&\cG(\phi;K)+\log\frac{
\int e^{\phi J\psi}\, e^{\lw\cW(\phi,\psi;K)\rw_{\psi,D_{j}(u;K)}}\, 
                                       d\mu_{C^{[j+1,\jbar)}_{u(K)}}(\psi)  }{
\int e^{\lw\cW(0,\psi;K)\rw_{\psi,D_{j}(u;K)}}\, 
                                       d\mu_{C^{[j+1,\jbar)}_{u(K)}}(\psi)}\cr
&\hskip1in=
\cG'(\phi;K')+ \log\frac{
\int e^{\phi J\psi}\, e^{\lw\cW'(\phi,\psi;K')\rw_{\psi,C_{j+1}(u';K')}}\, 
                                        d\mu_{C^{[j+1,\jbar)}_{u'(K')}}(\psi)}{
\int e^{\lw \cW'(0,\psi;K')\rw_{\psi,C_{j+1}(u';K')}}\, 
                                        d\mu_{C^{[j+1,\jbar)}_{u'(K')}}(\psi)}\cr
}\EQN\eqnNPoutin$$
The map $\cO_{j}$ moves the two--point part of $\cW$ into the covariance,
through $u'$, and updates the Wick ordering covariance (for a precise statement, see (\:\eqnNPKuprime) below). The map $\ren_{j,j+1}$ introduces 
the contribution of the current scale into the counterterm.

Using the maps $\Om_j$ of Definition \defNPinoutmap\ and the maps $\cO_j$
and $\ren_{j,j+1}$ we can describe the renormalization group flow. We start by 
choosing an arbitrary but fixed $j_0\ge 2$ and integrate out all scales 
from $1$ to $j_0$ to arrive at the initial effective interaction triple
$(\cW^\out_{j_0},\cG^\out_{j_0},u_{j_0})\in \cD^{(j_0,\form)}_{\rm out}$ with
$$\eqalign{
\cW^\out_{j_0}&=\tilde\Om_{C^{(\le j_0)}_{-K}}(\widetilde\cV)(\phi,\psi)
-\tilde\Om_{C^{(\le j_0)}_{-K}}(\widetilde\cV)(\phi,0)\cr
\cG^\out_{j_0}&=\tilde\Om_{C^{(\le j_0)}_{-K}}(\widetilde\cV)(\phi,0)\cr
u_{j_0}&=-K\cr
}$$
The renormalization group flow is the concatenation of the maps
$\cO_{j_0}$, $\Om_{j_0+1}$, $\cO_{j_0+1}$, $\Om_{j_0+2}$, $\cdots$,
$\Om_{j}$, $\cO_{j}$, $\cdots$ applied to the initial datum. 
 Set
$$\eqalign{
(\cW^\inp_j,\cG^\inp_j,u_j)   &=\cO_{j-1}\circ\Om_{j-1}\circ\cO_{j-2}\circ\cdots\circ\Om_{j_0+1}\circ \cO_{j_0}
(\cW^\out_{j_0},\cG^\out_{j_0},u_{j_0})   \in \cD^{(j,\form)}_{\rm in}\cr
(\cW^\out_j,\cG^\out_{j},u_j)   &=\Om_j \circ\cO_{j-1}\circ\Om_{j-1}\circ\cO_{j-2}\circ\cdots\circ\Om_{j_0+1}\circ \cO_{j_0}
(\cW^\out_{j_0},\cG^\out_{j_0},u_{j_0})   \in \cD^{(j,\form)}_{\rm out} \cr
}\EQN\eqnNPinouttriple$$
We recursively define maps 
$\ren_{i,j} :\fK^\form_{j}\rightarrow \fK^\form_{i}, \ \ j_0\le i\le j\ $ by
$$\eqalign{
\ren_{j,j}(K) & = K \cr
\ren_{i,j} (K) & =\ren_{i,j-1}\big(\ren_{j-1,j}(K)\big) \quad {\rm for\ } j>i \cr
}$$
We define for $K\in \fK^\form_{j}$
$$
\de e_j(K) = \ren_{j_0,j}(K)
$$
and show that, for the generating function of the connected Green's 
functions at scale $\jbar$ of Theorem \theoremNPmainthI,
$$\eqalign{
\cG_\jbar(\phi,\bar\phi;\de  e_j(K)) 
 &= \cG^\inp_{j}(\phi;K) +
\log\frac{\int e^{\phi J\psi}\, e^{\lw\cW^\inp_j(\phi,\psi;K)\rw_{\psi,C_{j}(u_j;K)}}\, d\mu_{C^{[j,\jbar)}_{u_j(K)}}(\psi)}{
\int e^{\lw \cW^\inp_j(0,\psi;K)\rw_{\psi,C_{j}(u_j;K)}}\, d\mu_{C^{[j,\jbar)}_{u_j(K)}}(\psi)}\cr
& = \cG^\out_{j}(\phi;K) +
\log\frac{\int e^{\phi J\psi}\, e^{\lw \cW^\out_j(\phi,\psi;K)\rw_{\psi,D_{j}(u_j;K)}}\,
 d\mu_{C^{[j+1,\jbar)}_{u_j(K)}}(\psi)}{
\int e^{\lw \cW^\out_j(0,\psi;K)}\rw_{\psi,D_{j}(u_j;K)}\, d\mu_{C^{[j+1,\jbar)}_{u_j(K)}}(\psi)}
}\EQN\eqnNPgenfnrengrp$$
for $K\in\fK^\form_{j}$ and $j_0\le j\le\jbar-2$.
When $\jbar=\infty$, (\eqnNPgenfnrengrp) holds with $\cG_\infty$ being the 
formal generating function of the connected Green's 
functions (\eqnNPforgenfn).

\noindent
Equation (\eqnNPgenfnrengrp) is proven by induction on $j$, 
combining Remark \remNPinoutmap\ and (\eqnNPoutin). To start the induction, observe that 
$\cG^\out_{j_0}(\phi;K)+\cW^\out_{j_0}(\phi,\psi;K)
=\tilde\Om_{C^{(\le j_0)}_{-K}}(\widetilde\cV)$,
so that, by the semigroup property (\eqnNPsemigrp),
$$\eqalign{
\cG_\jbar(\phi,\bar\phi;K) 
& = \cG^\out_{j_0}(\phi;K) 
+\log\frac{\int e^{\phi J\psi}\, e^{ \cW^\out_{j_0}(\phi,\psi;K)}\,
 d\mu_{C^{(j_0,\jbar)}_{-K}(\psi)}}{
\int e^{\cW^\out_{j_0}(0,\psi;K)}\, d\mu_{C^{(j_0,\jbar)}_{-K}(\psi)}}\cr
& = \cG^\out_{j_0}(\phi;K) +
\log\frac{\int e^{\phi J\psi}\, e^{\lw \cW^\out_{j_0}(\phi,\psi;K)\rw_{\psi,D_{j_0}(u_{j_0};K)}}\,
 d\mu_{C^{(j_0,\jbar)}_{u_{j_0}(K)}}(\psi)}{
\int e^{\lw \cW^\out_{j_0}(0,\psi;K)}\rw_{\psi,D_{j_0}(u_{j_0};K)}\, d\mu_{C^{(j_0,\jbar)}_{u_{j_0}(K)}}(\psi)}\cr
}$$
with $D_{j_0}=0$.

In Theorem \:\theoremNPinduction, we shall prove bounds that show that the limits
$\de e =\lim\limits_{j\rightarrow\infty}\de e_j(0)$ and
$\cG(\phi,\bar\phi;\de  e)
=\lim\limits_{j\rightarrow\infty}\cG^\out_{j}(\phi;0)
=\lim\limits_{j\rightarrow\infty}\cG^\inp_{j}(\phi;0)$ 
exist. To prove Theorem \theoremNPmainthI\ we show that $\lim\limits_{j\rightarrow\infty}
\big(\cG_{j}(\phi,\bar\phi)-\cG^\out_{j}(\phi;0)\big)=0$.

\vskip0.3cm

We now describe the passage from output data to input data, that is the maps 
$$
\cO_{j}:\cD_{\rm out}^{(j,\form)}\rightarrow \cD_{\rm in}^{(j+1,\form)}
\qquad
\ren_{j,j+1}(\ \cdot\ ,\cW,u):\fK^\form_{j+1}\rightarrow\fK^\form_{j}
$$
Let 
$(\cW,\cG,u)\in \cD^{(j,\form)}_{\rm out}$ and write
$$
\cW(0,\psi;K) =\sum\limits_{m\ge 2} 
\int d\xi_1\cdots d\xi_m\ W_{0,m}(\xi_1,\cdots ,\xi_m;K)
\ \psi(\xi_1)\cdots \psi(\xi_m)
$$
with each  $W_{0,m}(\xi_1,\cdots ,\xi_m;K)$ antisymmetric under permutation of
the $\xi_i$'s. 

To perform the reWick ordering, observe that, if $E$ is any covariance and 
$$
\lw \cW(\phi,\psi;K)\rw_{\psi,D_{j}(u;K)}
=\lw \tilde \cW(\phi,\psi;K)\rw_{\psi,D_{j}(u;K)+E}
$$
then, by Lemma \propBII.i of [FKTr1],
$$
\tilde \cW(\phi,\psi;K)
=\int \cW(\phi,\psi+\psi';K)\ d\mu_{E}(\psi')
$$
We wish to choose the covariance $E$ such that, after we reWick order $\lw \cW(\phi,\psi;K)\rw_{\psi,D_{j}(u;K)}$ to 
$\lw \tilde \cW(\phi,\psi;K)\rw_{\psi,D_{j}(u;K)+E}$ and move the quadratic
part of $\tilde \cW(\phi,\psi;K)$ into the covariance, replacing $u(K)$ by $u'(K')$,
then the Wick ordering covariance $D_{j}(u;K)+E$ is exactly
$C_{j+1}(u';K')$. When we replace $u(K)$ by $u'(K')$, we choose $K\in\fK^\form_{j}$ as a 
function of $K'\in\fK^\form_{j+1}$ in such a way that $u'(K')$ fulfills the third condition of Definition \defNPeffinttriple. This function will be denoted 
$K(K')=\ren_{j,j+1}(K',\cW,u)=K'+\de K(K')$.

The unknowns in the scheme outlined in the last paragraph are $E$ and $\de K$.
They are determined implicitly by the requirements of the last paragraph. We choose to express $E$ and $\de K$ in terms of one function $q(\xi_1,\xi_2;K')$, 
with $\half q$ being the kernel of $\tilde \cW_{0,2}$.  Once, $q$ is determined,
we set
$$\eqalign{
 \de \check K(\k;K';q) &= \check q\big((0,\k);K'\big)\,
\nu^{(\ge j+1)}((0,\k))\cr 
K(K';q) &= K'+\de K(K';q) \cr
\check u'(k;K';q) &=  \check u(k;K(K';q))+ \check q(k;K') \nu^{(\ge j+1)}(k)\cr
E(K';q)&=C_{j+1}\big( u'(\,\cdot\,;q);K' \big) - D_{j}( u;K(K';q))\cr
}\EQN\eqnNPKuprime$$
Set
$$
\tilde \cW(\phi,\psi;K';q)
=\int \cW(\phi,\psi+\psi';K(K';q))\ d\mu_{E(K';q)}(\psi')
$$
and expand
$$
\tilde\cW(0,\psi;K';q) =\sum\limits_{m\ge 0} 
\int d\xi_1\cdots d\xi_m\ \tilde W_{0,m}(\xi_1,\cdots ,\xi_m;K';q)
\ \psi(\xi_1)\cdots \psi(\xi_m)
$$
The requirement that $\half q$ be the kernel of $\tilde \cW_{0,2}$
is now an implicit equation.

\lemma{\STM\lemNPformalselfconsistent}{
There is a unique formal power series $q_0(\xi_1,\xi_2;K')$ in $\la$ that
solves the equation
$$\eqalign{
\half q(K') &=\tilde W_{0,2}\big(K';q(K')\big)\cr
}\EQN\eqnNPformalselfconsistent$$
The coefficient of $\la^0$ in $q_0$  vanishes.
}
\prf 
Equation (\eqnNPformalselfconsistent) is the form $q=F(\la,q)$, with $F$ 
being $C^\infty$ in $\la$ and $q$ and with $F(\la,0)$ being of order at 
least $\la$. An easy formal power series argument yields the result.
\endproof

\noindent
We define, for each $K'\in \fK^\form_{j+1}$
$$\eqalign{
\tilde \cW(\phi,\psi;K')
&=\tilde \cW(\phi,\psi;K';q_0(K'))\cr
\cW'(\phi,\psi;K') &=\tilde \cW\big(\phi,\psi;K'\big)
         -\tilde \cW\big(\phi,0;K'\big)
         -\half\int d\xi_1 d\xi_2\ q_0(\xi_1,\xi_2;K')\psi(\xi_1)\psi(\xi_2) \cr
\cG'(\phi;K') &= \cG(\phi;K(K'))  +\tilde\cW(\phi,0;K') -\tilde \cW(0,0;K')\cr
u'(K') &=  u'(K';q_0(K'))\cr
}$$
and
$$
\cO_{j}(\cW,\cG,u)=(\cW',\cG',u')
$$
We also define
$$
\ren_{j,j+1}(K',\cW,u)=K(K')= K(K';q_0(K'))\in \fK^\form_{j}
$$

\proposition{\STM\propNPoutIn}{
Let $j\ge j_0$, 
$(\cW,\cG,u)\in \cD^{(j,\form)}_{\rm out}$ and $(\cW',\cG',u')=\cO_{j}(\cW,\cG,u)$. Then
\Item a)
$$
(\cW',\cG',u')\in \cD^{(j+1,\form)}_{\rm in}
$$
\Item b) 
If $K'\in\fK_{j+1}^\form$ and $K=\ren_{j,j+1}(K',\cW,u)$ then, 
formally, ignoring the problems engendered by the infrared singularity,
$$\eqalign{
&\cG(\phi;K)+\log\frac{
\int e^{\phi J\psi}\, e^{\lw\cW(\phi,\psi;K)\rw_{\psi,D_{j}(u;K)}}\, 
                                       d\mu_{C^{[j+1,\jbar)}_{u(K)}}(\psi)  }{
\int e^{\lw\cW(0,\psi;K)\rw_{\psi,D_{j}(u;K)}}\, 
                                       d\mu_{C^{[j+1,\jbar)}_{u(K)}}(\psi)}\cr
&\hskip1in=
\cG'(\phi;K')+ \log\frac{
\int e^{\phi J\psi}\, e^{\lw\cW'(\phi,\psi;K')\rw_{\psi,C_{j+1}(u';K')}}\, 
                                        d\mu_{C^{[j+1,\jbar)}_{u'(K')}}(\psi)}{
\int e^{\lw \cW'(0,\psi;K')\rw_{\psi,C_{j+1}(u';K')}}\, 
                                        d\mu_{C^{[j+1,\jbar)}_{u'(K')}}(\psi)}\cr
}$$
if $j+1<\jbar\le\infty$.
}
\prf 
Let $K'\in \fK^\form_{j+1}$ and set $K = \ren_{j,j+1}(K',\cW,u)$.
\Item{a)}
We first verify that $(\cW',\cG',u')$ is a formal interaction triple. 
The only condition of Definition \defNPeffinttriple\ that is not trivially satisfied
is
$$\eqalign{
\check u'\big((0,\k);K'\big) 
&=\check u\big((0,\k);K\big(K';q_0(K')\big)\big)
+ \check q_0((0,\k);K') \nu^{(\ge j+1)}(0,\k)\cr
&=-\check K(\k;K';q_0(K'))+ \de \check K(\k;K';q_0(K'))\cr
&= - \check K'(\k)
}$$
We now verify the conditions of Definition \defNPinputData. 
 That $W'_{0,2}$ vanishes amounts to
$\tilde W_{0,2}=\half q_0$ which is  Lemma \lemNPformalselfconsistent.
Condition (ii) of Definition \defNPinputData\ is trivially fulfilled.
\Item b) 
Observe that 
$$\eqalign{
C^{[j+1,\jbar)}_{u'(K')} &= C^{[j+1,\jbar)}_{u(K(K'))+q_0(K')\nu^{(\ge j+1)}} \cr
}\EQN\eqnOutII $$
Set $\cU(K')=\half\int d\xi_1 d\xi_2\ q_0(\xi_1,\xi_2;K') \lw\psi(\xi_1)\psi(\xi_2)\rw_{C_{j+1}\!(u';K')}\ $. 
By Lemma \lemOSappGrassII\ of [FKTo2] and (\eqnOutII)
$$\eqalignno{
\cG(\phi;&K)+
\log\int e^{\phi J\psi}e^{\lw\cW(\phi,\psi;K)\rw_{\psi,D_{j}(u;K)}}\,
            d\mu_{C^{[j+1,\jbar)}_{u(K)}}(\psi)\cr
& =\cG'(\phi;K') +
 \log\int e^{\phi J\psi} 
        e^{\lw\tilde \cW(\phi,\psi;K)\rw_{\psi,C_{j+1}(u';K')}
        -\cU+\cU +(\cG(\phi;K)-\cG'(\phi;K')) }\, 
        d\mu_{C^{[j+1,\jbar)}_{u(K)}}(\psi)\cr
& =\cG'(\phi;K') +
  \log\int e^{\phi J\psi}\,
      e^{\lw\cW'(\phi,\psi;K')\rw_{\psi,C_{j+1}(u';K')}+\cU}\, 
     d\mu_{C^{[j+1,\jbar)}_{u(K)}}(\psi) +\const \cr
& =\cG'(\phi;K') +
   \log\int e^{\phi J\psi} 
    e^{\lw\cW'(\phi,\psi;K')\rw_{\psi,C_{j+1}(u';K')}}\,
    d\mu_{C^{[j+1,\jbar)}_{u(K)+q_0(K')\nu^{(\ge j+1)}}}(\psi) +\const \cr
& =\cG'(\phi;K') +
   \log\int e^{\phi J\psi} 
    e^{\lw\cW'(\phi,\psi;K')\rw_{\psi,C_{j+1}(u';K')}}\, 
    d\mu_{C^{[j+1,\jbar)}_{u'(K')}}(\psi) +\const \cr
}$$
Subtracting the same equation with $\phi=0$ gives part b).
\endproof

\vfill\eject

    \def\Mom{{\rm Mom}}
     \def\FC{{F}}

\appendix{\APappModelComp}{Model Computations}\PG\pgNPA

This appendix provides a number of model computations that illustrate
important features of the present construction. Various other model computations
are given in the introductory sections of this paper and other papers
in this series. Here is a table of model computations and their locations.

\def\gap{0.05in}

\vskip.1in\centerline{
\vbox{\offinterlineskip
\hrule
\halign{\vrule#&
         \strut\hskip\gap\hfil#\hfil&
         \hskip\gap\vrule#\hskip\gap&
          \hfil#\hfil&
           \hskip\gap\vrule#\cr
height2pt&\omit&&\omit&\cr
&Topic&&Location&\cr
height2pt&\omit&&\omit&\cr
\noalign{\hrule}
height2pt&\omit&&\omit&\cr
&Overlapping loop volume improvement&&\S\CHintroOverview, subsection 4&\cr
height2pt&\omit&&\omit&\cr
\noalign{\hrule}
height2pt&\omit&&\omit&\cr
&Particle--particle bubble volume improvement&&\S\CHintroOverview, subsection 5&\cr
height2pt&\omit&&\omit&\cr
\noalign{\hrule}
height2pt&\omit&&\omit&\cr
&Particle--hole bubble sign cancellation&&[FKTl, Lemma \lemLADprimitivemanfred]&\cr
height2pt&\omit&&\omit&\cr
\noalign{\hrule}
height2pt&\omit&&\omit&\cr
&Sectorization and conservation of momentum&&Example \exNPsectorizebound&\cr
height2pt&\omit&&\omit&\cr
\noalign{\hrule}
height2pt&\omit&&\omit&\cr
&Power counting with sectorization&&\S\CHintroOverview, subsection 8&\cr
height2pt&\omit&&\omit&\cr
\noalign{\hrule}
height2pt&\omit&&\omit&\cr
&Overlapping Loops and Sectors&&Example \exNPloopsector&\cr
height2pt&\omit&&\omit&\cr
\noalign{\hrule}
height2pt&\omit&&\omit&\cr
&Sectorization and change of scale&&Example \exNPsectorizescalechange&\cr
height2pt&\omit&&\omit&\cr
\noalign{\hrule}
height2pt&\omit&&\omit&\cr
&Cancellations at high orders of perturbation theory&&\S\CHintroOverview, subsection 9&\cr
height2pt&\omit&&\omit&\cr
& &&[FKTr2, \S X]&\cr
height2pt&\omit&&\omit&\cr
& &&[FKTo1, \S V]&\cr
height2pt&\omit&&\omit&\cr
}\hrule}}
\vskip.1in


\example{\STM\exNPsectorizebound (Sectorization and Conservation of Momentum)}{
Sectors enable us to apply $L^1$ norms to functions
for which the $L^1$ norm well--approximates the $L^\infty$ norm of the
Fourier transform. We provide a simple illustration of the use of sectorization as a tool
for bounding $\|\ \|_{1,\infty}$ norms of Green's functions built from
$C^{(j)}$. 

Recall that $C^{(j)}(k)=\sfrac{\nu^{(j)}(k)}{ik_0-e(\k)}$ and that,
on the support of $\nu^{(j)}$, $\big|ik_0-e(\k)\big|\approx\sfrac{1}{M^j}$.
To make this example as explicit as possible, we suppress $k_0$, choose
$d=3$, choose $e(\k)=|\k|^2-1$, replace $ik_0-e(\k)$ by $\sfrac{|\k|}{M^j}$
and replace $\nu^{(j)}(k)$ by $\varphi\big(M^j(|\k|-1)\big)$ where 
$\varphi\in C^\infty([-1,1])$ is real and even. So we define
$$
c^{(j)}(\k)=\sfrac{M^j}{|\k|}\varphi\big(M^j(|\k|-1)\big)
\qquad
c^{(j)}(\x,\x')=\int\sfrac{d^d\k}{(2\pi)^d}\ e^{i\k\cdot(\x-\x')} c^{(j)}(\k)
$$
We have rigged things
so that we can compute $c^{(j)}(\x,\x')$ relatively explicitly using 
the Fourier transform 
$\int_{S^{d-1}}d\si(\k')\ \ e^{ir\k'\cdot(\x-\x')}
= 2^{{d\over 2}-1}\Ga\big(\sfrac{d}{2}\big)\om_d\ (r|\x-\x'|)^{1-{d\over 2}}
J_{{d\over 2}-1}(r|\x-\x'|)$ of the unit sphere in $\bbbr^d$. Here $\Ga$,
$\om_d$ and $J_\nu$ are the Gamma function, the surface area of $S^{d-1}$ and
the Bessel function of order $\nu$, respectively. The answer is
$$
c^{(j)}(\x,\x')=\sfrac{1}{2\pi^2|\x-\x'|}\ \sin\big(|\x-\x'|\big)\hat\varphi\big(\sfrac{|\x-\x'|}{M^j}\big)
$$
In particular
$$\eqalign{
\big\|c^{(j)}(\x,\x')\big\|_{1,\infty}
=\int d^3\x\ \sfrac{1}{2\pi^2|\x|}\ \Big|\sin\big(|\x|\big)\hat\varphi\big(\sfrac{|\x|}{M^j}\big)\Big|
=\sfrac{2}{\pi}\int_0^\infty \!\!\!dr\ r \Big|\sin(r)\hat\varphi\big(\sfrac{r}{M^j}\big)\Big|
}$$
For large $j$, $|\sin(r)|$ is much more rapidly varying than $\hat\varphi\big(\sfrac{r}{M^j}\big)$
and replacing $|\sin(r)|$ by its average value, $\sfrac{2}{\pi}$, introduces
only a small error.
$$\eqalign{
\big\|c^{(j)}(\x,\x')\big\|_{1,\infty}
=\Big(\sfrac{4}{\pi^2}+O\big(\sfrac{1}{M^j}\big)\Big)\int_0^\infty \!\!\!dr\ r \Big|\hat\varphi\big(\sfrac{r}{M^j}\big)\Big|
=M^{2j}\Big(\sfrac{4}{\pi^2}+O\big(\sfrac{1}{M^j}\big)\Big)\int_0^\infty \!\!\!dr\ r \big|\hat\varphi(r)\big|
}\EQN\eqnSMPlone$$
Observe that $\big\|c^{(j)}(\x,\x')\big\|_{1,\infty}$ is a factor of about
$M^j$ larger than $\sup_\k \big|c^{(j)}(\k)\big|\sim M^j$.

Now introduce a sectorization $\Si$ of the Fermi surface 
$\set{\k\in\bbbr^3}{|\k|=1}$ as in subsection 8 of \S\CHintroOverview\ 
(for details, see Definition \defNPsectors) using 
approximately square sectors of side $\sfrac{1}{M^{j/2}}$. We may construct,
also as in subsection 8 of \S\CHintroOverview\ (for details, see 
(\eqnOSpartunit) of [FKTo3] and Lemma \lemOSsectpartunit\ of [FKTo3]), 
a partition of unity, $\chi_s,\ s\in\Si$, of the support of $\varphi\big(M^j(|\k|-1)\big)$ such that
$$
c^{(j)}_s(\k)=\sfrac{M^j}{|\k|}\varphi\big(M^j(|\k|-1)\big)\chi_s(\k)
\qquad
c^{(j)}_s(\x,\x')
=\int\sfrac{d^d\k}{(2\pi)^d}\ e^{i\k\cdot(\x-\x')} c^{(j)}_s(\k)
$$
obeys
$
\big\|c^{(j)}_s(\x,\x')\big\|_{1,\infty}
\le\abcst\, M^j
$.
The proof of this bound was sketched in subsection 7 of \S\CHintroOverview.
For details, see Proposition \propOSGenDecay\  and Lemma  \lemOSsectorderiv\ 
of [FKTo3]. Observe that, with sectorization,
$$ 
\big\|c^{(j)}_s(\x,\x')\big\|_{1,\infty}
\le\abcst\, \sup_\k \big|c^{(j)}_s(\k)\big| 
$$
Had we chosen sectors of side $\sfrac{1}{M^{\aleph j}}$ with $\aleph<\half$,
this would no longer be the case. Also observe that, since $|\Si|$ is of order
$\big(M^{j/2}\big)^2$,
$$
\big\|c^{(j)}(\x,\x')\big\|_{1,\infty}
\le \sum_{s\in\Si}\big\|c^{(j)}_s(\x,\x')\big\|_{1,\infty}
\le\abcst\, \big(M^{j/2}\big)^2 M^{j}
=\abcst\,  M^{2j}
$$
recovers (\eqnSMPlone), up to an unimportant constant, by a technique that extends to nonround Fermi surfaces, for
which explicit computations of $c^{(j)}(\x,\x')$ are not available. 

Finally, consider 
$$
G_2(\x,\x')=\int d\y\ c^{(j)}(\x,\y)c^{(j)}(\y,\x')
$$
This would be a (first order) contribution to a model with covariance 
$c^{(j)}$ and an ultralocal quadratic interaction. If we attempt to bound
$\big\|G_2(\x,\x')\big\|_{1,\infty}$ just using $\big\|c^{(j)}(\x,\x')\big\|_{1,\infty}$,
we get
$$\eqalign{
\big\|G_2(\x,\x')\big\|_{1,\infty}
&=\int d\x\ \big|G_2(\x,\0)\big|
=\int d\x\Big|\int d\y\ c^{(j)}(\x,\y)c^{(j)}(\y,\0)\Big|\cr
&\le\int d\x \int d\y\ \big|c^{(j)}(\x,\y)c^{(j)}(\y,\0)\big|
=\big\|c^{(j)}(\x,\x')\big\|_{1,\infty}^2
\sim \const M^{4j}\cr
}$$
which is  a terrible answer: $G_2(\x,\x')$ is the Fourier transform of
$c^{(j)}(\k)^2=\sfrac{M^{2j}}{|\k|^2}\varphi({\sst M^j(|\k|-1)})^2$
 and
 $\sfrac{1}{|\k|^2}\varphi({\sst M^j(|\k|-1)})^2$ is very much like
$\sfrac{1}{|\k|}\varphi({\sst M^j(|\k|-1)})$, so the real behaviour of
$\big\|G_2(\x,\x')\big\|_{1,\infty}$ is $M^{3j}$. We may recover this real
behaviour using sectors.
$$\eqalign{
\big\|G_2(\x,\x')\big\|_{1,\infty}
&=\int d\x\ \Big|\int d\y\ c^{(j)}(\x,\y)c^{(j)}(\y,\0)\Big|
=\int d\x\ \Big|\sum_{s,s'\in\Si}\int d\y\ c^{(j)}_s(\x,\y)c^{(j)}_{s'}(\y,\0)\Big|\cr
&\le\sum_{s,s'\in\Si}\int d\x\ \Big|\int d\y\ c^{(j)}_s(\x,\y)c^{(j)}_{s'}(\y,\0)\Big|\cr
}$$
But $\int d\y\ c^{(j)}_s(\x,\y)c^{(j)}_{s'}(\y,\0)$ is the Fourier transform
of $c^{(j)}_s(\k)c^{(j)}_{s'}(\k)=c^{(j)}(\k)^2\chi_s(\k)\chi_{s'}(\k)$,
which vanishes identically unless the supports of $\chi_s$ and $\chi_{s'}$
overlap. For each fixed $s\in\Si$, there are at most $9$ sectors $s'\in\Si$
that overlap with $s$. Hence
$$\eqalign{
\big\|G_2(\x,\x')\big\|_{1,\infty}
&\le 9|\Si|\max_{s,s'\in\Si}
\int d\x\ \Big|\sum_{s,s'\in\Si}\int d\y\ c^{(j)}_s(\x,\y)c^{(j)}_{s'}(\y,\0)\Big|\cr
&\le 9|\Si|\max_{s,s'\in\Si}
\big\|c^{(j)}_s(\x,\x')\big\|_{1,\infty}\big\|c^{(j)}_{s'}(\x,\x')\big\|_{1,\infty}
\le \const |\Si| M^{2j}\cr
&\le \const  M^{3j}\cr
}$$
as desired.

}

\goodbreak

\example{\STM\exNPloopsector (Overlapping Loops and Sectors)}{

Let $\Ga$ be the diagram of subsection 4 in \S\CHintroOverview\ with vertices $w_4$ fulfilling the bounds
$$
\tn\hat\om_{4}\tn_{1,\Si}= O\big( \sfrac{1}{\fl} \big), \qquad \qquad
\tn\hat\om_{4}\tn_{3,\Si}= O\big(1)
$$
of (\eqnOVsectorvertexreq) and (\eqnOVsectorvertexthreereq). The naive
power counting bound (\eqnOVsectorvertexreq) gives that 
$\tn\hat\Ga\tn_{1,\Si}= O\big( \sfrac{1}{\fl} \big)$. We show that exploiting 
overlapping loop volume improvement leads to the bounds 
$\tn\hat\Ga \tn_{1,\Si}= O(1)$ and $\tn\hat\Ga \tn_{3,\Si}= O(\fl)$.

Let $\Ga'(p_1,p_2,p_3',q_1,q_2',q_3')$ be the six legged subgraph consisting of the left two vertices of $\Ga$.

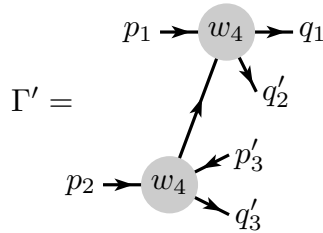
\centerline{\figput{dbubble8a}}

\noindent
As in (\eqnOVsectcontrbnd),
$$\deqalign{
\tn\hat\Ga'\tn_{1,\Si}
&\le  9 \ \tn\hat\om_4\tn^2_{1,\Si}\ \max_{s\in \Si} \|C^{(j)}_s(x)\|_1
&\le O\big(\sfrac{M^j}{\fl^2}\big)\cr
\tn\hat\Ga'\tn_{3,\Si}
&\le  9 \ \tn\hat\om_4\tn_{3,\Si}\tn\hat\om_4\tn_{1,\Si}\ \max_{s\in \Si} \|C^{(j)}_s(x)\|_1
&\le O\big(\sfrac{M^j}{\fl}\big)\cr
}$$
Clearly, $\Ga$ is obtained by joining $\Ga'$ to $w_4$ with three lines.

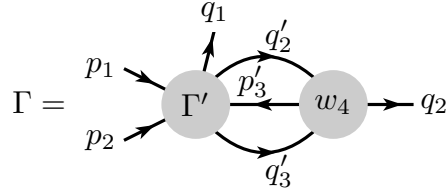
\centerline{\figput{dbubble8b}}

\noindent
As in subsection 8 of \S\CHintroOverview, by conservation of momentum,
$$\eqalign{
&\hat \Ga ({\sst (x_1,s_1),(x_2,s_2),(x_3,s_3),(x_4,s_4) }) \cr
&\hskip 1cm =\sum_{\si_i,\si_i',\si_i'' \in \Si  
\atop{\si_i\cap \si_i' \cap \si_i'' \ne \emptyset
\atop {\rm for\ } i=1,2,3 } }
\int {\sst dy_1 dy_2 dy_3\,dz_1 dz_2  dz_3}\ 
\hat \Ga' ({\sst (x_1,s_1),(x_2,s_2),(y_1,\si_1 ),
(x_3,s_3),(y_2,\si_2 ),(y_3,\si_3 ) })\cr
&\hskip 4cm 
C_{\si_1'}^{(j)}{\sst (y_1-z_1)}\,C_{\si_2'}^{(j)}{\sst  (z_2-y_2)} 
\,C_{\si_3'}^{(j)}{\sst  (z_3-y_3)} \
\hat \om_4 ({\sst (z_2,\si_2'' ),(z_3,\si_3'' ),(z_1,\si_1'' ),(x_4,s_4) })
}$$
so that
$$\eqalign{
&\big\| \hat \Ga ({\sst (\cdot,s_1),(\cdot,s_2),(\cdot,s_3),(\cdot,s_4) })
\big\|_{1,\infty}  \cr
&\hskip 1cm \le \sum_{\si_i,\si_i',\si_i'' \in \Si  
\atop{\si_i\cap \si_i' \cap \si_i'' \ne \emptyset
\atop {\rm for\ } i=1,2,3 } }
\| \hat \Ga' ({\sst (\cdot,s_1),(\cdot,s_2),(\cdot,\si_1 ),
(\cdot,s_3),(\cdot,\si_2 ),(\cdot,\si_3 ) }) \|_{1,\infty} \cr
&\hskip 3cm  \| C_{\si_1'}^{(j)}{\sst(x )}\|_1 
 \| C_{\si_2'}^{(j)}{\sst(x )}\|_\infty 
 \| C_{\si_3'}^{(j)}{\sst(x )}\|_\infty 
 \ \| \hat \om_4 ({\sst (\cdot,\si_2'' ),(\cdot,\si_3'' ),
(\cdot,\si_1'' ),(\cdot,s_4) }) \|_{1,\infty}
}$$
For fixed sectors $s_1,s_2,s_3$
$$\eqalign{
&\sum_{s_4}\big\| \hat \Ga ({\sst (\cdot,s_1),(\cdot,s_2),(\cdot,s_3),(\cdot,s_4) })
\big\|_{1,\infty}  \cr
&\hskip 1cm \le 3^6 \sum_{\si_1,\si_2,\si_3 \in \Si }
\| \hat \Ga' ({\sst (\cdot,s_1),(\cdot,s_2),(\cdot,\si_1 ),
(\cdot,s_3),(\cdot,\si_2 ),(\cdot,\si_3 ) }) \|_{1,\infty}
\Big(  \max_{\si_1'\in \Si}  \| C_{\si_1'}^{(j)}{\sst(x )}\|_1 \Big)\cr
&\hskip 3cm 
\Big(  \max_{\si'\in \Si}  \| C_{\si'}^{(j)}{\sst (x )}\|_\infty \Big)^2  \ \max_{\si_1'', \si_2'',\si_3'' \in \Si}\sum_{s_4}\| \hat \om_4 ({\sst (\cdot,\si_2'' ),(\cdot,\si_3'' ),
(\cdot,\si_1'' ),(\cdot,s_4) }) \|_{1,\infty}\cr
&\hskip 1cm\le O\big(M^j\big(\sfrac{\fl}{M^j}\big)^2\big)
\tn\hat\om_{4}\tn_{3,\Si}
 \sum_{\si_1,\si_2,\si_3 \in \Si }
\| \hat \Ga' ({\sst (\cdot,s_1),(\cdot,s_2),(\cdot,\si_1 ),
(\cdot,s_3),(\cdot,\si_2 ),(\cdot,\si_3 ) }) \|_{1,\infty}
}$$
since, for a given sector $\si_i$, there are three sectors $\si_i'$ and three sectors $\si_i''$ with $\si_i \cap \si_i' \ne \emptyset$, 
$\si_i \cap \si_i''\ne \emptyset$. 
We see that the contribution to $\tn\hat\Ga \tn_{3,\Si}$ with 
fixed $s_1,s_2,s_3$ is bounded by 
$$
O\big(M^j\big(\sfrac{\fl}{M^j}\big)^2\big)
\tn\hat\om_{4}\tn_{3,\Si}\tn\Ga'\tn_{3,\Si}
=O\big( M^j\ \sfrac{\fl^2}{M^{2j}}\ 1\  \sfrac{M^j}{\fl} \big) 
=O(\fl)
$$
and, taking the sector sum over $s_2,s_3$,
the contribution to $\tn\hat\Ga \tn_{1,\Si}$ with fixed $s_1$ is bounded by 
$$
O\big(M^j\big(\sfrac{\fl}{M^j}\big)^2\big)
\tn\hat\om_{4}\tn_{3,\Si}\tn\Ga'\tn_{1,\Si}
=O\big( M^j\ \sfrac{\fl^2}{M^{2j}}\ 1\  \sfrac{M^j}{\fl^2} \big) 
=O(1)
$$
The contributions with other sectors fixed are estimated in the same way. 
}

\goodbreak

\example{\STM\exNPsectorizescalechange (Sectorization and Change of Scale)}{

Each time the  renormalization group flows to a new scale, the associated
sector decomposition changes. Therefore, in the notation of \S\CHintroOverview,
we have to choose new sectorized representatives for $w_2,\ w_4,\ \cdots$ and compare the norms of these new
sectorized representatives with respect to the new sectorization to the  
norms of the old sectorized representatives with respect to the old
sectorization. To isolate this problem, suppose that $j'>j$, that $\Si$ and $\Si'$, respectively, are sectorizations of scale $j$ and $j'$, respectively,
of length $\fl$ and $\fl'$, respectively and that $\fl'<\fl$. Furthermore,
let $\om_{2n}({\sst (p_1,s_1),\ \cdots,\ (p_n,s_n),\ (q_1, s_{n+1})
,\ \cdots,\ (q_n, s_{2n})})$ be a $\Si$--sectorized representative of $w_{2n}$.
Using the partition of unity $\{\chi_{s'}\}_{s'\in\Si'}$ subordinate so $\Si'$,
one constructs the $\Si'$--sectorized representative
$$
\om'_{2n}({\sst (p_1,s'_1),\ \cdots,\ (q_n, s'_{2n})})
=\sum_{{s'_i\in\Si'\atop s'_i\cap s_i\ne\emptyset}\atop 1\le i\le 2n}
\chi_{s'_1}(p_1)\cdots\chi_{s'_{2n}}(q_n)\ 
\om_{2n}({\sst (p_1,s_1),\ \cdots,\ (q_n, s_{2n})})
$$
of $w_{2n}$. The norm $\TN \hat\om'_{2n}\TN _{1,\Si'}$ 
of (\eqnOVtriponesi) is defined
in terms of a supremum over $s'\in\Si'$ of a sum over
$$\eqalign{
\Mom_i(s')=\big\{\,(s_1',\cdots,s_{2n}')\in\Si'^{2n}\,\big|\,
s_i'=s' &\hbox{ and there exist }p_\ell\in s'_\ell,\ q_\ell\in s'_{n+\ell},
\ 1\le \ell\le {n}\cr
&\hbox{ such that }p_1+\cdots+p_n=q_1+\cdots+q_n\,\big\}\cr
}$$
It is natural to partition the sum over $\Mom_i(s')$ into a sum over
$(s_1,\cdots,s_{2n})\in\Si^{2n}$ followed by a sum over elements of 
$\Mom_i(s')$ that obey $s_\ell\cap s'_\ell\ne\emptyset$ for all
$ 1\le \ell\le {2n}$.
Now, we will try to motivate that, ``morally'', for any fixed 
$s_1,\ \cdots\,s_{2n}\in\Si$, there are at most 
$\ \big[\const\sfrac{\fl}{\fl'}\big]^{{2n}-3}\ $ elements of
$\Mom_{i}(s')$ obeying $s_\ell\cap s'_\ell\ne\emptyset$ for all
$ 1\le \ell\le {2n}$. We may assume that $i=1$. Then $s'_1$ must be $s'$.
Denote by $I_\ell$ the interval  on the Fermi curve $\FC$ that has length $\fl+2\fl'$ and is centered on $s_\ell\cap\FC$. 
If $s'\in\Si'$ intersects $s_\ell$, then
$s'\cap\FC$ is contained in $I_\ell$.
Every sector in $\Si'$ contains an interval of $\FC$ of
length $\sfrac{3}{4}\fl'$ that does not intersect any other sector in $\Si'$.
(The specific number $\sfrac{3}{4}$ comes from Definition \defNPsectors. 
It is not important.)
 At most $[\sfrac{4}{3}\sfrac{\fl+2\fl'}{\fl'}]$ of these 
``hard core'' intervals can be contained in $I_\ell$. Thus there are at most
$[\sfrac{4}{3}\sfrac{\fl}{\fl'}+3]^{{2n}-3}$ choices for $s_i',\ i\ne 1,n,2n$. 

Fix $s_i',\ i\ne n,2n$. Once $s'_n$ is chosen, $s'_{2n}$ is essentially
uniquely determined by conservation of momentum. But the desired bound
demands more. It says, roughly speaking, that both
$s'_{n}$ and $s'_{2n}$ are essentially uniquely determined. 
As $p_\ell$ and $q_\ell$ run over $s'_\ell$ and $s'_{n+\ell}$,
respectively, for $1\le\ell\le n-1$,
the sum $\p_1+\cdots+\p_{n-1}-\q_1-\cdots-\q_{n-1}$ runs over a small set 
centered on some point $\k$. In order for $(s'_1,\cdots,s'_{2n})$ 
to be in $\Mom_1(s')$, there must exist 
$\p'\in s'_{n}\cap\FC$ and $\q'\in s'_{2n}\cap\FC$ with 
$\q'-\p'$ very close to $\k$. But $\q'-\p'$ is a secant joining 
two points of the Fermi curve $\FC$. We have assumed that $\FC$ is strictly
convex.%
\vadjust{\hfil\figput{convsect}\hfill} 
Consequently, for any given $\k\ne 0$ in $\bbbr^2$ there exist at most two
pairs $(\p',\q')\in\FC^2$ with $\q'-\p'=\k$. So, if $\k$ is not near the 
origin, $s'_{n}$ and $s'_{2n}$ are almost uniquely determined. If $\k$ is
close to zero, then $\p_1+\cdots+\p_{n-1}-\q_1-\cdots-\q_{n-1}$ must
be close to zero and the number of allowed $s_i',\ i\ne n,2n$
is reduced. Careful application of these types of arguments yields
(Proposition \propOSresectorI\ of [FKTo4])
$$\eqalign{
\TN \hat\om'_{2n} \TN_{1,\Si'}
&\le\abcst^n\,\big[\sfrac{\fl}{\fl'}\big]^{2n-3}
\Big( \tn \hat\om_{2n}\tn_{1,\Si} + \sfrac{1}{\fl}\, \tn \hat\om_{2n}\tn_{3,\Si} \Big)\cr
\tn \hat\om'_{2n} \tn_{3,\Si'}
&\le\abcst^n\,\big[\sfrac{\fl}{\fl'}\big]^{2n-4}
 \tn \hat\om_{2n}\tn_{3,\Si} \cr
}\EQN\eqnSMPresector$$
For this reason, we usually estimate the combination 
$\tn \hat\om_{2n}\tn_{1,\Si} + \sfrac{1}{\fl}\, \tn \hat\om_{2n}\tn_{3,\Si}$.
The analog of (\eqnOVsectorvertexreq) and (\eqnOVsectorvertexthreereq)
for this combination is 
$$
\tn \hat\om_{2n}\tn_{1,\Si} + \sfrac{1}{\fl}\, \tn \hat\om_{2n}\tn_{3,\Si}
 \ \ {\rm is\ of\ order\ } 
\sfrac{M^{j(n-2)}}{\fl^{n-1}}\ \ {\rm for\ all\ }n 
$$
Thanks to (\eqnSMPresector), this bound is preserved, for $n>2$, when 
the sector decomposition is refined.
$$\eqalignno{
&M^{-(j+1)(n-2)}{\fl'}^{n-1}
\Big(\TN \hat\om'_{2n} \TN_{1,\Si'}
+ \sfrac{1}{\fl'}\, \TN \hat\om'_{2n}\TN_{3,\Si'} \Big)\cr
&\le\abcst^n\, M^{-(j+1)(n-2)}{\fl'}^{n-1}
\big[\sfrac{\fl}{\fl'}\big]^{2n-3}
\Big( \tn \hat\om_{2n}\tn_{1,\Si} + \sfrac{1}{\fl}\, \tn \hat\om_{2n}\tn_{3,\Si} \Big)\cr
&=\abcst^n\, M^{-(n-2)}
\big[\sfrac{\fl}{\fl'}\big]^{n-2}\ \ 
 M^{-j(n-2)}\fl^{n-1}
\Big( \tn \hat\om_{2n}\tn_{1,\Si} + \sfrac{1}{\fl}\, \tn \hat\om_{2n}\tn_{3,\Si} \Big)\cr
&=\abcst^n\, M^{-(1-\aleph)(n-2)}\ \ 
 M^{-j(n-2)}\fl^{n-1}
\Big( \tn \hat\om_{2n}\tn_{1,\Si} + \sfrac{1}{\fl}\, \tn \hat\om_{2n}\tn_{3,\Si} \Big)\cr
}$$
We even have a small factor, $ M^{-(1-\aleph)(n-2)}$, available for eating
up constants like $\abcst^n$.

}

\vfill\eject

\titlea{References}\PG\pgNPIref

\item{[AM]} N. W. Ashcroft, N. D. Mermin, {\it Solid State Physics},
Saunders College (1976).
\smallskip%
\item{[BG]} G. Benfatto and G. Gallavotti, {\it Renormalization Group},
Physics Notes, Vol. 1, Princeton University Press (1995).
\smallskip%
\item{[C]} E. R. Caianello, {\bf Number of Feynman Diagrams and Convergence},
Nuovo Cimento {\bf 3}, 223-225 (1956).
\smallskip%
\item{[DMR]} M. Disertori, J. Magnen, V. Rivasseau, 
{\bf  Interacting Fermi liquid in three dimensions at finite temperature: 
Part I: Convergent Contributions}, preprint cond-mat/0012270. 
\smallskip%
\item{[DR1]} M. Disertori, V. Rivasseau, 
{\bf  Interacting Fermi liquid in two dimensions at finite temperature. 
Part I: Convergent Attributions}, Communications in Mathematical Physics  {\bf 215}, 251--290 (2000). 
\smallskip%
\item{[DR2]} M. Disertori, V. Rivasseau, 
{\bf  Interacting Fermi liquid in two dimensions at finite temperature. 
Part II: Renormalization}, Communications in Mathematical Physics  {\bf 215}, 291--341 (2000). 
\smallskip%
\item{[FKLT1]} J. Feldman, H. Kn\"orrer, D. Lehmann, E. Trubowitz, 
{\bf Fermi Liquids in Two-Space Dimensions}, in {\it Constructive Physics},
 V. Rivasseau ed., 
Springer Lecture Notes in Physics 446, 267-300 (1995).
\smallskip%
\item{[FKLT2]} J. Feldman, H. Kn\"orrer, D. Lehmann, E. Trubowitz, 
{\bf Are There Two Dimensional Fermi Liquids?}, in {\it Proceedings of the XIth
International Congress of Mathematical Physics}, D. Iagolnitzer ed., 
440-444 (1995).
\smallskip%
\item{[FKLT3]} J. Feldman, H. Kn\"orrer, D. Lehmann, E. Trubowitz, 
{\bf A Class of Fermi Liquids}, in {\it Particles and Fields}, G. Semenoff
and L. Vinet eds., CRM Series in Mathematical Physics, 35-62 (1999),
Springer-Verlag, New York.
\smallskip%
\item{[FKTa]} J. Feldman, H. Kn\"orrer, E. Trubowitz, 
{\bf Asymmetric Fermi Surfaces for Magnetic Schr\"odinger Operators}, 
      Communications in Partial Differential Equations {\bf 25} (2000),
      319-336.
\smallskip%
\item{[FKTcf]} J. Feldman, H. Kn\"orrer, E. Trubowitz,  
{\bf A Nonperturbative Representation for Fermionic Correlation Functions},
Communications in Mathematical Physics,  {\bf 195}, 465-493 (1998).
\smallskip%
\item{[FKTf2]} J. Feldman, H. Kn\"orrer, E. Trubowitz, 
{\bf A Two Dimensional Fermi Liquid, Part 2: Convergence}, preprint.
\smallskip%
\item{[FKTf3]} J. Feldman, H. Kn\"orrer, E. Trubowitz, 
{\bf A Two Dimensional Fermi Liquid, Part 3: The Fermi Surface}, preprint.
\smallskip%
\item{[FKTffi]} J. Feldman, H. Kn\"orrer, E. Trubowitz, 
     {\bf Fermionic Functional Integrals and the Renormalization Group},
     Andr\'e Aisenstadt Monograph Series, to appear.
\smallskip%
\item{[FKTl]} J. Feldman, H. Kn\"orrer, E. Trubowitz, 
 {\bf Particle--Hole Ladders}, preprint.
\smallskip%
\item{[FKTo1]} J. Feldman, H. Kn\"orrer, E. Trubowitz, 
{\bf Single Scale Analysis of Many Fermion Systems, Part 1: Insulators}, preprint.
\smallskip%
\item{[FKTo2]} J. Feldman, H. Kn\"orrer, E. Trubowitz, 
{\bf Single Scale Analysis of Many Fermion Systems, Part 2: The First Scale}, preprint.
\smallskip%
\item{[FKTo3]} J. Feldman, H. Kn\"orrer, E. Trubowitz, 
{\bf Single Scale Analysis of Many Fermion Systems, Part 3: Sectorized Norms}, preprint.
\smallskip%
\item{[FKTo4]} J. Feldman, H. Kn\"orrer, E. Trubowitz, 
{\bf Single Scale Analysis of Many Fermion Systems, Part 4: Sector Counting}, preprint.
\smallskip%
\item{[FKTr1]} J. Feldman, H. Kn\"orrer, E. Trubowitz, 
{\bf Convergence of Perturbation Expansions in Fermionic Models, Part 1: Nonperturbative Bounds}, preprint.
\smallskip%
\item{[FKTr2]} J. Feldman, H. Kn\"orrer, E. Trubowitz, 
{\bf Convergence of Perturbation Expansions in Fermionic Models, Part 2: Overlapping Loops}, preprint.
\smallskip%
\item{[FMRT]} J. Feldman, J. Magnen, V. Rivasseau and E.Trubowitz,
{\bf An Infinite Volume Expansion for Many Fermion Green's Functions},
Helvetica Physica Acta, {\bf 65} (1992) 679-721.
\smallskip%
\item{[FST1]} J.\ Feldman, M.\ Salmhofer, and E.\ Trubowitz,
{\bf Perturbation Theory around Non--Nested Fermi Surfaces, 
I. Keeping the Fermi Surface fixed}, Journal of Statistical Physics,
 {\bf 84} (1996) 1209-1336.
\smallskip%
\item{[FST2]} J.\ Feldman, M.\ Salmhofer, and E.\ Trubowitz,
{\bf Regularity of the Moving Fermi Surface: RPA Contributions}, 
Communications on Pure and Applied Mathematics,  {\bf LI}, 1133-1246 (1998).
\smallskip%
\item{[FST3]} J.\ Feldman, M.\ Salmhofer, and E.\ Trubowitz,
{\bf  Regularity of Interacting Nonspherical Fermi Surfaces: The Full 
Self--Energy}, Communications on Pure and Applied Mathematics, {\bf LII},
 273-324  (1999).
\smallskip%
\item{[FST4]} J.\ Feldman, M.\ Salmhofer, and E.\ Trubowitz,
{\bf An inversion theorem in Fermi surface theory}, 
Communications on Pure and Applied Mathematics, {\bf LIII}, 1350--1384 (2000).
\smallskip%
\item{[FW]} A.L. Fetter and J.D. Walecka, {\it Quantum Theory of Many-Particle
Systems}, McGraw-Hill, 1971.
\smallskip%
\item{[MR]} J. Magnen, V. Rivasseau, {\bf  A Single Scale Infinite 
Volume Expansion for Three-Dimensional Many Fermion Green's
Functions}, Mathematical Physics Electronic Journal {\bf 1}, No. 3 (1995).
\smallskip%
\item{[MCD]} W. Metzner, C. Castellani and C. Di Castro, 
{\bf  Fermi Systems with Strong Forward Scattering}, 
Adv. Phys. {\bf 47}, 317-445 (1998).
\smallskip%
\item{[PT]}  J. P\"oschel, E. Trubowitz, {\it Inverse Spectral Theory}, 
Academic Press (1987). 
\item{[S]}  M. Salmhofer, {\bf Continuous renormalization for Fermions
and Fermi liquid theory}, Communications in Mathematical Physics  {\bf 194}, 249--295 (1998).

\vfill\eject

\hoffset=-0.2in
\titlea{Notation}\PG\pgNPInot
\vfil
\centerline{
\vbox{\offinterlineskip
\hrule
\halign{\vrule#&
         \strut\hskip0.05in\hfil#\hfil&
         \hskip0.05in\vrule#\hskip0.05in&
          #\hfil\hfil&
         \hskip0.05in\vrule#\hskip0.05in&
          #\hfil\hfil&
           \hskip0.05in\vrule#\cr
height2pt&\omit&&\omit&&\omit&\cr
&Not'n&&Description&&Reference&\cr
height2pt&\omit&&\omit&&\omit&\cr
\noalign{\hrule}
height2pt&\omit&&\omit&&\omit&\cr
&$\cE$&&counterterm space&&Definition \defNPCTMSpace&\cr
height2pt&\omit&&\omit&&\omit&\cr
&$r_0$&&number of $k_0$ derivatives tracked&&following (\eqnNPinteraction)&\cr
height2pt&\omit&&\omit&&\omit&\cr
&$r$&&number of $\k$ derivatives tracked&&following (\eqnNPinteraction)&\cr
height2pt&\omit&&\omit&&\omit&\cr
&$M$&&scale parameter, $M>1$&&before Definition \defNPscales&\cr
height2pt&\omit&&\omit&&\omit&\cr
&$\nu^{(j)}(k)$&&$j^{\rm th}$ scale function&&Definition \defNPscales&\cr
height2pt&\omit&&\omit&&\omit&\cr
&$\nu^{(\ge j)}(k)$&&$\smsum_{i\ge j}\nu^{(j)}(k)$&&Definition \defNPscales&\cr
height2pt&\omit&&\omit&&\omit&\cr
&$n_0$&&degree of asymmetry&&Definition \defNPstrongasymm&\cr
height2pt&\omit&&\omit&&\omit&\cr
&$\tn \ \cdot\ \tn_{1,\Si}$&&no derivatives, all but $1$ sector summed
&&(\eqnOVtriponesi)&\cr
height4pt&\omit&&\omit&&\omit&\cr
&$\tn \ \cdot\ \tn_{3,\Si}$&&no derivatives, all but $3$ sectors summed
&& (\eqnOVtripthreesi) &\cr
height2pt&\omit&&\omit&&\omit&\cr
&$J$&&particle/hole swap operator&&(\eqnNPjdef)&\cr
height2pt&\omit&&\omit&&\omit&\cr
&$\Om_S(\cW)(\phi,\psi)$
&&$\log\sfrac{1}{Z} \int  e^{\cW(\phi,\psi+\ze)}\,d\mu_{S}(\ze)$
&&Definition \defNPrengroupmap&\cr
height2pt&\omit&&\omit&&\omit&\cr
&$\tilde \Om_C(\cW)(\phi,\psi)$
&&$\log \sfrac{1}{Z}\int e^{\phi J\ze}\,e^{\cW(\phi,\psi +\ze)} d\mu_C(\ze)$
&&Definition \defNPrengroupmap&\cr
height2pt&\omit&&\omit&&\omit&\cr
}\hrule}}

\end